\documentclass[a4paper,12pt]{amsart}

\usepackage[final]{graphicx}
\usepackage{lscape}
\usepackage{natbib}
\usepackage{amssymb}
\usepackage[pdftex]{thumbpdf}

\theoremstyle{plain}

\theoremstyle{definition}

\def\aap{A\&A}

\def\apj{ApJ}
\def\apjs{ApJS}

\def\apjl{ApJ}

\def\mnras{MNRAS}
\def\nat{Nature}
\def\pasp{PASP}

\def\gca{Geochim.\,Cosmochim.\,Acta}

\def\planss{P\&SS}
\def\araa{Ann.Rev.Astron.Astroph.}

 \title[Astrochemistry: the issue of molecular complexity]{Astrochemistry: the issue of molecular complexity in astrophysical environments$^\dagger$}
 \thanks{$^\dagger$Based partly on lectures given at the University of Li\`ege since January 2007.}

 \author[M. De Becker]{Micha\"el De Becker}
 \thanks{}
 \address[M. De Becker]{\newline
 Micha\"el De Becker\newline 
 Department of Astrophysics, Geophysics \& Oceanography\newline
 \&\newline
 Li\`ege Space Research Institute (LiSRI)\newline
 University of Li\`ege\newline
 17, all\'ee du 6 Ao\^ut, B\^at. B5c\newline
 B-4000 Li\`ege (Sart-Tilman)\newline
 Belgium\newline
 }
 \email{Michael.DeBecker@ulg.ac.be} 

 \date{22 May 2013}

\begin{document}
\maketitle

\begin{abstract}
Astrochemistry aims at studying chemical processes in astronomical environments. This discipline -- located at the crossroad between astrophysics and chemistry -- is rapidly evolving and explores the issue of the formation of molecules of increasing complexity in particular physical conditions that deviate significantly from those frequently encountered in chemistry laboratories. The main goal of this paper is to provide an overview of this discipline. So far, about 170 molecules have been identified in the interstellar medium (ISM). The presence of this molecular diversity constitutes a firm evidence that efficient formation processes are at work in the interstellar medium. This paper aims at summarizing most of present ideas that are explored by astrochemists to investigate the chemistry taking place in various astronomical environments, with emphasis on the particular conditions which are met in space (including radiation fields, cosmic-rays, low densities...). The more ambitious question of the molecular complexity is addressed following two approaches presented to be converging. The first approach considers the growing complexity starting from the most simple chemical species in interstellar environments, and the second approach envisages successive precursors of the most complex species commonly found on Earth, and in particular in our biochemistry. The issue of molecular complexity constitutes one of the main modern scientific questions addressed by astrochemistry, and it is used as a guideline across this paper.\\

\end{abstract}

\section{Astrochemistry in the context of modern sciences}

\subsection{What is astrochemistry?}
Astrochemistry is the {\it science devoted to the study of the chemical processes at work in astrophysical environments, including the interstellar medium, comets, circumstellar and circumplanetary regions}.\\

It is important to insist on the difference between two fields that are often confused, at least in the astrophysicists community, i.e. chemistry and molecular spectroscopy. Chemistry is a science interested in processes transforming molecular species, and molecular spectroscopy is a technique used to make a qualitative and quantitative census of molecules in a given environment. In astronomy, the ongoing exploration of the sky leads to the discovery of many molecules, mainly in the interstellar medium and in circumstellar environments. These discoveries are made through molecular spectroscopy studies. Beside that, researchers are trying to understand the processes likely to lead to the formation of these molecules, as a function of the properties of their environment, along with the potential filiations that exist between chemical species found at a given place. This is typically what astrochemists are doing, on the basis of the very valuable information gathered by molecular spectroscopists.\\

Astrochemistry constitutes therefore some kind of overlap between chemistry and astrophysics, its two parent disciplines. It should also be kept in mind that this inter-relation is not univoque. On the one hand, astrochemistry extends the domains where chemistry can be applied. Chemistry is not limited to a laboratory science, and astrochemistry opens the door of astronomy, in a similar way as biochemistry opens that of living organisms. On the other hand, astrochemical processes are expected to be important for many issues of astrophysics in the sense that chemical processes can be used as valuable tracers for physical properties of astronomical environments to which they are intimately dependent. Astrochemistry is the natural pluridisciplinary answer that modern science has developed to address scientific questions that are totally out of the scope of existing individual scientific disciplines.

\subsection{The challenges of astrochemistry}
In short, the aim of astrochemistry is to understand the chemical processes at work in space. However, some more specific objectives are worth discussing briefly before going through the details in the next sections.

One of the tasks of astrochemists is to understand the impact of the astrophysical environment on chemical processes responsible for the formation of molecules. In this context, their studies aim at understanding the level of complexity that can be reached by molecular species in space. This is important in the sense that chemical complexity is a necessary condition for the development of life in the Universe.

What are the most complex molecules that can be formed in space? How do they form? Are these molecules likely to resist to the collapse of protostellar clouds leading to the formation of stellar systems? What is the relation between the molecular populations found in comets and in the interstellar medium? Is the interstellar medium the reactor where some molecules important for life are formed? These questions summarize the main scientific issues that are motivating research in astrochemistry.

The circumstances under which astrochemistry is taking place are somewhat different from those of laboratory chemistry, and these differences deserve a few words. Astrochemistry investigates indeed chemical processes under unusual conditions (low densities, low temperatures, exposition to strong radiation fields and to high energy particles often called cosmic rays...). These conditions are difficult to reproduce in laboratory, and this constitutes a severe problem if one wants to explore the reactivity on the basis of laboratory experiments. In addition, the study of chemical reactions in space is based on an uncomplete census of potential reaction partners. Molecular spectroscopy studies allow us to obtain many information on the molecular populations in space, but the task in not easy. Molecular spectroscopists have to deal with blends of molecules whose spectroscopic identification is often difficult. As molecular complexity increases, this difficulty increases rapidly, and we do not have the opportunity to separate compounds in order to facilitate their identification. The dependence on a lot of physical parameters which are not at all controlled is an issue as well. The physical parameters can not be tuned at will in order to study the influence of a parameter or another on reactivity.

\subsection{Temporary census of astromolecules \label{census}}

Molecular spectroscopy studies led to the detection of many molecules in the interstellar medium. Most of them have been discovered via their rotational signature from radio to far-infrared frequencies, even though some have also been detected in the visible and near-infrared domains. The first interstellar molecules identified thanks to their sharp bands in gas phase were CH \citep{ch-swings}, CN \citep{ism-cn} and CH$^+$ \citep{ch+-douglas}, and the first polyatomic molecule (NH$_3$) was discovered in the end of the 60's \citep{cheung-nh3}. Since then, mostly thanks to significant progresses in millimeter astronomy, many other molecules of increasing complexity have been firmly identified in the ISM. So far, this amounts up to about 170 molecules, and much more are certainly waiting to be discovered in the future. Lists of identified molecules, sorted by increasing number of constituting atoms, are given Table\,\ref{molcensus}. It should be noted that, up to now, most of the molecules that have been detected in the interstellar medium or elsewhere in space are neutral species. Some cations have also been found, along with only a few anions whose discovery is quite recent.

\begin{table}[h]
\begin{center}
\caption{Temporary census of interstellar molecules.\label{molcensus}}
\vspace*{3mm}
\begin{tabular}{ll}
\hline
2 atoms & AlCl, AlF, AlO, C$_2$, CF$^+$, CH, CH$^+$, CN, CN$^-$, CO,\\
 & CO$^+$, CP, CS, FeO, H$_2$, HCl, HCl$^+$, HF, NH, KCl\\
 & N$_2$, NO, NS, NaCl, O$_2$, OH, OH$^+$, PN, SH, SH$^+$, \\
 & SO, SO$^+$, SiC, SiN, SiO, SiS, PO \\
\hline
3 atoms & AlNC, AlOH, C$_3$, C$_2$H, C$_2$O, C$_2$P, C$_2$S,CO$_2$, H$_3^+$, \\
 & CH$_2$, H$_2$Cl$^+$ , H$_2$O, H$_2$O$^+$, HO$_2$ H$_2$S, HCN, HCO, \\
 & HCO$^+$, HCS$^+$, HCP, HNC, HN$_2^+$, HNO, HOC$^+$, \\
 & KCN, MgCN, NH$_2$, N$_2$H$^+$, N$_2$O, NaCN, OCS, SO$_2$, \\
 & c-SiC$_2$, SiCN, SiNC, FeCN \\
\hline
4 atoms & C$_2$H$_2$, l-C$_3$H, c-C$_3$H, C$_3$N, C$_3$O, C$_3$S, H$_3$O$^+$, H$_2$O$_2$,\\
 & H$_2$CN, H$_2$CO, H$_2$CS, HCCN, HCNH$^+$, HCNO, HOCN, \\
 & HOCO$^+$, HNCO, HNCS, HSCN, NH$_3$, SiC$_3$, PH$_3$ \\
\hline
5 atoms & C$_5$, CH$_4$, c-C$_3$H$_2$, l-C$_3$H$_2$, H$_2$CCN, H$_2$C$_2$O,\\
 & H$_2$CNH, H$_2$COH$^+$, C$_4$H, C$_4$H$^-$, HC$_3$N, HCCNC,\\
 &  HCOOH, NH$_2$CN, SiC$_4$, SiH$_4$, HCOCN,\\
 &  HC$_3$N$^-$, HNCNH, CH$_3$O \\
\hline
6 atoms & c-H$_2$C$_3$O, C$_2$H$_4$, CH$_3$CN, CH$_3$NC, CH$_3$OH, \\
 & CH$_3$SH, l-H$_2$C$_4$, HC$_3$NH$^+$, HCONH$_2$, C$_5$H, HC$_2$CHO,\\
 & HC$_4$N, CH$_2$CNH, C$_5$N$^-$, HNCHCN \\
\hline
7 atoms &  c-C$_2$H$_4$O, CH$_3$C$_2$H, H$_3$CNH$_2$,  CH$_2$CHCN, H$_2$CHCOH,\\
 & C$_6$H, C$_6$H$^-$, HC$_4$CN, CH$_3$CHO, HC$_5$N$^-$\\
\hline
8 atoms & H$_3$CC$_2$CN, H$_2$COHCOH, CH$_3$OOCH, CH$_3$COOH,\\
 & C$_6$H$_2$, CH$_2$CHCHO, CH$_2$CCHCN, C$_7$H,\\
 &  NH$_2$CH$_2$CN, CH$_3$CHNH \\
\hline
9 atoms & CH$_3$C$_4$H, CH$_3$OCH$_3$, CH$_3$CH$_2$CN, CH$_3$CONH$_2$,\\
 &  CH$_3$CH$_2$OH, C$_8$H, HC$_6$CN, C$_8$H$^-$, CH$_2$CHCH$_3$ \\
\hline
$>$\,9 atoms &  CH$_3$COCH$_3$, CH$_3$CH$_2$CHO, CH$_3$C$_5$N, \\
 & HC$_8$CN, CH$_3$C$_6$H, CH$_3$OC$_2$H$_5$, HC$_{10}$CN, \\
 & C$_6$H$_6$, C$_2$H$_5$OCHO, C$_3$H$_7$CN, C$_{60}$, C$_{70}$, C$_{60}^+$ \\
\hline
Deuterated & HD, H$_2$D$^+$, HDO, D$_2$O, DCN, DCO, DNC, N$_2$D$^+$,\\
 &  NHD$_2$, ND$_3$, HDCO, D$_2$CO, CH$_2$DCCH, CH$_3$CCD, \\
 & D$_2$CS \\
\hline
\end{tabular}
\end{center}
\end{table}

\subsection{Molecular complexity \label{molcomp}}

The present (but still evolving) census\footnote{The census of firmly identified interstellar molecules presented here is partly taken from the following website {\tt http://www.astrochymist.org}, created and maintained by D.E. Woon.} of interstellar molecules leads automatically to a question which lies at the core of astrochemistry, i.e. the question of {\it molecular complexity}. Such a concept deserves a few words before exploration in the context of astrochemistry. 

We know that the primordial nucleosynthesis, at the beginning of the Universe, produced mainly hydrogen and helium, along with some lithium. Heavier elements are produced by the nucleosynthetic activity of stars (nuclear fusion, fast a slow neutron captures followed by $\beta$-decay) and are expelled in the interstellar medium through various episodes of their evolution. On the one hand, we know that some of these atoms react to form simple molecules, with the two most abundant being H$_2$ and CO. Chemical reactions are likely to produce increasingly complex molecules, as illustrated in the census of astromolecules given in Section\,\ref{census}. On the other hand, we know from modern chemistry and biochemistry that chemical processes can reach a very high level of complexity, with the synthesis of large and multi-functionalized species presenting elaborate properties. The most simple astromolecules are located at the bottom of some kind of ladder of chemical complexity, and several rungs of the ladder are already occupied by many polyatomic species identified in several astronomical environments. However, a huge gap still has to be filled if one wants to understand the processes leading to the top of the ladder. Even though substantial steps towards the highest chemical complexity levels were probably made at the prebiotic ages of the Earth, recent findings provided by astrochemistry suggest that at least some parts of the gap could have been filled in space.\\

The concept of complexity by itself is hard to define, as exemplified by the numerous approaches adopted to qualify it. In the context of this paper, we will adopt a vision of molecular complexity that is in agreement with that introduced by \citet{milman}. This measure of complexity, based on information theory, takes into consideration the main elements that should a priori contribute to the intuitive idea that one could develop on what should be a complex molecular species: the number of constituting atoms, the number and diversity of functional groups, and the presence (if any) of stereogenic centers. Even though such a measure of complexity, in the form of molecular complexity indices, was motivated mainly by structural description considerations, it is important to emphasize the strong correlation that exists between structural complexity and the diversity in the capability of a molecule to be involved efficiently in chemical reactions of any kind.

The exploration of the successive levels of the molecular complexity scale coincides indeed with the increase of emerging properties of chemical species. For instance, it is quite common sense that molecular hydrogen is {\it less complex} than methane that is made of five atoms (with respect to only two for molecular hydrogen). On the other hand, it is tentative to attribute a higher level of complexity to CO with respect to H$_2$. Both species contain the same number of atoms, but carbone monoxide is not homonuclear: as a result, a dipolar moment appears, therefore affecting significantly its reactivity. In addition, the molecular orbitals of CO are populated by a larger number of electrons, likely to be involved in processes leading to the formation of new chemical bondings. For the same reason, methane is viewed as {\it less complex} than methanol which includes an oxygen atom in its structure (one additional atom, plus a permanent dipolar moment, leading both to an increased complexity). The existence of a functional group in organic compounds opens the door for more specific reactivities (see for instance the specific reactivity of carbonyl compounds dominantly influenced by the electric dipole in the double bond between carbon and oxygen), and the simultaneous presence of more than one functional group in a given molecule gives undoubtedly access to higher levels of molecular complexity (see for instance the successive formation of peptidic bonds between amino acids leading to the formation of proteins). As a further step, the inclusion of several heteroatoms in organic compounds increases the possibility to form stereogenic centers, adding another level of diversity in molecular species. As a result, different enantiomers can participate differently in enantiospecific processes, undoubtedly increasing the level of molecular complexity. As an example, the simultaneous presence of several functional groups in carbohydrates generates stereogenic centers, justifying for instance the pivotal discrimination between D and L sugars, and the same is true for amino acids.

At least qualitatively, and to some extent intuitively, the exploration of molecular populations in various astronomical environments follows some kind of {\it complexity line} along which molecular species can be located as function of their content and properties. In this context, one may consider two different approaches while exploring molecular complexity:
\begin{enumerate}
\item[1.] {\bf Bottom-up approach.} It consists of exploring molecular complexity from the most simple species to the more complex ones. This is typically the approach followed when basic processes involving atoms are explored in order to produce simple diatomic species, likely in turn involved in the formation of polyatomic species. These processes are reviewed in Sect.\,\ref{BU}.
\item[2.] {\bf Top-down approach.} This approach starts with high complexity species, and aks the question of the likely precursor of such molecules in order to successively identify probable filiations connecting complex molecules with much more simple ones. A discussion based on this approach is given in Sect.\,\ref{TD}.
\end{enumerate}
It is obvious that these two approaches are by no means independent, as they address the same issue but starting from opposite initial conditions. This is why one should see them as {\it two converging approaches}. The main body of this paper will consider successively these two approaches (in Sections\,\ref{BU} and \ref{TD}, respectively), emphasizing the participation of the main components contributing to molecular complexity, i.e the number of constituting atoms, the number and diversity of functional groups, and the potential presence of stereogenic centers.

\section{The bottom-up approach}\label{BU}

The interstellar medium is not empty. It is populated by matter in different forms: neutral atoms, atomic ions, electrons, molecules (neutral and ionic), and dust particles. From the physical chemistry point of view, the ISM can be seen as a gas phase volume, where gas phase kinetics can be applied. In such circumstances, it is obvious that the low particle densities constitute an issue. In addition, interstellar temperatures can be very low, which is not in favor of fast chemical kinetics. As a consequence, one can expect reaction time-scales to be very long and it is a priori difficult to envisage the formation of complex molecules. However, the current census of interstellar molecules provides strong evidence that efficient chemical processes are at work in space, despite the a priori unfavorable gas phase kinetics. This apparent contradiction is lifted when the catalytic effect of dust grains is taken into account in astrochemical models. Dust particles constitute indeed a solid-state component in the ISM (especially in dense clouds, \citealt{pagani2010}), whose surface can act as catalyst responsible for a considerable activation of interstellar chemical kinetics. Both approaches -- gas phase and grain surface -- are discussed below before being combined in order to provide a more complete view of the physical chemistry that is governing astrochemical processes.

\subsection{Gas-phase processes}
During a chemical -- or photochemical -- process, chemical species are destroyed and other chemical species are formed. We can consider formation and destruction rates for various kinds of reactions:
\begin{enumerate}
\item[-] unimolecular reactions: A $\rightarrow$ C
$$
-\frac{d\,\mathrm{n(A)}}{d\,\mathrm{t}} = \mathrm{k}\,\mathrm{n(A)} = \frac{d\,\mathrm{n(C)}}{d\,\mathrm{t}}
$$
\item[-] bimolecular reactions: A + B $\rightarrow$ C
$$
-\frac{d\,\mathrm{n(A)}}{d\,\mathrm{t}} = \mathrm{k}\,\mathrm{n(A)}\,\mathrm{n(B)} = \frac{d\,\mathrm{n(C)}}{d\,\mathrm{t}}
$$
\item[-] t(h)ermolecular reactions: A + B + M $\rightarrow$ C + M
$$
-\frac{d\,\mathrm{n(A)}}{d\,\mathrm{t}} = \mathrm{k}\,\mathrm{n(A)}\,\mathrm{n(B)}\,\mathrm{n(M)} = \frac{d\,\mathrm{n(C)}}{d\,\mathrm{t}}
$$
\end{enumerate}

In these relations, the temporal derivatives are reaction rates, and the factors (k) preceding the numerical densities are reaction rate coefficients, also called kinetic constants. The reaction rates are always expressed in cm$^{-3}$\,s$^{-1}$, whilst the units of reaction rate coefficients depend on the type of reaction. For instance, for bimolecular reactions, k is expressed in cm$^3$\,s$^{-1}$, while for unimolecular reactions it is expressed in s$^{-1}$.\\

It should be emphasized that when one is dealing with chemical kinetics, two factors are essential:
\begin{enumerate}
\item[-] {\it Density.} Typically, densities expressed in particles per cm$^{3}$ are quite low, with respect to terrestrial standards. To give an idea, diffuse and dense molecular clouds in the interstellar medium are characterized by densities of the order of 10$^2$\,cm$^{-3}$ and 10$^4$ to 10$^6$\,cm$^{-3}$ respectively, and the densest circumstellar clouds have densities of the order of 10$^{11}$-10$^{12}$\,cm$^{-3}$; but at sea level, the atmosphere of the Earth has a density of the order of 2.5\,$\times$\,10$^{19}$\,cm$^{-3}$. The immediate consequence is a low probability of interaction between reaction partners, translating into low reaction rates.
\item[-] {\it Temperature.} In molecular clouds, i.e. the densest interstellar region where most interstellar molecules have been identified, temperatures may be as low as a few K, up to several tens of K. Other parts of the interstellar medium where elements are ionized may be characterized by much higher temperatures, but such environments are permeated by strong radiation fields that compromise at least the existence of small molecules. In short, the only places in the interstellar medium which are significantly populated by molecules are very cold media. Gas phase kinetics, in the context of the perfect gas approximation, teaches us that reaction rate coefficients (k) depend intimately on the temperature (T) according to the following proportionality relation: $\mathrm{k} \propto \mathrm{T}^{1/2}\,\exp\,\Big(-\,\frac{\mathrm{E_a}}{\mathrm{k_B\,T}}\Big)$. In this relation\footnote{This kind of proportionality relation is of frequent use in parametric form to express reaction rate coefficients, as described for most elementary processes below. In practise, some differences are however expected because of deviations from the perfect gas approximation.}, the first factor is related to the properties of the gas and to their consequences on the interaction between reaction partners, and the second factor expresses the need to cross an activation barrier $\mathrm{E_a}$. These two factors clearly show that the low temperature of molecular clouds leads to low values for k, which is not in favor of fast chemical kinetics. 
\end{enumerate}

This points to an especially strong constraint for astrochemistry, at least in interstellar environments: one is dealing with slow kinetics, and this is {\it a priori} a good reason to consider that the interstellar molecular complexity will suffer from stringent limitations.\\

In the context of astrochemical studies, it is also not relevant to establish complex reaction mechanisms or to derive heavy analytical expressions for reaction rate equations. Even when simple situations are considered, with only a small number of first or second order processes, analytical relations become rapidly too complex to be manipulated. The approach adopted consists therefore to derive numerical solutions for systems of equations involving many {\it elementary processes}, i.e. processes for which there is no identified intermediate process, occurring consecutively and/or in parallel. The complete problem consisting of determining abundances of many chemical species as a function of time is therefore split into a large number of simple problems, for which the solutions can be derived much more easily. For a given population of chemical species, a census of elementary processes has therefore to be made, with a priori knowledge of the related rate coefficients. Several kinds of elementary processes of common use in astrochemistry are discussed below. A discussion of some of these processes can be found for instance in \citet{tielens}.

\subsubsection{Photodissociation}

In the ISM, the dominant destruction agent for small molecules is the far ultraviolet (FUV) radiation field from early-type stars.  These stars are not the most abundant ones, but their brightness in the ultraviolet and visible domains has a strong impact on their environment \citep[see e.g.][]{stromgren,capriotti2001,freyer2003}. Typical chemical bonding energies are of the order of 5-10\,eV, corresponding to wavelengths of about 3000\,\AA\, and shorter, in the FUV domain.

A photodissociation reaction is typically a reaction of the type:

$$ \mathrm{AB} + h\nu \rightarrow \mathrm{A} + \mathrm{B} $$

The photodissociation rate in the diffuse ISM can be expressed by

$$ \mathrm{k_{pd}} = \int_{\nu_\mathrm{d}}^{\nu_\mathrm{H}} 4\,\pi\,\mathrm{J_{IS}}\,\alpha(\nu)\,d\nu) $$

where $\mathrm{J_{IS}}$ is the mean intensity of the interstellar radiation field and $\alpha(\nu)$ is the photodissociation cross section at a given frequency $\nu$. The integration is considered from the dissociation limit ($\nu_\mathrm{d}$), as the photon has to be energetic enough to break the chemical bonding, to the hydrogen photo-ionization limit ($\nu_\mathrm{H}$), as it is generally true that higher energy photons are completely absorbed by atomic hydrogen that is ubiquitous and very abundant. In the interstellar mean radiation field ($\sim$\,10$^{8}$ FUV photon\,cm$^{-2}$\,s$^{-1}$\,sr$^{-1}$), this translates into a rate of 10$^{-9}$-10$^{-10}$\,s$^{-1}$.

In the presence of dust in the ISM, the FUV radiation field will be attenuated. As the absorption and scattering by dust is wavelength dependent, the frequency distribution of the FUV radiation field will vary as a function of the depth into the cloud. In this context, it is therefore important to distinguish two reaction rates: the unshielded one (in the absence of dust), and the attenuated one (in the presence of dust). The latter can be expressed as the former multiplied by an exponential factor depending on the visual extinction (A$_{v}$) due to dust. As we are mostly dealing with photodissociation due to the UV radiation field, one has to take into account the increased extinction of dust at ultraviolet wavelengths via the following relation: 

$$ \mathrm{k_{pd}} = \alpha\,\exp\,[-\,\gamma\,\mathrm{A}_v] $$

Intensive radiative transfer studies allowed to estimate $\alpha$ and $\gamma$ for many photodissociation reactions. A few examples, taken from the UMIST database for astrochemistry ({\tt http://www.udfa.net}: \citealt{umist}) are provided in Table\,\ref{photoreactions}.

\begin{table}[h]
\begin{center}
\caption{$\alpha$ and $\gamma$ factors for a few important photodissociation reactions \citep[UMIST database:][]{umist}. \label{photoreactions}}
\vspace*{3mm}
\begin{tabular}{lcc}
\hline
Reaction	& $\alpha$	& $\gamma$ \\
\hline
HCN + $h\nu$ $\longrightarrow$  CN + H & 1.3\,$\times$\,10$^{-9}$ & 2.1 \\
HCO + $h\nu$ $\longrightarrow$  CO + H & 1.1\,$\times$\,10$^{-9}$ & 0.8 \\
H$_2$O + $h\nu$ $\longrightarrow$  OH + H & 5.9\,$\times$\,10$^{-10}$ & 1.7 \\
CH + $h\nu$ $\longrightarrow$  C + H & 8.6\,$\times$\,10$^{-10}$ & 1.2 \\
CH$^+$ + $h\nu$ $\longrightarrow$  C + H$^+$ & 2.5\,$\times$\,10$^{-10}$ & 2.5 \\
CH$_2$ + $h\nu$ $\longrightarrow$  CH + H & 7.2\,$\times$\,10$^{-11}$ & 1.7 \\
C$_2$ + $h\nu$ $\longrightarrow$  C + C & 1.5\,$\times$\,10$^{-10}$ & 2.1 \\
C$_2$H + $h\nu$ $\longrightarrow$  C$_2$ + H & 5.1\,$\times$\,10$^{-10}$ & 1.9 \\
C$_2$H$_2$ + $h\nu$ $\longrightarrow$  C$_2$H + H & 7.3\,$\times$\,10$^{-9}$ & 1.8 \\
C$_2$H$_3$ + $h\nu$ $\longrightarrow$  C$_2$H$_2$ + H & 1.0\,$\times$\,10$^{-9}$ & 1.7 \\
CO + $h\nu$ $\longrightarrow$  C + O & 2.0\,$\times$\,10$^{-10}$ & 2.5 \\
N$_2$ + $h\nu$ $\longrightarrow$  N + N & 2.3\,$\times$\,10$^{-10}$ & 3.8 \\
NH + $h\nu$ $\longrightarrow$  N + H & 5.0\,$\times$\,10$^{-10}$ & 2.0 \\
NO + $h\nu$ $\longrightarrow$  N + O & 4.3\,$\times$\,10$^{-10}$ & 1.7 \\
O$_2$ + $h\nu$ $\longrightarrow$  O + O & 6.9\,$\times$\,10$^{-10}$ & 1.8 \\
OH + $h\nu$ $\longrightarrow$  O + H & 3.5\,$\times$\,10$^{-10}$ & 1.7 \\
\hline
\end{tabular}
\end{center}
\end{table}

In the astronomical context, where abundant molecules such as H$_2$ (or even CO) are considered, we may envisage three situations when one wants to quantify the photodissociation process:
\begin{enumerate}
\item[-] the photodissociation occurs in a diffuse cloud in the ISM because of the presence of a population of massive stars in the vicinity, responsible for the production of a strong FUV radiation field. Here, the photodissociation rate can be considered as constant across the cloud, provided that the level population distribution is constant.
\item[-] the medium permeated by the FUV photons is not only made of gas, but also contains dust. The radiation field undergoes therefore a cumulative attenuation as deeper layers are considered. As a consequence, the photodissociation rate depends on the dust distribution in the cloud. 
\item[-] the gas is dense, and the radiation field is further attenuated by the molecules themselves, in addition to the attenuation due to dust. This is what is called self-shielding. In this case, the photodissociation rate depends on the abundance of the molecule and the level population distribution as a function of depth in the cloud.
\end{enumerate}

When dust absorption and self-shielding are considered, in the case of H$_2$, the photodissociation rate coefficient can be expressed as follows:
$$ \mathrm{k_{UV}(H_2)} =  \mathrm{k_{o}(H_2)}\,\xi_{SS}\,\exp\,(-\tau_d) $$
where $\mathrm{k_{o}(H_2)}$ is the unshielded photodissociation rate in the absence of dust, $\xi_{SS}$ is the self-shielding factor, and $\tau_d$ is the opacity due to dust in the cloud. It should also be mentioned that ISM clouds are not homogenous. The influence of dust and of self-shielding will therefore depend intimately on the line of sight.

Carbon monoxide (CO) is a very stable molecule. The bonding energy is 11.09\,eV, corresponding to 1118\,\AA\,. Considering the fact that interstellar photons with energies higher than 13.6\,eV are almost completely used for the ionization of abundant hydrogen atoms, the photodissociation of CO can only occur through photons with wavelengths between 912 and 1118\,\AA\,.

\subsubsection{Photoionization}
Photons permeating interstellar clouds can lead to the singificant direct ionization of atoms or molecules, provided wealth of photons are available. For instance, in diffuse clouds (low density clouds which are not opaque to UV radiation), photoionization of C to yield carbon cations is the triggerring step of carbon chemistry. This is a consequence of the fact that the ionization potential of C is 11.26\,eV, i.e. lower than the ionization potential of atomic hydrogen. As the ionization potentials of O and N are higher than 13.6\,eV (respectively, 13.62 and 14.53\,eV), these elements can not be photoionized and mainly exist in neutral form in diffuse clouds. A few examples of photoionization reactions are given in Table\,\ref{photoion}, with their $\alpha$ and $\gamma$ parameters carrying the same meaning as in the case of photodissociation reactions.

\begin{table}[h]
\begin{center}
\caption{$\alpha$ and $\gamma$ factors for a few important photoionization reactions \citep[UMIST database:][]{umist}. \label{photoion}}
\vspace*{3mm}
\begin{tabular}{lcc}
\hline
Reaction	& $\alpha$	& $\gamma$ \\
\hline
C + $h\nu$ $\longrightarrow$  C$^+$ + e$^-$ & 3.0\,$\times$\,10$^{-10}$ & 3.0 \\
C$_2$ + $h\nu$ $\longrightarrow$  C$_2^+$ + e$^-$ & 4.1\,$\times$\,10$^{-10}$ & 3.5 \\
CH + $h\nu$ $\longrightarrow$  CH$^+$ + e$^-$ & 7.6\,$\times$\,10$^{-10}$ & 2.8 \\
NH + $h\nu$ $\longrightarrow$  NH$^+$ + e$^-$ & 1.0\,$\times$\,10$^{-11}$ & 2.0 \\
OH + $h\nu$ $\longrightarrow$  OH$^+$ + e$^-$ & 1.6\,$\times$\,10$^{-12}$ & 3.1 \\
\hline
\end{tabular}
\end{center}
\end{table}

\subsubsection{Neutral-neutral reactions}\label{sect-neu-neu}

This kind of reaction is typically of the type

$$ \mathrm{A} + \mathrm{B} \rightarrow \mathrm{C} + \mathrm{D} $$

Reaction rate coefficient are considered to be of the form

$$ \mathrm{k = \alpha\,(T/300)^\beta\,\exp\,[-\gamma/T]} $$
with the parameters $\alpha$, $\beta$ and $\gamma$ given in Table\,\ref{neu-neu} for a few reactions. It should be noted that the values attributed to these parameters are valid only for a given temperature range. In addition, these values may differ according to the approach adopted to derive them (see the UMIST database for details: \citealt{umist}). One should also note the significant deviation of $\beta$ with respect to the 0.5 value expected for perfect gases.

\begin{table}[h]
\begin{center}
\caption{$\alpha$, $\beta$ and $\gamma$ parameters for a few examples of neutral-neutral reactions \citep[UMIST database:][]{umist}. \label{neu-neu}}
\vspace*{3mm}
\begin{tabular}{lccc}
\hline
Reaction	& $\alpha$	& $\beta$ & $\gamma$ \\
\hline
H + OH $\longrightarrow$  O + H$_2$ & 7.0\,$\times$\,10$^{-14}$ & 2.8 & 1950 \\
H + CH$_4$ $\longrightarrow$  CH$_3$ + H$_2$ & 5.9\,$\times$\,10$^{-13}$ & 3.0 & 4045 \\
H + NH$_3$ $\longrightarrow$  NH$_2$ + H$_2$ & 7.8\,$\times$\,10$^{-13}$ & 2.4 & 4990 \\
H + H$_2$CO $\longrightarrow$  HCO + H$_2$ & 4.9\,$\times$\,10$^{-12}$ & 1.9 & 1379 \\
H$_2$ + CN $\longrightarrow$  HCN + H & 4.0\,$\times$\,10$^{-13}$ & 2.9 & 820 \\
C + OH $\longrightarrow$  CH + O & 2.3\,$\times$\,10$^{-11}$ & 0.5 & 14800 \\
C + NH $\longrightarrow$  N + CH & 1.7\,$\times$\,10$^{-11}$ & 0.5 & 4000 \\
CH + O $\longrightarrow$  OH + C & 2.5\,$\times$\,10$^{-11}$ & 0.0 & 2381 \\
CH + O$_2$ $\longrightarrow$  HCO + O & 1.4\,$\times$\,10$^{-11}$ & 0.7 & 3000 \\
CH + N$_2$ $\longrightarrow$  HCN + N & 5.6\,$\times$\,10$^{-13}$ & 0.9 & 10128 \\
CH$_3$ + OH $\longrightarrow$ CH$_4$ + O & 3.3\,$\times$\,10$^{-14}$ & 2.2 & 2240 \\
N + HCO $\longrightarrow$ CO + NH & 5.7\,$\times$\,10$^{-12}$ & 0.5 & 1000 \\
N + HNO $\longrightarrow$ N$_2$O + H & 1.4\,$\times$\,10$^{-12}$ & 0.5 & 1500 \\
N + O$_2$ $\longrightarrow$ NO + O & 2.3\,$\times$\,10$^{-12}$ & 0.9 & 3134 \\
NH + CN $\longrightarrow$  HCN + N & 2.9\,$\times$\,10$^{-12}$ & 0.5 & 1000 \\
NH + OH $\longrightarrow$  NH$_2$ + O & 2.9\,$\times$\,10$^{-12}$ & 0.1 & 5800 \\
NH + O$2$ $\longrightarrow$  HNO + O & 6.9\,$\times$\,10$^{-14}$ & 2.7 & 3281 \\
O + CN $\longrightarrow$  CO + N & 4.4\,$\times$\,10$^{-11}$ & 0.5 & 364 \\
O + HCN $\longrightarrow$  CO + NH & 7.3\,$\times$\,10$^{-13}$ & 1.1 & 3742 \\
O + C$_2$H$_4$ $\longrightarrow$ H$_2$CCO + H$_2$ & 5.1\,$\times$\,10$^{-14}$ & 1.9 & 92 \\
OH + OH $\longrightarrow$  H$_2$O + O & 1.7\,$\times$\,10$^{-12}$ & 1.1 & 50 \\
O$_2$ + SO $\longrightarrow$  SO$_2$ + O & 1.1\,$\times$\,10$^{-14}$ & 1.9 & 1538 \\
\hline
\end{tabular}
\end{center}
\end{table}

The attractive interaction is due to van der Waals forces that is effective only at very short distances (interaction potential $\propto$ 1/r$^6$). Neutral-neutral processes are generally reactions with large activation barriers because of the necessity to break chemical bonds during the molecular rearrangement. As a consequence, such reactions can be of importance where the gas is warm enough (i.e. possessing the requested energy): for instance in stellar ejecta, in hot cores associated to protostars, in dense photodissociation regions associated with luminous stars, and in clouds crossed by hydrodynamic shocks. In the colder conditions of dark clouds, only neutral-neutral reactions involving atoms or radicals can occur significantly because they have only small or no activation barriers. In the latter case, the reaction rate coefficient has the following form
$$ \mathrm{k = \alpha\,(T/300)^\beta} $$

\subsubsection{Ion-molecule reactions}\label{sect-ion-mol}

In this case, we consider reactions of the type
$$ \mathrm{A^+} + \mathrm{B} \rightarrow \mathrm{C^+} + \mathrm{D} $$

This kind of reaction occurs generally more rapidly than neutral-neutral because of the strong polarization-induced interaction potential ($\propto$ 1/r$^4$). Typically, the reaction rate coefficient is of the order of 10$^{-9}$\,cm$^3$\,s$^{-1}$ with no temperature dependence. This is generally at least one or two orders of magnitude faster than neutral-neutral reactions. For this reason, even a small amount of ions in a given medium can have a strong impact on interstellar chemistry. In addition, if the reaction involves an ion and a neutral species with a permanent dipole, the efficiency of the interaction can reach much higher levels, with improvement of the kinetics of one or two additional orders of magnitude. A few examples of ion-molecule reactions are given in Table\,\ref{ion-mol}.

\begin{table}[h]
\begin{center}
\caption{Reaction rate coefficients for a few examples of ion-molecule reactions \citep[UMIST database:][]{umist}.\label{ion-mol}}
\vspace*{3mm}
\begin{tabular}{lc}
\hline
Reaction	& k \\
\hline
H$_2^+$ + OH $\longrightarrow$  H$_2$O$^+$ + H & 7.6\,$\times$\,10$^{-10}$ \\
H$_2$ + OH$^+$ $\longrightarrow$  H$_2$O$^+$ + H & 1.1\,$\times$\,10$^{-9}$ \\
H$_2$O$^+$ + H$_2$ $\longrightarrow$  H$_3$O$^+$ + H & 6.1\,$\times$\,10$^{-10}$ \\
H$_2$O$^+$ + CO $\longrightarrow$  HCO$^+$ + OH & 5.0\,$\times$\,10$^{-10}$ \\
H$_2$O$^+$ + C$_2$ $\longrightarrow$  C$_ 2$H$^+$ + OH & 4.7\,$\times$\,10$^{-10}$ \\
C$^+$ + OH $\longrightarrow$  CO$^+$ + H & 7.7\,$\times$\,10$^{-10}$ \\
C$^+$ + CH$_3$ $\longrightarrow$  C$_2$H$_2^+$ + H & 1.3\,$\times$\,10$^{-9}$ \\
C$^+$ + C$_2$H$_2$ $\longrightarrow$  C$_3$H$^+$ + H & 2.2\,$\times$\,10$^{-9}$ \\
C$^+$ + C$_2$H$_4$ $\longrightarrow$  C$_3$H$_2^+$ + H$_2$ & 3.4\,$\times$\,10$^{-10}$ \\
CH$^+$ + OH $\longrightarrow$  CO$^+$ + H$_2$ & 7.5\,$\times$\,10$^{-10}$ \\
CH$_2^+$ + O$_2$ $\longrightarrow$  HCO$^+$ + OH & 9.1\,$\times$\,10$^{-10}$ \\
CH$_3^+$ + OH $\longrightarrow$  H$_2$CO$^+$ + H$_2$ & 7.2\,$\times$\,10$^{-10}$ \\
O$^+$ + CH$_4$ $\longrightarrow$  CH$_3^+$ + OH & 1.1\,$\times$\,10$^{-10}$ \\
OH$^+$ + CN $\longrightarrow$  HCN$^+$ + O & 1.0\,$\times$\,10$^{-9}$ \\
OH$^+$ + NH$_3$ $\longrightarrow$  NH$_4^+$ + O & 1.0\,$\times$\,10$^{-9}$ \\
\hline
\end{tabular}
\end{center}
\end{table}

\subsubsection{Dissociative electron recombination reactions}\label{sect-der}

In such a process, an electron is captured by an ion to form a neutral species in an excited electronic state, then follows a dissociation of the neutral compound: 
$$ \mathrm{A^+} + \mathrm{e^-} \rightarrow \mathrm{A^*}  \rightarrow \mathrm{C} + \mathrm{D} $$

Rate coefficients are typically of the order of 10$^{-7}$\,cm$^{3}$\,s$^{-1}$. Some examples of reaction rate coefficients, along with their dependence on the temperature ($\mathrm{k = \alpha(T/300)}^\beta$) are given in Table\,\ref{dissrec}. This process constitutes a significant production channel for many small neutral molecules that could not easily be formed through the single addition of atomic species in gas phase. Many neutral chemical species are significantly produced through ion-molecule processes, terminated by a dissociative electron recombination reaction with the loss a of fragment. For instance, several H-bearing molecules could be formed via mutliple reactions of cations with H$_2$ molecules, terminated by a dissociative electronic recombination:  e.g. the formation of ammonia initiated by N$^+$, and the formation of methane initiated by C$^+$. One may also consider the growth of larger neutral aliphatic hydrocarbons through the successive addition of C-bearing molecules to a cationic hydrocarbon, followed by the reaction with an electron. Considering the large value of the rate coefficients of this process, it constitutes the main terminator of cationic processes.

\begin{table}[h]
\begin{center}
\caption{$\alpha$ and $\beta$ parameters for a few examples of dissociative electron recombination reactions \citep[UMIST database:][]{umist} .\label{dissrec}}
\vspace*{3mm}
\begin{tabular}{lcc}
\hline
Reaction	& $\alpha$	& $\beta$ \\
\hline
H$_2^+$ + e$^-$ $\longrightarrow$ H + H & 1.6\,$\times$\,10$^{-8}$ & -0.43 \\
H$_2$O$^+$ + e$^-$ $\longrightarrow$ O + H + H & 3.1\,$\times$\,10$^{-7}$ & -0.5 \\
H$_2$O$^+$ + e$^-$ $\longrightarrow$ OH + H & 8.6\,$\times$\,10$^{-8}$ & -0.5 \\
H$_3$O$^+$ + e$^-$ $\longrightarrow$ OH + H$_2$ & 6.0\,$\times$\,10$^{-8}$ & -0.5 \\
H$_3$O$^+$ + e$^-$ $\longrightarrow$ O + H + H$_2$ & 5.6\,$\times$\,10$^{-9}$ & -0.5 \\
H$_3^+$ + e$^-$ $\longrightarrow$ H$_2$ + H & 2.3\,$\times$\,10$^{-8}$ & -0.52 \\
HCN$^+$ + e$^-$ $\longrightarrow$ CN + H & 2.0\,$\times$\,10$^{-7}$ & -0.5 \\
HCO$^+$ + e$^-$ $\longrightarrow$ CO + H & 2.4\,$\times$\,10$^{-7}$ & -0.69 \\
HNO$^+$ + e$^-$ $\longrightarrow$ NO + H & 3.0\,$\times$\,10$^{-7}$ & -0.69 \\
C$_2^+$ + e$^-$ $\longrightarrow$ C + C & 3.0\,$\times$\,10$^{-7}$ & -0.5 \\
CH$^+$ + e$^-$ $\longrightarrow$ C + H & 1.5\,$\times$\,10$^{-7}$ & -0.42 \\
CH$_2^+$ + e$^-$ $\longrightarrow$ CH + H & 1.4\,$\times$\,10$^{-7}$ & -0.55 \\
CH$_2^+$ + e$^-$ $\longrightarrow$ C + H + H & 4.0\,$\times$\,10$^{-7}$ & -0.6 \\
CH$_3^+$ + e$^-$ $\longrightarrow$ CH + H + H & 2.0\,$\times$\,10$^{-7}$ & -0.4 \\
CH$_3^+$ + e$^-$ $\longrightarrow$ CH + H$_2$ & 2.0\,$\times$\,10$^{-7}$ & -0.5 \\
CH$^+$ + e$^-$ $\longrightarrow$ C + H & 1.5\,$\times$\,10$^{-7}$ & -0.42 \\
CN$^+$ + e$^-$ $\longrightarrow$ C + N & 1.8\,$\times$\,10$^{-7}$ & -0.5 \\
NH$_4^+$ + e$^-$ $\longrightarrow$ NH$^2$ + H$_2$ & 1.5\,$\times$\,10$^{-7}$ & -0.47 \\
NO$_2^+$ + e$^-$ $\longrightarrow$ NO + O & 3.0\,$\times$\,10$^{-7}$ & -0.5 \\
OH$^+$ + e$^-$ $\longrightarrow$ O + H & 3.8\,$\times$\,10$^{-8}$ & -0.5 \\
OCS$^+$ + e$^-$ $\longrightarrow$ CS + O & 4.9\,$\times$\,10$^{-8}$ & -0.62 \\
\hline
\end{tabular}
\end{center}
\end{table}

\subsubsection{Cosmic-ray induced reactions}
A particularly important example of reaction occurring in the interstellar medium is the interaction of chemical species in molecular clouds with cosmic-rays. Cosmic-rays are high energy charged particles accelerated in quite extreme astrophysical environments \citep[see e.g.][]{hillas1984,drury2012}. The most probable source of Galactic cosmic-rays are supernova remnants \citep{koyama1995,aharonian2005,helderpasnr2012}, but other astrophysical environments presenting for instance strong hydrodynamic shocks likely to accelerate particles may also contribute \citep{RPR2006,debeckerreview}. Cosmic-ray induced processes are especially efficient in dense molecular clouds, where the probability of an interaction of the cosmic-ray particle with a partner is significant.

Cosmic-rays can interact with molecules and lead to their dissociation, as illustrated by the examples in Table\,\ref{cr-diss}. Such a process can be an efficient agent of destruction of molecules in clouds that are too dense to allow significant photodissociation by the interstellar UV radiation field. The dissociating particles may be the primary cosmic-ray, or even secondary ions produced by its interaction with the cloud material, such as helium cations.

\begin{table}[h]
\begin{center}
\caption{Reaction rate coefficients for a few examples of cosmic-ray induced dissociation reactions \citep[UMIST database:][]{umist}.\label{cr-diss}}
\vspace*{3mm}
\begin{tabular}{lc}
\hline
Reaction	& k \\
\hline
H$_2$ + CR $\longrightarrow$ H$^+$ + H + e$^-$ & 1.2\,$\times$\,10$^{-17}$ \\
He$^+$ + CO $\longrightarrow$ C$^+$ + O + He & 1.6\,$\times$\,10$^{-9}$ \\
He$^+$ + OH $\longrightarrow$ O$^+$ + H + He & 1.1\,$\times$\,10$^{-9}$ \\
He$^+$ + H$_2$O $\longrightarrow$  H$^+$ + OH + He & 2.0\,$\times$\,10$^{-10}$ \\
\hline
\end{tabular}
\end{center}
\end{table}

Cosmic-ray can also directly ionize neutral species. Such a process can be viewed as equivalent to a classical photoionization, but it is induced by the interaction with a cosmic-ray. Examples of cosmic-ray ionization reactions are given in Table\,\ref{cr-ion}. It is noticeable that rate coefficients for such cosmic-ray ionization reactions are rather weak, suggesting a global weak efficiency of such processes. It should however be reminded that in dark clouds opaque to radiation fields, cosmic-rays constitute the unique ion source, even though this is a weak source. In other words, without their presence, almost no ion would be produced in dark clouds. In addition, one should not neglect the fact that the typcial length of the trajectory of cosmic-rays in interstellar clouds is long, admitting the possibility that a significant number of interactions actually occur along their path. If the cross section of such high energy particles was much larger, cosmic-rays would lose most of their energy before reaching the deepest parts of dark clouds, therefore inhibitting their capability to enhance interstellar chemistry in these regions. As a result, ion-neutral processes would almost be forbidden in the densest moelcular clouds that constitute the starting material for stellar and planetary systems formation.

\begin{table}[h]
\begin{center}
\caption{Reaction rate coefficients for a few examples of cosmic-ray ionization reactions \citep[UMIST database:][]{umist}.\label{cr-ion}}
\vspace*{3mm}
\begin{tabular}{lc}
\hline
Reaction	& k \\
\hline
H + CR $\longrightarrow$  H$^+$ + e$^-$ & 6.0\,$\times$\,10$^{-18}$ \\
He + CR $\longrightarrow$  He$^+$ + e$^-$ & 6.5\,$\times$\,10$^{-18}$ \\
H$_2$ + CR $\longrightarrow$  H$_2^+$ + e$^-$ & 1.2\,$\times$\,10$^{-17}$ \\
C + CR $\longrightarrow$  C$^+$ + e$^-$ & 2.3\,$\times$\,10$^{-17}$ \\
CO + CR $\longrightarrow$  CO$^+$ + e$^-$ & 3.9\,$\times$\,10$^{-17}$ \\
Cl + CR $\longrightarrow$  Cl$^+$ + e$^-$ & 3.9\,$\times$\,10$^{-17}$ \\
N + CR $\longrightarrow$  N$^+$ + e$^-$ & 2.7\,$\times$\,10$^{-17}$ \\
O + CR $\longrightarrow$  O$^+$ + e$^-$ & 3.4\,$\times$\,10$^{-17}$ \\
\hline
\end{tabular}
\end{center}
\end{table}

Finally, it has been demonstrated that the electrons generated by the interaction of cosmic-rays with chemical species can collisionally excite hydrogen molecules in dense clouds. These excited hydrogen molecules (in the Lyman and Werner systems) relax through the emission of UV photons \citep{PT-CR}. This process constitutes a significant provider of UV photons in inner parts of clouds which are too dense to be reached by the interstellar UV radiation field. Several examples of cosmic-ray induced photoreactions (ionization and dissociation) are given in Table\,\ref{crip}, for reaction rate coefficients of the form

$$ \mathrm{k = \alpha\,(T/300)^\beta\,\gamma/(1 - \omega)} $$
where $\alpha$ is the cosmic-ray photoreaction rate, $\beta$ is a parameter governing the temperature dependence of the process (often equal to 0.0), $\gamma$ is an efficiency factor, and $\omega$ is the average albedo of dust grains that contribute significantly to the decrease of the local UV radiation field produced by cosmic-rays (typically equal to 0.6 at 150\,nm).

\begin{table}[h]
\begin{center}
\caption{Reaction rate coefficients for a few examples of cosmic-ray induced photoreactions \citep[UMIST database:][]{umist}.\label{crip}}
\vspace*{3mm}
\begin{tabular}{lccc}
\hline
Reaction	& $\alpha$ & $\beta$ & $\gamma$ \\
\hline
H$_2$O + CR $\longrightarrow$  OH + H & 1.3\,$\times$\,10$^{-17}$ & 0.0 & 486 \\
HCO + CR $\longrightarrow$  HCO$^+$ + e$^-$ & 1.3\,$\times$\,10$^{-17}$ & 0.0 & 585 \\
HCO + CR $\longrightarrow$  CO + H & 1.3\,$\times$\,10$^{-17}$ & 0.0 & 211 \\
C + CR $\longrightarrow$  C$^+$ + e$^-$ & 1.3\,$\times$\,10$^{-17}$ & 0.0 & 155 \\
CH + CR $\longrightarrow$  C + H & 1.3\,$\times$\,10$^{-17}$ & 0.0 & 365 \\
C$_2$ + CR $\longrightarrow$  C + C & 1.3\,$\times$\,10$^{-17}$ & 0.0 & 120 \\
CO + CR $\longrightarrow$  C + O & 3.9\,$\times$\,10$^{-17}$ & 1.17 & 105 \\
NH + CR $\longrightarrow$  N + H & 1.3\,$\times$\,10$^{-17}$ & 0.0 & 250 \\
O$_2$ + CR $\longrightarrow$  O + O & 1.3\,$\times$\,10$^{-17}$ & 0.0 & 376 \\
\hline
\end{tabular}
\end{center}
\end{table}

The latter process is especially important in the sense that it is capable of recovering neutral carbon atoms from its main reservoir (i.e. CO), in the absence of significant UV interstellar radiation field from the neighboring stars. The efficiency of this process is also expected to be higher than that of the dissociative charge transfer due to He$^+$ followed by neutralization of C$^+$ through electronic recombination, which is not very efficient. The combination of these two processes to recover neutral carbon is at least one order of magnitude less efficient than the cosmic-ray induced photodissociation. Considering this process appears therefore very important in order to understand quite large values of the n(C)/n(CO) ratio as measured in the case of several dense clouds \citep[e.g.][]{philhugg}. Furthermore, this UV production process in the inner part of dense clouds can play a significant role in the desorption of chemical species adsorbed on dust grains, in order to release them in the gas phase. Grain surface processes are discussed in Sect.\,\ref{grain}.

Beside the fact that cosmic-ray induced reactions introduce ions in a medium that is too dense to be ionized through the interstellar radiation field ({\it which is important for kinetics purpose}), these high-energy particles vehiculate large amounts of energy that is distributed to the products of these cosmic-ray induced reactions ({\it which is important for thermodynamics purpose}). For instance, one may consider the case of a possible way to produce ammonia in interstellar clouds. This process starts with the ion-molecule reaction of N$^+$ with molecular hydrogen, in a so-called H-atom abstraction reaction, producing NH$^+$. Consecutive similar reactions lead to the NH$_3^+$ + H$_2$ reaction, with the formation of the ammonium cation. The latter species is likely to undergo a dissociative electronic recombination reaction resulting in the formation of neutral ammonia. However, this chain of processes is initiated by the reaction of N$^+$ with molecular hydrogen, which is an endoergic process by about $\Delta$E/k$_\mathrm{B}$\,$\sim$\,85\,K. If N$^+$ cations are at thermal energy in cold clouds ($\sim$\,10\,K), the process described above for the formation of ammonia could not start \citep{ASM-kinexc}. However, the main source of nitrogen cations is expected to be the cosmic-ray induced dissociation of N$_2$, with the resulting nitrogen cation possessing an excess energy that is sufficient to overcome the endoergicity of the ion-molecule process. This example emphasizes the important role of kinetically excited ions in the gas-phase chemistry of cold interstellar clouds.

\subsubsection{Charge transfer reactions}

Another kind of reaction worth considering is the following:
$$ \mathrm{A^+} + \mathrm{B} \rightarrow \mathrm{A} + \mathrm{B^+} $$
In this reaction, there is no break of chemical bonds. The only transformation is the transfer of an electric charge. The rate coefficient can reach values of the order of 10$^{-9}$\,cm$^{3}$\,s$^{-1}$. 

\begin{table}[h]
\begin{center}
\caption{Reaction rate coefficients for a few examples of charge transfer reactions \citep[UMIST database:][]{umist}.\label{ch-tr}}
\vspace*{3mm}
\begin{tabular}{lc}
\hline
Reaction	& k \\
\hline
H$^+$ + OH $\longrightarrow$ H + OH$^+$ & 2.1\,$\times$\,10$^{-9}$ \\
N$_2^+$ + H$_2$O $\longrightarrow$ N$_2$ + H$_2$O$^+$ & 2.3\,$\times$\,10$^{-9}$ \\
C$^+$ + CH $\longrightarrow$ C + CH$^+$ & 3.8\,$\times$\,10$^{-10}$ \\
C$^+$ + CH$_2$ $\longrightarrow$ C + CH$_2^+$ & 5.2\,$\times$\,10$^{-10}$ \\
C$^+$ + C$_2$H$_4$ $\longrightarrow$ C + C$_2$H$_4^+$ & 3.0\,$\times$\,10$^{-10}$ \\
N$^+$ + CH$_2$ $\longrightarrow$ N + CH$_2^+$ & 1.0\,$\times$\,10$^{-9}$ \\
O$^+$ + CH $\longrightarrow$ O + CH$^+$ & 3.5\,$\times$\,10$^{-10}$ \\
O$^+$ + H$_2$O $\longrightarrow$ O + H$_2$O$^+$ & 3.2\,$\times$\,10$^{-9}$ \\
\hline
\end{tabular}
\end{center}
\end{table}

Charge transfer involving atoms are crucial to set the ionization balance in the ISM. An important example is the ionization balance of trace species in HII regions (parts of the interstellar medium that are mainly populated by ionized hydrogen), with in particular the charge exchange reaction between O and H$^+$. This latter reaction is important in interstellar chemistry because it transfers ionization to oxygen which can then participate in the chemistry, and therefore allow the insertion of oxygen in molecules. This reaction can be characterized by a large rate coefficient if the ionization potentials are close. In the case of ion-molecule charge transfer reactions, the rate coefficient may be rather large because of the larger number of electronic states available of molecules as compared to atoms.

\subsubsection{Radiative association reactions} \label{rar}

In such a process, the excited reaction product is stabilized through the emission of a photon. Such a process can be illustrated by the following equation:
$$ \mathrm{A} + \mathrm{B} \rightarrow \mathrm{AB}^{*} \rightarrow \mathrm{AB} + h\nu$$

In this context, some typical time-scales are worth discussing. The radiative lifetime for an allowed transition is $\tau_{rad} \sim 10^{-7}$\,s. The collision time-scale ($\tau_{col}$) is of the order of 10$^{-13}$\,s. This leads to an efficiency for the radiation process of the order of 10$^{-6}$, i.e. in a population of 10$^{6}$ molecules produced by the collision only one will be stabilized through the emission of radiation. The result of the collision between the two partners can be seen as an activated complex that can then follow three different reaction pathways:
\begin{enumerate}
\item[1.] The activated complex relaxes through the emission of radiation to yield a stable reaction product.
\item[2.] The most probable scenario is that the activated complex redissociates.
\item[3.] If collisions are significant, other collision partners can take away the excess energy and stabilize the reaction product. However, in order to become significant, this pathway requires density conditions that are unlikely met in the interstellar medium. 
\end{enumerate}
Only long-lived activated complexes will lead to efficient radiative association reactions. A few examples of radiative association reactions are given in Table\,\ref{radass}.

\begin{table}[h]
\begin{center}
\caption{Reaction rate coefficients for a few examples of radiative association reactions \citep[UMIST database:][]{umist}.\label{radass}}
\vspace*{3mm}
\begin{tabular}{lc}
\hline
Reaction	& k \\
\hline
C + H $\longrightarrow$ CH + $h\nu$& 1.0\,$\times$\,10$^{-17}$ \\
C + C $\longrightarrow$ C$_2$ + $h\nu$ & 4.4\,$\times$\,10$^{-18}$ \\
C + N $\longrightarrow$ CN + $h\nu$ & 1.4\,$\times$\,10$^{-18}$ \\
C + H$_2$ $\longrightarrow$ CH$_2$ + $h\nu$ & 1.0\,$\times$\,10$^{-17}$ \\
C$^+$ + H $\longrightarrow$  CH$^+$ + $h\nu$ & 1.7\,$\times$\,10$^{-17}$ \\
C$^+$ + H$_2$ $\longrightarrow$  CH$_2^+$ + $h\nu$ & 4.0\,$\times$\,10$^{-16}$ \\
O + O $\longrightarrow$ O$_2$ + $h\nu$ & 4.9\,$\times$\,10$^{-20}$ \\
\hline
\end{tabular}
\end{center}
\end{table}

The lifetime of the activated complex is the main factor governing the kinetics of such processes. The excess energy of the activated complex is temporarily stored in vibrational form. The density of states where the energy can be stored is a increasing function of the number of atoms in a chemical species, as this number is directly related to the number of vibrational modes over which this energy can be distributed. Larger species will therefore lead to a longer-lived activated complex, and hence to a higher radiative association rate. For this reason, this process is very important for large molecular species such as polycyclic aromatic hydrocarbons in the interstellar medium.

\subsubsection{Associative detachment reactions} \label{assdetreac}

In this process, an anion and an atom collide. The neutral product is then stabilized thanks to the emission of an electron that takes away the energy excess:
$$ \mathrm{A^-} + \mathrm{B} \rightarrow \mathrm{AB} + e^- $$

The most important example is the following reaction:
$$ \mathrm{H^-} + \mathrm{H} \rightarrow \mathrm{H_2} + e^- $$
whose rate coefficient is 1.3\,$\times$\,10$^{-9}$\,cm$^3$\,s$^{-1}$. This process is strongly believed to have been responsible for the production of molecular hydrogen in the early Universe, before its enrichment in heavier elements due to stellar activity. Since then, molecular hydrogen is indeed produced in the presence of dust particles, as discussed in section\,\ref{formhomodia}. Only the densest astronomical environments are likely to be populated by a significant amount of H$^-$.

Such a process should not be very relevant for most astrophysical environments. For instance, the radiative combination of atomic carbon with an electron to produce C$^-$ has a reaction rate constant of 2.3\,$\times$\,10$^{-15}$\,cm$^3$\,s$^{-1}$, which is very weak compared to the rate of photoionization of C$^-$, that is of the order of 10$^{-7}$\,s$^{-1}$. As a result, potential anions are very short-lived species, which have very low probabilities to combine with neutral partners before undergoing photoionization. In very dense astronomical environment, the low electronic association rate could be compensated by large densities of electrons and neutral species, favoring the existence of anions. However, the larger densities will allow more frequent collisions likely to remove the additional electrons from the chemical species, leading to its neutralization. So far, only a few anions have been identified in the interstellar medium (typically, anionic polyines). Only polyatomic species able to stabilize the negative charge have indeed a significant chance to exist, and to be detected in interstellar clouds. However, it is interesting to emphasize the recent detection of the CN$^-$ anion in the circumstellar envelope of IRC\,+10216 \citep{cnanion}.

\subsubsection{Collisional association reactions} \label{thermolecular}
One may consider processes where three bodies are interacting according to the following way:
$$ \mathrm{A} + \mathrm{B} + \mathrm{M} \rightarrow \mathrm{AB} + \mathrm{M} $$
Such processes are generally called termolecular processes because they involve three partners. They are sometimes called thermolecular processes, with the prefix `ther' justified by the thermal role played by the third partner, as the latter takes away the excess energy in order to stabilize the reaction product that would otherwise dissociate back to the reactants.

Such a process is unlikely to occur significantly in many astrophysical environments for a very simple reason: the probability of such a three body process is negligible due to the low densities found in the interstellar medium. However, it may become significant for the dense gas near stellar photospheres on in circumstellar disks characterized by large densities (i.e. 10$^{11}$\,cm$^{-3}$ or larger). This process is also significantly contributing to the chemistry of planetary atmospheres.

\subsubsection{Collisional dissociation reactions} \label{colldiss}
The breaking of chemical bonds can also be the consequence of the collision of a given molecule with any other partner. Such a process is especially efficient in high temperature environments (several kK), where the colliding partners have kinetic energy that is sufficient to promote electrons up to high $v$ levels ($v$ is the vibrationnal quantum number), and the dissociation continuum of the molecule is easily reached.

An example worth discussing is the comparison of H$_2$ to CO, the two most abundant molecules in the Universe. In hot environments, both are significantly destroyed by collisions. Main contributors to this collisional dissociation are H, He and H$_2$; i.e. the most abundant species. As molecular hydrogen is homonuclear, it does not possess a permanent dipole. The radiative relaxation of high $v$ electrons is therefore forbidden by vibrational selection rules, and there is accumulation of vibrationally excited H$_2$ molecules. Consequently, its dissociation is fast. On the contrary, CO has a permanent dipole, and the radiative relaxation of this molecule is permitted. There is no accumulation of vibrationally excited CO molecules and their dissociation is therefore much slower than that of H$_2$ in the same environment.

\subsection{Grain-surface processes}\label{grain}
Particle densities in space are much lower than in the Earth's atmosphere. In such conditions, the probability of an interaction between reaction partners in the gas phase of the interstellar medium is quite weak. Moreover, we know that the association of some molecules and/or atoms can lead to activated complexes that have to radiate their excess energy in order to be stabilized, but the dissociation of the activated complex back to the reaction partners is often much more likely. In order to be efficient, associations may require a third partner able to take away the excess energy (see Sect.\,\ref{thermolecular}). However the probability of such a three-body process is very weak considering usual interstellar particle densities. In this context, dust grains can play the role of the third body, allowing therefore to increase substantially the efficiency of the association between chemical species. Taking into account grain-surface processes appears thus to be very important in order to understand the formation of some common molecules -- including H$_2$ -- along with that of more complex molecules that have been found in space.

\subsubsection{Schematic view of a grain-surface process}

Two different surface mechanisms can be considered: the Langmuir-Hinshelwood mechanism and the Eley-Rideal mechanism. In short, a Langmuir-Hinshelwood grain-surface process can involve four major steps:
\begin{enumerate}
\item[]{\bf Accretion.} Particles (atoms or molecules) migrate from the gas phase to the surface of dust grains and are adsorbed onto it.
\item[]{\bf Migration.} Once the particles (adsorbates) are bound to the surface, they can migrate or diffuse over the surface. This mobility is important in order to allow the adsorbates to come together and to react.
\item[]{\bf Reaction.} The reaction between adsorbates occurs when two reaction partners come close enough in order to interact and create new chemical bondings.
\item[]{\bf Ejection.} The reaction products can be ejected from the dust grain and populate the gas phase.
\end{enumerate}
In the case of the Eley-Rideal mechanism, there is no migration and it is considered that the accreting species interact directly with previously adsorbed species, with immediate reaction. In other words, the reaction is considered to occur through collision of a gaseous species with a stationary species held on the surface of a solid.\\

{\bf Accretion.} Atoms or molecules from the gas phase can adsorb on a grain surface. The accretion rate coefficient on grains can be expressed as follows \citep{tielens}:
$$ \mathrm{k_{ac} = n_d\,\sigma_d\,v\,S} $$
where $\mathrm{n_d}$ is the number density of dust grains, $\mathrm{\sigma_d}$ is the cross-section of dust grains, v is the mean speed of particles, and S is the sticking factor translating the probability that a particle colliding with the dust grain accretes onto it. The sticking factor -- that is dependent on the dust and gas temperatures -- is generally close to 1. In the particular case of H, it is evaluated to be equal to 0.8 at 10\,K, and to 0.5 at 100\,K.\\

Two adsorption processes must be considered:
\begin{enumerate}
\item[-] {\it Physisorption}: the weak interaction is that of van der Waals forces. These forces are due to the mutually induced dipole moments in the electron shells of the adsorbate and of the surface of the dust grain. The typical interaction energy is of the order of a few 0.1\,eV. This interaction energy increases when the fixation site corresponds to an kink or a corner in the structure of the dust grain.
\item[-] {\it Chemisorption}: in this case, the interaction can be about one order of magnitude stronger than in the case of physisorption. Forces responsible for this reflect the overlap between the respective wave functions of the adsorbate and the surface, i.e. a chemical bond is created. These forces act on much smaller length scales. The interaction energy is of the order of a few eV.
\end{enumerate}

The binding energy can also be influenced by the composition of the dust grain, and by the presence of adsorbates on the surface that can strengthen or weaken the surface bond. Assessment of adsorption energies can therefore become exceedingly complicated.\\

{\bf Surface migration.} From the energetic point of view, the dust grain surface can be considered as an irregular surface made of hills and valleys of different depth, where the adsorbates can deposit. In this analogy, chemisorption sites correspond to deep holes, and physisorption sites are represented by shallow cavities. On this surface, the adsorbates may migrate from one site to another following some kind of random walk between the moment they accrete and the moment they desorb. The mobility of adsorbates can proceed following two different approaches. First, in the context of our analogy, one can envisage that adsorbate climb up the hills provided they possess enough energy to do it before settling in other cavities. In this case, we are dealing with a thermal hopping migration process. However, even at very low temperature, the mobility of the lightest adsorbates is not completely annealed. In this latter case, the mobility is supported by the quantum effect called tunneling \citep[see e.g.][]{cazaux}.

We may therefore consider three different kinds of motion on the surface: (a) a transition between a physisorbed site and a chemisorbed site, (b) a transition from a chemisorbed site to another chemisorbed site, and (c) and a transition from a physisorbed site to another physisorbed site. Generally speaking, heavier species are less mobile that lighter ones. For this reason, efficient reactions will mostly involve at least one light (and therefore mobile) species such as H, and also C, N, and O.

The migration time-scale dependence on the dust temperature can be expressed as follows:
$$ \tau_{m} \propto \exp\,\mathrm{\Big(\frac{E_m}{k\,T_d}\Big)} $$
where $\mathrm{E_m}$ is the energy barrier against migration (typically of the order of one third of the binding energy), and $\mathrm{T_d}$ is the dust temperature.\\

{\bf Reaction.} During their random walk along the grain surface, adsorbates (potential reaction partners) may come close enough from each other to interact and create (and/or break) chemical bonds. Such interactions will allow new chemical species to be formed, before being rejected in the gas phase.

We may first consider reactions involving radicals -- therefore reactions without activation barrier. Such reactions will involve mostly H, C, N, and O with other potentially heavier radical partners. Beside these radical reactions, reactions with activation barrier are also worth considering. As discussed earlier in this paper, such reactions are inhibited in gas phase. However, the prolongated residence time on grain surfaces allows such reactions to occur significantly. This is one of the main distinguishing characteristics of grain surface chemistry with respect to gas phase.\\

Let us consider a mobile species adsorbed on a site where it can react with a neighboring species with a given barrier (E$_a$). The probability for a reaction before evaporation ($p_r$) as given by \citet{tielens} is then
$$ p_r = \tau\,\theta\,p $$
where $p$ is the probability for the penetration of the reaction activation barrier. It has been assumed that the reactant has much more chance to migrate than to penetrate the reaction barrier. This latter relation clearly shows that the probability that the reaction occurs depends directly on the overall time spent by the atom on the grain surface ($\tau$), and on the surface coverage of potential coreactants ($\theta$). The dependence with respect to the activation barrier of the reaction is included in the factor $p$. A few examples of reactions with their activation barrier (expressed in K) are given in Table\,\ref{actbar-grain}.

\begin{table}[h]
\begin{center}
\caption{Examples of grain surface reactions with activation barriers involving H atoms.\label{actbar-grain}}
\vspace*{3mm}
\begin{tabular}{lclc}
\hline
Reactants	&	& products & E$_a$(K)\\
\hline
H + CO & $\longrightarrow$ & HCO & 1000 \\
H + O$_2$ & $\longrightarrow$ & HO$_2$ & 1200 \\
H + H$_2$O$_2$ & $\longrightarrow$ & H$_2$O + OH & 1400 \\
H + O$_3$ & $\longrightarrow$ & O$_2$ + OH & 450 \\
H + C$_2$H$_2$ & $\longrightarrow$ & C$_2$H$_3$ & 1250 \\
H + C$_2$H$_4$ & $\longrightarrow$ & C$_2$H$_5$ & 1100 \\
H + H$_2$S & $\longrightarrow$ & SH + H$_2$ & 860 \\
H + N$_2$H$_2$ & $\longrightarrow$ & N$_2$H + H$_2$ & 650 \\
H + N$_2$H$_4$ & $\longrightarrow$ & N$_2$H$_3$ + H$_2$ & 650 \\
\hline
\end{tabular}
\end{center}
\end{table}

{\bf Evaporation.} The residence time of an adsorbed chemical species on the surface of a dust grain can be expressed as follows:
$$ \tau_{ev} = \nu_o^{-1}\,\exp\,\mathrm{\Big(\frac{E_b}{k\,T_d}\Big)} $$
where $\mathrm{E_b}$ is the binding energy of the adsorbed species, and $\mathrm{T_d}$ is the dust temperature. $\nu_o$ is the vibrational frequency of the adsorbed species on the grain surface. The residence time is the typical time-scale for a species to acquire sufficient energy through thermal fluctuations to evaporate. This time-scale is very sensitive to temperature elevations of the dust grain. Heavier species will benefit from longer residence times, and this dependence on the mass of the particle is provided by the factor $\nu_o$.

As an example, let us consider the case of the adsorption of atomic H on dust grains. For a typical value of $\nu_o$ of 10$^{-12}$, and a bonding energy ($\mathrm{E_b/k}$) of the order of 500\,K for a physisorbed site, the residence time of atomic H is of the order of 2\,$\times$\,10$^{5}$\,s at 10\,K, and 2\,$\times$\,10$^{-8}$\,s at 40\,K. For a chemisorbed site with a typical bonding energy of the order of 20000\,K, the residence time tends to $\infty$, even at 100\,K. The conclusion is that, in the absence of significantly strong adsorption site, a temperature of a few tens of K is too large to allow a reaction of adsorbed hydrogen atoms with any other adsorbed partners during its short residence time. At such temperatures, chemisorbed sites are necessary to envisage surface reactions of atomic hydrogen.

\subsubsection{The formation of homonuclear diatomic species \label{formhomodia}}

As we are dealing with surface processes, we will not adopt densities expressed in cm$^{-3}$, but we will talk about monolayer surface coverage translating the fraction of the total dust grain surface that is covered by a given species. The reaction rates are therefore expressed in units of monolayer per second. In the discussion below, moslty inspired by the work of \citet{cazaux} devoted to H$_2$ formation, only the Langmuir-Hinshelwood mechanism is considered.

Let us first consider the surface reaction process of A atoms leading to the formation of A$_2$:
$$ \mathrm{A + A \rightarrow A_2} $$
\noindent Here, we consider that A atoms accrete onto the grain surface with a rate $F$. The surface coverages of physisorbed and chemisorbed reactants are noted A$_P$ and A$_C$ respectively, and that of the reaction product is noted A$_2$. We will also make use of diffusion coefficients noted $\alpha_i$, with the subscript $i$ depending on the nature of the starting and arrival sites, i.e. physisorbed (P) or chemisorbed (C). The desorption rate coefficients will be noted $\beta_{A_P}$, $\beta_{A_C}$ and $\beta_{A_2}$ respectively for physisorbed A, chemisorbed A and A$_2$ (assumed to occupy only physisorbed sites\footnote{This assumption is correct if, for instance, molecular hydrogen is considered. The only two electrons in such a molecule are involved in the intramolecular chemical bonding, and could therefore not be made available for any chemical bonding with the surface. Chemisorption is therefore inhibitted for such molecules.}). We can write the rate equations of physisorbed and chemisorbed A atoms, taking into account every gain and loss processes likely to affect their surface coverage:

\begin{eqnarray*}
\mathrm{\dot{A}}_P & = & F\,(1 - \mathrm{A}_P - \mathrm{A}_2) - \alpha_{PC}\,\mathrm{A}_P\,(1 - \mathrm{A}_C)- \alpha_{PC}\,\mathrm{A}_P\,\mathrm{A}_C - 2\,\alpha_{PP}\,\mathrm{A}_P^2 \\
 & & + \alpha_{CP}\,\mathrm{A}_C\,(1 - \mathrm{A}_P - \mathrm{A}_2) - \alpha_{CP}\,\mathrm{A}_C\,\mathrm{A}_P - \beta_{A_P}\,\mathrm{A}_P
\end{eqnarray*}

\begin{eqnarray*}
\mathrm{\dot{A}}_C & = & \alpha_{PC}\,\mathrm{A}_P\,(1 - \mathrm{A}_C) - \alpha_{PC}\,\mathrm{A}_P\,\mathrm{A}_C - \alpha_{CP}\,\mathrm{A}_C\,(1 - \mathrm{A}_P - \mathrm{A}_2)\\
 & & - \alpha_{CP}\,\mathrm{A}_C\,\mathrm{A}_P - 2\,\alpha_{CC}\,\mathrm{A}_C^2 - \beta_{A_C}\,\mathrm{A}_C
\end{eqnarray*}

\begin{figure}[h]
\begin{center}
\includegraphics[width=8cm,angle=0]{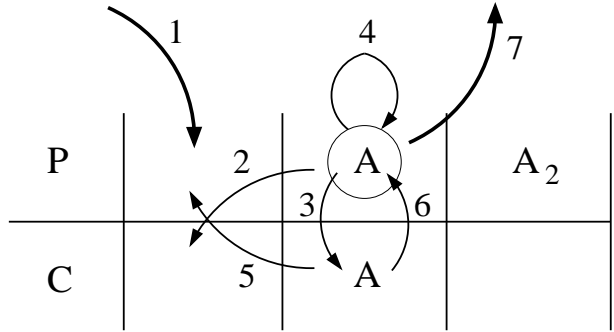}
\caption{Elementary processes diagram illustrating the census of elementary processes affecting the surface coverage of physisorbed A species (surrounded by a circle), for the formation of A$_2$. The upper and lower parts of the diagram stand respectively for physisorbed and chemisorbed sites. The boxes in each row illustrate the occupation of adsorption sites: they can be empty, or occupied by atomic or molecular species.\label{EPD}}
\end{center}
\end{figure}

A convenient way to establish such rate equations is based on elementary processes diagrams (EPD), such as displayed in Fig.\,\ref{EPD} for the rate equation of physisorbed A. Each elementary process is illustrated by a unique arrow, and each arrow translates into one term in the right hand side of the rate equation. The first term is the rate of incoming atoms that are staying on the surface ({\it arrow number 1 in the EPD}). It is considered indeed that the accretion occurs only on physisorbed sites (direct chemisorption is unlikely), taking into account that a fraction of physisorption sites are already occupied by atoms or molecules (here, only species made of A are taken into account). The second term expresses the rate of physisorbed atoms moving to free chemisorption sites ({\it arrow number 2}). The third term stands for the recombination of moving physisorbed A atoms with chemisorbed A atoms ({\it arrow number 3}). The fourth term stands for the recombination between physisorbed A atoms ({\it arrow number 4}). The fifth and sixth terms represent respectively the migration of chemisorbed atoms to free physisorbed sites ({\it arrow number 5}) and the reaction of chemisorbed atoms with physisorbed ones ({\it arrow number 6}). The last term expresses the desorption of physisorbed A atoms ({\it arrow number 7}).

A similar approach can be adopted for the rate equation of chemisorbed A. The first term expresses the rate of A atoms coming from physisorbed sites and migrating into free chemisorbed sites. The second term stands for the recombination between a moving physisorbed atom and a chemisorbed atom. The third term expresses the migration of a chemisorbed atom into free physisorbed sites. The fourth term represents recombination of moving chemisorbed A atoms with physisorbed A atoms. The fifth term expresses the rate of recombination between two chemisorbed A atoms. The last term stands for the desorption of chemisorbed A atoms.

The desorption of the A$_2$ molecules can proceed following two approaches:
\begin{enumerate}
\item[-] a first order desorption where the physisorbed molecule leaves the dust grain surface.
\item[-] a second order desorption where the molecule evaporates from the surface during the formation process. Typically, the energy released by the formation of the molecule is used to leave the energy well of the physisorption site. It is considered that only a fraction $\mu$ of the molecules stays in the physisorption site after its formation. The complementary fraction (1 - $\mu$) desorbs spontaneously during the formation process.
\end{enumerate}

Taking into account these considerations, we may write the rate equation for A$_2$:

$$ \mathrm{\dot{A}}_2 = \mu\,(\alpha_{PP}\,\mathrm{A}_P^2 + \alpha_{CP}\,\mathrm{A}_C\,\mathrm{A}_P + \alpha_{PC}\,\mathrm{A}_P\,\mathrm{A}_C + \alpha_{CC}\,\mathrm{A}_C^2) - \beta_{A_2}\,\mathrm{A}_2 $$

\vspace{0.2cm}
and the total desorption rate is

$$ \mathrm{\dot{A}}_2 = (1 -\mu)\,(\alpha_{PP}\,\mathrm{A}_P^2 + \alpha_{CP}\,\mathrm{A}_C\,\mathrm{A}_P + \alpha_{PC}\,\mathrm{A}_P\,\mathrm{A}_C + \alpha_{CC}\,\mathrm{A}_C^2) + \beta_{A_2}\,\mathrm{A}_2 $$
where the two terms stand respectively for the second order and first order desorption rates.\\

The production of an A$_2$-like species in the interstellar medium is of course not so simple, but this discussion gives a idea of the process that is now believed to be responsible for the formation of homonuclear diatomic molecules. The most straightforward example is that of molecular hydrogen, but this process may hold also for other diatomic species such as O$_2$ or N$_2$, for instance (even though for species with atomic numbers superior to that of hydrogen, the chemisorption of diatomic molecules may also be considered). 

\subsubsection{The formation of heteronuclear diatomic species \label{formheterodia}}

If the formation of heteronuclear diatomic species is envisaged, one has to consider a process of the type:
$$ \mathrm{A + B \rightarrow AB} $$
\noindent Once again, for the sake of simplicity, we will consider that molecular species can only occupy physisorbed sites. As a first step, let's consider that species A and B can only form AB molecules (i.e. homunuclear species A$_2$ and B$_2$ are not formed). Following the approach described above, we derive the following rate equations for physisorbed and chemisorbed A and B:

\begin{eqnarray*}
\mathrm{\dot{A}}_P & = & F\,(1 - \mathrm{A}_P - \mathrm{B}_P - \mathrm{AB}) - \alpha_{PC}\,\mathrm{A}_P\,(1 - \mathrm{A}_C - \mathrm{B}_C)  \\
 & & - \alpha_{PC}\,\mathrm{A}_P\,\mathrm{B}_C - \alpha_{PP}\,\mathrm{A}_P\,\mathrm{B}_P - \alpha_{CP}^\prime\,\mathrm{B}_C\,\mathrm{A}_P - \alpha_{PP}^\prime\,\mathrm{B}_P\,\mathrm{A}_P \\
 & & + \alpha_{CP}\,\mathrm{A}_C\,(1 - \mathrm{A}_P - \mathrm{B}_P - \mathrm{AB})  - \beta_{A_P}\,\mathrm{A}_P
\end{eqnarray*}

\begin{eqnarray*}
\mathrm{\dot{A}}_C & = & \alpha_{PC}\,\mathrm{A}_P\,(1 - \mathrm{A}_C - \mathrm{B}_C) - \alpha_{PC}^\prime\,\mathrm{B}_P\,\mathrm{A}_C \\
 & & - \alpha_{CP}\,\mathrm{A}_C\,(1 - \mathrm{A}_P - \mathrm{B}_P - \mathrm{AB}) - \alpha_{CP}\,\mathrm{A}_C\,\mathrm{B}_P\\
 & & - \alpha_{CC}\,\mathrm{A}_C\,\mathrm{B}_C - \alpha_{CC}^\prime\,\mathrm{B}_C\,\mathrm{A}_C - \beta_{A_C}\,\mathrm{A}_C
\end{eqnarray*}

\begin{eqnarray*}
\mathrm{\dot{B}}_P & = & F\,(1 - \mathrm{A}_P - \mathrm{B}_P - \mathrm{AB}) - \alpha_{PC}^\prime\,\mathrm{B}_P\,(1 - \mathrm{A}_C - \mathrm{B}_C)  \\
 & & - \alpha_{PC}^\prime\,\mathrm{B}_P\,\mathrm{A}_C - \alpha_{PP}\,\mathrm{A}_P\,\mathrm{B}_P - \alpha_{CP}\,\mathrm{A}_C\,\mathrm{B}_P - \alpha_{PP}^\prime\,\mathrm{B}_P\,\mathrm{A}_P \\
 & & + \alpha_{CP}^\prime\,\mathrm{B}_C\,(1 - \mathrm{A}_P - \mathrm{B}_P - \mathrm{AB}) - \beta_{B_P}\,\mathrm{B}_P
\end{eqnarray*}

\begin{eqnarray*}
\mathrm{\dot{B}}_C & = & \alpha_{PC}^\prime\,\mathrm{B}_P\,(1 - \mathrm{A}_C - \mathrm{B}_C) - \alpha_{PC}\,\mathrm{A}_P\,\mathrm{B}_C \\
 & & - \alpha_{CP}^\prime\,\mathrm{B}_C\,(1 - \mathrm{A}_P - \mathrm{B}_P - \mathrm{AB}) - \alpha_{CP}^\prime\,\mathrm{B}_C\,\mathrm{A}_P\\
 & & - \alpha_{CC}^\prime\,\mathrm{B}_C\,\mathrm{A}_C - \alpha_{CC}\,\mathrm{A}_C\,\mathrm{B}_C - \beta_{B_C}\,\mathrm{B}_C
\end{eqnarray*}

Finally, the rate equations for the reaction product AB, respectively for adsorbed and desorbed species, are
\begin{eqnarray*}
\mathrm{\dot{AB}} & = & \mu\,(\alpha_{PP}\,\mathrm{A}_P\,\mathrm{B}_P + \alpha_{CP}\,\mathrm{A}_C\,\mathrm{B}_P + \alpha_{PC}\,\mathrm{A}_P\,\mathrm{B}_C + \alpha_{CC}\,\mathrm{A}_C\,\mathrm{B}_C \\
 & & + \alpha_{PP}^\prime\,\mathrm{B}_P\,\mathrm{A}_P + \alpha_{CP}^\prime\,\mathrm{B}_C\,\mathrm{A}_P + \alpha_{PC}^\prime\,\mathrm{B}_P\,\mathrm{A}_C + \alpha_{CC}^\prime\,\mathrm{B}_C\,\mathrm{A}_C) \\
 & & - \beta_{AB}\,\mathrm{AB}
\end{eqnarray*}

\begin{eqnarray*}
\mathrm{\dot{AB}} & = & (1 -\mu)\,(\alpha_{PP}\,\mathrm{A}_P\,\mathrm{B}_P + \alpha_{CP}\,\mathrm{A}_C\,\mathrm{B}_P + \alpha_{PC}\,\mathrm{A}_P\,\mathrm{B}_C + \alpha_{CC}\,\mathrm{A}_C\,\mathrm{B}_C \\
 & & + \alpha_{PP}^\prime\,\mathrm{B}_P\,\mathrm{A}_P + \alpha_{CP}^\prime\,\mathrm{B}_C\,\mathrm{A}_P + \alpha_{PC}^\prime\,\mathrm{B}_P\,\mathrm{A}_C + \alpha_{CC}^\prime\,\mathrm{B}_C\,\mathrm{A}_C) \\
 & & + \beta_{AB}\,\mathrm{AB}
\end{eqnarray*}
One should note that in the particular case where reactant B is the same as reactant A, the rate equations given above take the form given in Section\,\ref{formhomodia}. The situation becomes however significantly more complex when the simultaneous formation of homonuclear diatomic species is considered as (i) additional terms should be included to take into account homonuclear processes and (ii) the surface coverage of the additional species should not be neglected. In such circumstances, the rate equations have the following form:

\begin{eqnarray*}
\mathrm{\dot{A}}_P & = & F\,(1 - \mathrm{A}_P - \mathrm{B}_P - \mathrm{AB} - \mathrm{A}_2 - \mathrm{B}_2) - \alpha_{PC}\,\mathrm{A}_P\,(1 - \mathrm{A}_C - \mathrm{B}_C)\\
 & & - \alpha_{PC}\,\mathrm{A}_P\,\mathrm{A}_C - 2\,\alpha_{PP}\,\mathrm{A}_P^2 - \alpha_{PP}\,\mathrm{A}_P\,\mathrm{B}_P - \alpha_{PP}^\prime\,\mathrm{B}_P\,\mathrm{A}_P\\
 & & - \alpha_{PC}\,\mathrm{A}_P\,\mathrm{B}_C - \alpha_{CP}^\prime\,\mathrm{B}_C\,\mathrm{A}_P - \alpha_{CP}\,\mathrm{A}_C\,\mathrm{A}_P\\
 & & + \alpha_{CP}\,\mathrm{A}_C\,(1 - \mathrm{A}_P - \mathrm{B}_P - \mathrm{AB} - \mathrm{A}_2 - \mathrm{B}_2) - \beta_{A_P}\,\mathrm{A}_P
\end{eqnarray*}

\begin{eqnarray*}
\mathrm{\dot{B}}_P & = & F\,(1 - \mathrm{A}_P - \mathrm{B}_P - \mathrm{AB} - \mathrm{A}_2 - \mathrm{B}_2) - \alpha_{PC}^\prime\,\mathrm{B}_P\,(1 - \mathrm{A}_C - \mathrm{B}_C)\\
 & & - \alpha_{PC}^\prime\,\mathrm{B}_P\,\mathrm{A}_C - 2\,\alpha_{PP}^\prime\,\mathrm{B}_P^2 - \alpha_{PP}^\prime\,\mathrm{B}_P\,\mathrm{A}_P - \alpha_{PP}\,\mathrm{A}_P\,\mathrm{B}_P\\
 & & - \alpha_{PC}^\prime\,\mathrm{B}_P\,\mathrm{B}_C - \alpha_{CP}^\prime\,\mathrm{B}_C\,\mathrm{B}_P - \alpha_{CP}\,\mathrm{A}_C\,\mathrm{B}_P\\
 & & + \alpha_{CP}^\prime\,\mathrm{B}_C\,(1 - \mathrm{A}_P - \mathrm{B}_P - \mathrm{AB} - \mathrm{A}_2 - \mathrm{B}_2) - \beta_{B_P}\,\mathrm{B}_P
\end{eqnarray*}

 \begin{eqnarray*}
\mathrm{\dot{A}}_C & = & \alpha_{PC}\,\mathrm{A}_P\,(1 - \mathrm{A}_C - \mathrm{B}_C) - \alpha_{CP}\,\mathrm{A}_C\,(1 - \mathrm{A}_P - \mathrm{B}_P - \mathrm{AB} - \mathrm{A}_2 - \mathrm{B}_2)\\
 & & - \alpha_{CP}\,\mathrm{A}_C\,\mathrm{A}_P - \alpha_{CP}\,\mathrm{A}_C\,\mathrm{B}_P - \alpha_{PC}\,\mathrm{A}_P\,\mathrm{A}_C - \alpha_{PC}^\prime\,\mathrm{B}_P\,\mathrm{A}_C\\
 & & - 2\,\alpha_{CC}\,\mathrm{A}_C^2 - \alpha_{CC}\,\mathrm{A}_C\,\mathrm{B}_C - \alpha_{CC}^\prime\,\mathrm{B}_C\,\mathrm{A}_C - \beta_{A_C}\,\mathrm{A}_C
\end{eqnarray*}

\begin{eqnarray*}
\mathrm{\dot{B}}_C & = & \alpha_{PC}^\prime\,\mathrm{B}_P\,(1 - \mathrm{A}_C - \mathrm{B}_C) - \alpha_{CP}^\prime\,\mathrm{B}_C\,(1 - \mathrm{A}_P - \mathrm{B}_P - \mathrm{AB} - \mathrm{A}_2 - \mathrm{B}_2)\\
 & & - \alpha_{CP}^\prime\,\mathrm{B}_C\,\mathrm{B}_P - \alpha_{CP}^\prime\,\mathrm{B}_C\,\mathrm{A}_P - \alpha_{PC}^\prime\,\mathrm{B}_P\,\mathrm{B}_C - \alpha_{PC}\,\mathrm{A}_P\,\mathrm{B}_C\\
 & & - 2\,\alpha_{CC}^\prime\,\mathrm{B}_C^2 - \alpha_{CC}^\prime\,\mathrm{B}_C\,\mathrm{A}_C - \alpha_{CC}\,\mathrm{A}_C\,\mathrm{B}_C  - \beta_{B_C}\,\mathrm{B}_C
\end{eqnarray*}

\noindent With the following rate equations for the three diatomic species:

\begin{eqnarray*}
\mathrm{\dot{A}}_2 & = & \mu\,(\alpha_{PP}\,\mathrm{A}_P^2 + \alpha_{CP}\,\mathrm{A}_C\,\mathrm{A}_P + \alpha_{PC}\,\mathrm{A}_P\,\mathrm{A}_C + \alpha_{CC}\,\mathrm{A}_C^2) - \beta_{A_2}\,\mathrm{A}_2
\end{eqnarray*}

\begin{eqnarray*}
\mathrm{\dot{B}}_2 & = & \mu\,(\alpha_{PP}^\prime\,\mathrm{B}_P^2 + \alpha_{CP}^\prime\,\mathrm{B}_C\,\mathrm{B}_P + \alpha_{PC}^\prime\,\mathrm{B}_P\,\mathrm{B}_C + \alpha_{CC}^\prime\,\mathrm{B}_C^2) - \beta_{B_2}\,\mathrm{B}_2
\end{eqnarray*}

\begin{eqnarray*}
\mathrm{\dot{AB}} & = & \mu\,(\alpha_{PP}\,\mathrm{A}_P\,\mathrm{B}_P + \alpha_{CP}\,\mathrm{A}_C\,\mathrm{B}_P + \alpha_{PC}\,\mathrm{A}_P\,\mathrm{B}_C + \alpha_{CC}\,\mathrm{A}_C\,\mathrm{B}_C \\
 & & + \alpha_{PP}^\prime\,\mathrm{B}_P\,\mathrm{A}_P + \alpha_{CP}^\prime\,\mathrm{B}_C\,\mathrm{A}_P + \alpha_{PC}^\prime\,\mathrm{B}_P\,\mathrm{A}_C + \alpha_{CC}^\prime\,\mathrm{B}_C\,\mathrm{A}_C) \\
 & & - \beta_{AB}\,\mathrm{AB}
\end{eqnarray*}

Te latter rate equation is obviously the same as in the previous case, where only the formation of AB was considered. The most obvious example of reaction process obeying the rate equations established above is the formation of molecular hydrogen in the presence of deuterium, with synthesis of the three isotopologues H$_2$, HD and D$_2$.

\subsubsection{The formation of heteronuclear polyatomic species \label{formheteropoly}}

The situations becomes even more difficult to deal when first generation products are allowed to react with simple atomic species, reaching therefore higher level of molecular complexity. Let us consider for instance a mechanism described by the following surface processes:

\begin{eqnarray*}
A + B & \rightarrow & AB \\
A + A & \rightarrow & A_2 \\
B + B & \rightarrow & B_2 \\
AB + B & \rightarrow & AB_2 \\
AB_2 + B & \rightarrow & AB_3 \\
\end{eqnarray*}

All rate equations will not be detailed here, but as an example one may consider the rate equation of physisorbed B, assuming all molecular species are only physisorbed, and that their mobility is inhibitted.

\begin{eqnarray*}
\mathrm{\dot{B}}_P & = & F\,(1 - \mathrm{A}_P - \mathrm{B}_P - \mathrm{AB} - \mathrm{A}_2 - \mathrm{B}_2 - \mathrm{AB}_2 - \mathrm{AB}_3) \\
 & & + \alpha_{CP}^\prime\,\mathrm{B}_C\,(1 - \mathrm{A}_P - \mathrm{B}_P - \mathrm{AB} - \mathrm{A}_2 - \mathrm{B}_2 - \mathrm{AB}_2 - \mathrm{AB}_3) \\
 & & - \alpha_{PC}^\prime\,\mathrm{B}_P\,(1 - \mathrm{A}_C - \mathrm{B}_C) - 2\,\alpha_{PP}^\prime\,\mathrm{B}_P^2 - \alpha_{PC}^\prime\,\mathrm{B}_P\,\mathrm{A}_C \\
 & & - \alpha_{PP}^\prime\,\mathrm{B}_P\,\mathrm{A}_P - \alpha_{PP}\,\mathrm{A}_P\,\mathrm{B}_P - \alpha_{CP}\,\mathrm{A}_C\,\mathrm{B}_P - \alpha_{CP}^\prime\,\mathrm{B}_C\,\mathrm{B}_P \\
 & &  - \alpha_{PC}^\prime\,\mathrm{B}_P\,\mathrm{B}_C - \alpha_{PP}^\prime\,\mathrm{B}_P\,\mathrm{AB} - \alpha_{PP}^\prime\,\mathrm{B}_P\,\mathrm{AB}_2 - \beta_{B_P}\,\mathrm{B}_P
\end{eqnarray*}

This case is suitable to consider, for instance, the formation of NH$_3$ in the presence of N and H atoms on dust grains, with the formation of H$_2$, N$_2$, NH and NH$_2$ molecules as secondary and intermediate products. A similar process could be envisaged for the formation of the C$_2$, CH, CH$_2$ and CH$_3$ molecules, also known to significantly populate molecular clouds.\\

If one wants to refine the mechanisms considered above, several aspects of the approach described in this section should be upgraded:
\begin{enumerate}
\item[-] Only the Langmuir-Hinshelwood mechanism has been considered. If the surface coverage of reactants is high enough, additional terms in the rate equations should be included to take into account Eley-Rideal 0processes.
\item[-] If the density of the overlying gas phase is high enough, the accretion rate of chemical species can increase substantially and the approximation of the monolayer is not valid anymore.
\item[-] Only one typical migration coefficient has been considered for each kind of motion on the grain surface (C$\rightarrow$C, C$\rightarrow$P, P$\rightarrow$P or P$\rightarrow$C). It is obvious that, depending on the nature of the solid grain and on that of other accreted species, physisorption and adsorption interaction may vary significantly. This will have a strong impact on the migration coefficients that should be taken into account in the rate equations.
\item[-] Actually, the diversity of reaction partners adsorbed onto the surface is much larger than illustrated here, leading therefore to very diversified reaction products, among which molecules of increasing complexity. The incorporation of various atoms in the process (H, C, N, O) is the key to synthesize many interesting molecules of increasing complexity. In addition, successive addition reactions involving un-desorbed species lead to heavier products, as illustrated by the last exemple.
\end{enumerate}

\subsubsection{Synergies between gas and dust}

From here, it may be useful to introduce the concept of chemical network. Basically, a chemical network is a group of elementary processes governing the transformations of chemical species and reproducing the potential filiations between molecules in a given medium. When one has to deal with the chemistry of a given part of the interstellar medium, a census of molecular species and of their related transformation processes has to be made.

A chemical network can reach rapidly a high level of complexity. Many different kind of processes as described above are likely to take place, and all species present in the medium are most probably involved in several processes simultaneously, as reactant or as reaction product. Quantitatively, chemical networks are modelled using theoretical codes fed with a lot of data including mainly reaction rate coefficients (determined in laboratories or theoretically) and abundances when made available thanks to spectroscopic studies.\\

Several general rules based on chemical considerations govern gas-phase chemical networks:
\begin{enumerate}
\item[-] the most abundant molecules are the most probable reaction partners. As H$_2$ is the most abundant molecule in the Universe, reactions involving this latter molecule will dominate the chemistry. CO being the second most abundant molecule in the Universe, its role in astrochemistry should never be neglected.
\item[-] reaction rate coefficients for ion-neutral reaction are significantly higher than for neutral-neutral reactions. For this reason, ions will dominate the chemistry. Ionizing agents are therefore very important for gas-phase astrochemistry, whatever the environment considered. In diffuse parts of the ISM permeated by ionizing radiation from neighboring stars, ionization will occur through the action of ultraviolet light. In denser parts of the ISM, the interaction of high energy charged particles (cosmic-rays) with neutral species will ionize the medium. A probable channel for this ionization process is the ionization of abundant He atoms, followed by charge transfer to other neutral species. The action of cosmic-rays is unlikely in diffuse clouds because of the low probability of interaction with other species. In the latter circumstances, it may contribute but it will not dominate.
\item[-] the main loss channels for ions is the reaction with H$_2$ (the most abundant potential partners), or electron recombination (characterized by larger rate coefficients). One should not forget that free electrons are abundant in a medium populated by cations.
\item[-] the most abundant ions in the diffuse interstellar medium are C$^+$, H$^+$ and He$^+$. This is easily understood if one considers the abundance of their parent elements.
\item[-] in molecular clouds (significantly denser than diffuse clouds), carbon and hydrogen are mostly found in molecular species, and the dominating ions are H$_3^+$, HCO$^+$ and H$_3$O$^+$. Here again the role of He$^+$ should not be neglected.
\item[-] if there is a lack of dominating ions, neutral species will react most probably with small radicals (atoms or diatomic molecules).
\item[-] generally speaking, involving ions in the reaction network constitutes a crucial leap to molecular complexity, as it is the key element that activates gas phase chemistry.
\end{enumerate}

The molecular build-up in the interstellar medium starts from atomic gas. However, given the low rates for radiative association reactions, the first step combining two atoms into a diatomic species is an issue in the gas phase. The main implication is that gas phase chemistry is not enough to explain the chemical complexity of the interstellar medium. For this reason, grain surface processes must be taken into account. As dust grains act as catalysts, the kinetics of chemical processes can be substantially improved, and the combination of atomic species in order to build the first diatomic molecules can efficiently take place. Once this rate determining step has been overcome, small molecules may constitute valuable reaction partners involved both in gas phase and grain surface processes. The net result of this joint action of gas phase and surface processes is a significant synthesis of molecules of increasing complexity on time-scales of the order of -- or shorter than -- the typical evolution time-scales of astrophysical environments. The a priori idea that interstellar environments could not harbour significant amounts of diversified molecules is consequently obsolete, in agreement with the current census of interstellar molecules discussed in Sect.\,\ref{census}.

In order to illustrate the chemistry that is likely to take place in the ISM, the example of carbon monoxyde is of particular interest. Indeed, CO is the most abundant molecule of the Universe after molecular hydrogen. Its first interstellar detection comes from \citet{WJP-co} thanks to observations in the radio domain. CO is commonly used as a proxy for molecular hydrogen, as it is generally more easy to detect than H$_2$ and their abundances are often correlated \citep[see e.g.][]{dickman-h2-co}. This correlation comes from the fact that the radio emission of CO (mainly at 2.6 and 1.3\,mm) is caused by collisional excitation with H$_2$ molecules. Through this process, the emission lines of CO are good tracers of molecular hydrogen whose UV lines are often heavily obscured in dense clouds.

\begin{figure}[h]
\begin{center}
\includegraphics[width=11cm,angle=0]{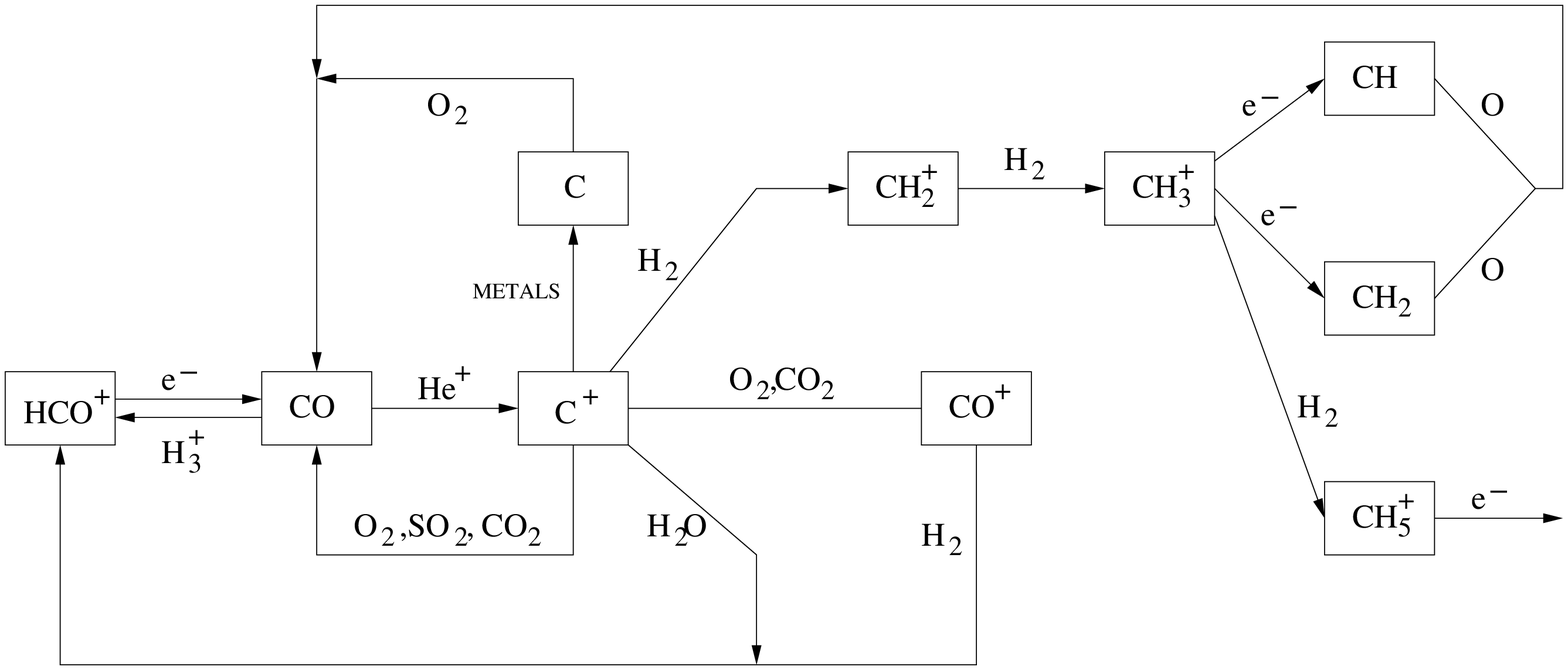}
\caption{Simplified gas-phase chemical network including the main reactions involved in the formation and destruction of CO (adapted from \citealt{prasad-cno}).\label{co-cn}}
\end{center}
\end{figure}

The formation of carbon monoxide is likely to occur rather efficiently in gas phase, through the radical reactions (neutral-neutral reactions) between C or O, with a small O- or N-bearing molecule, i.e. between C and OH, between C and O$_2$, between CH and O, or between CH$_2$ and O. The dissociative electronic recombination reaction involving the ion HCO$^+$ may also contribute. These reactions are included in the gas-phase chemical network presented in Fig.\,\ref{co-cn}. 

\begin{figure}[h]
\begin{center}
\includegraphics[width=15cm,angle=0]{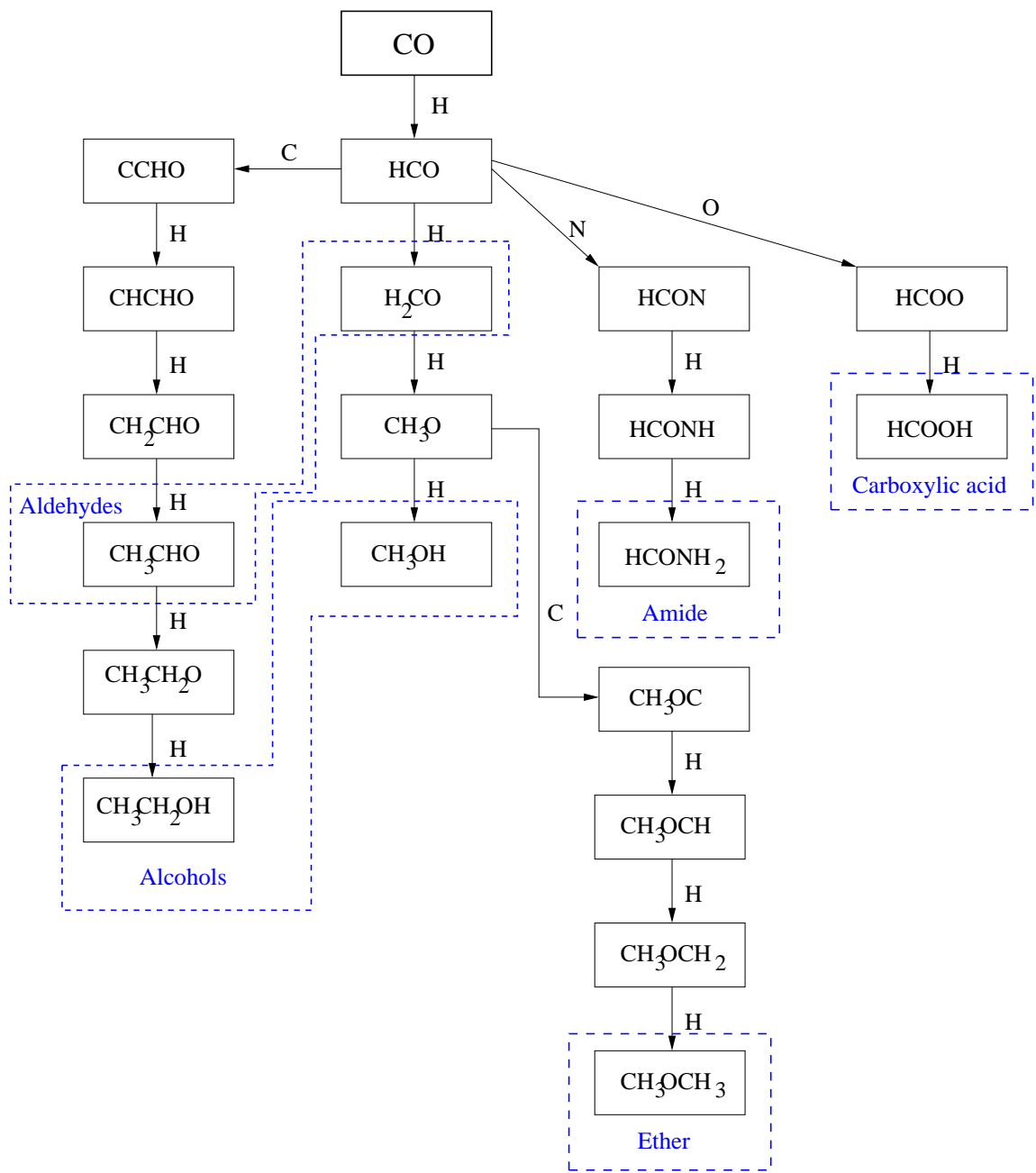}
\caption{Simplified grain surface chemical network including reactions involving CO, along with some of its derivatives, with abundant mobile atomic species.\label{grainco}}
\end{center}
\end{figure}

Carbon monoxide synthesized for instance in tha gas-phase could easily adsorb on dust grains and therefore be involved in surface processes. A straghtforward example is the hydrogenation of CO, leading to various chemical species belonging to the category of organic compounds (see Sect.\,\ref{sect-org}). As illustrated in Fig.\,\ref{grainco}, the simultaenous presence of CO and other lighter (and therefore more mobile) species such as H, N, C and O allows the possibility to increase the diversity and complexity of chemical compounds, leading to the formation of various functional groups of significant relevance in organic chemistry. The chemical network in Fig.\,\ref{grainco} is notably based on one of the more fundamental principles governing grain surface chemistry: reactions should mainly involve mobile species (the most abundant one being atomic hydrogen) with less mobile (heavier) species. A consequence of this principle is the successive hydrogenation of small organic compounds, leading to the formation for instance of alcohols, carboxylic acids, and other functional groups. The latter molecules could afterwards be rejected in gas-phase where they could be involved in other processes. The difficult transition from small species to somewhat heavier molecules has been successfully passed, and the molecular content of the interstellar cloud can continue to evolve through the joint action of gas-phase and grain surface processes.

\begin{figure}[h]
\begin{center}
\includegraphics[width=10cm,angle=0]{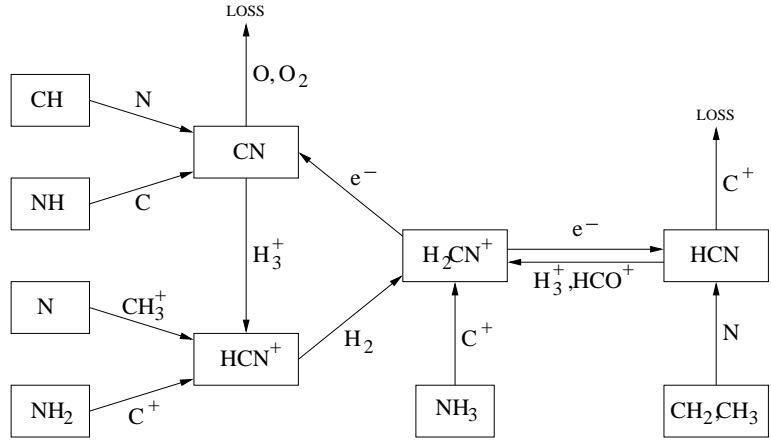}
\caption{Simplified chemical network including the main reactions involved in the formation and destruction of cyanide and hydrogen cyanide (adapted from \citealt{prasad-cno}).\label{cn-cn}}
\end{center}
\end{figure}

Another important example is that of the cyanide radical (CN). This is one of the first molecules identified in the interstellar medium. The first interstellar lines originating from CN were reported by \citet{ism-cn}.The chemistry of cyanide is likely related to that of hydrogen cyanide molecule (HCN) whose first detection in the interstellar medium is due to \citet{snyder-hcn}. A simplified chemical network describing the chemistry of CN and of HCN is shown in Fig.\,\ref{cn-cn}. Here again, one could envisage similar processes such as those considered in the case of CO on dust grains, leading to the formation of various amine functional groups.

Several molecules containing a cyano group have been found in the interstellar medium: NaCN \citep{turner-nacn}, MgCN \citep{ziurys-mgcn}, SiCN \citep{guelin-sicn}. In addition, organic compounds containing CN groups have been detected as well (see Sect.\,\ref{sect-org}).

So far, we have seen how the combined action of gas-phase and grain-surface processes could lead to the formation of the most simple molecules, and how such processes could lead to the growth of chemical species in order to increase molecular complexity. The same approach will now be followed to explore the increase of molecular complexity in the particular -- and very important -- class of organic comounds.

\subsection{From simple hydrocarbons to (poly-)functionalized organic compounds}

\subsubsection{Aliphatic and aromatic hydrocarbons} 

Carbon chemistry occupies a priviledged position in chemistry, as it constitutes the basement of organic compounds, including the building blocks of life. Historically, the most simple molecule containing only hydrogen and carbon discovered in the ISM is the CH radical, and this discovery has been reported at the end of the 30's by \citet{ch-swings}. Since then, many polyatomic hyrocarbons characterized by various degrees of unsaturation have been identified. In this discussion, we will mostly address the cases of the two main classes of compounds: {\it aliphatic hydrocarbons} and {\it aromatic hydrocarbons}. A few words will also be given on other carbonaceous compounds (likely to be) found in the ISM.

\begin{figure}[ht]
\begin{center}
\includegraphics[width=11cm,angle=0]{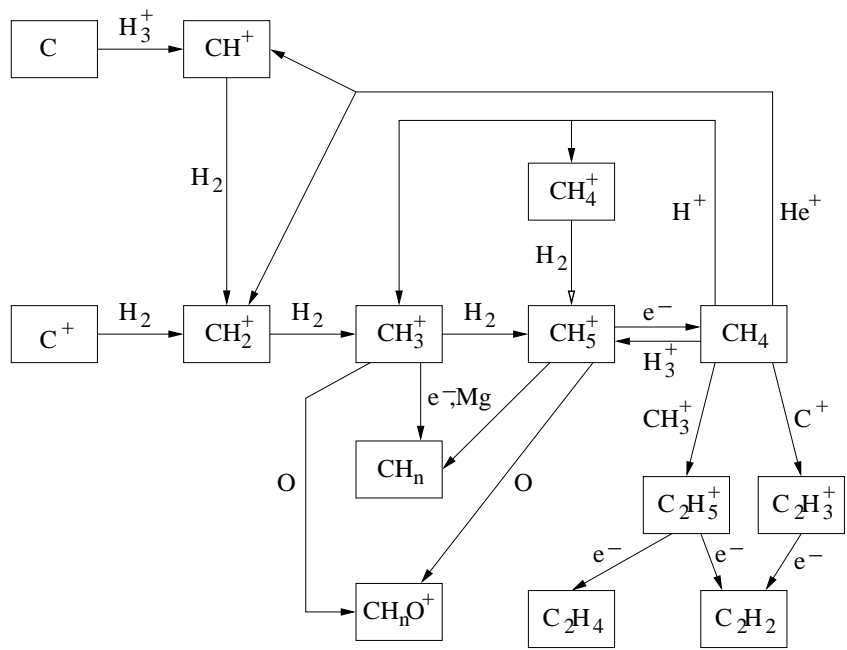}
\caption{Simplified chemical network including the main reactions involved in the formation of methane in interstellar clouds (adapted from \citealt{prasad-cno2}).\label{methane-cn}}
\end{center}
\end{figure}

In order to illustrate the main chemical processes leading to the formation of organic compounds, we may first consider the chemical reactions involved in the formation of the simplest saturated hydrocarbon, i.e. methane (CH$_4$), that has been detected in the interstellar medium for the first time by \citet{lacy-ch4}. The corresponding gas phase chemical network, as proposed by \citet{prasad-cno2}, is illustrated in Fig.\,\ref{methane-cn}. 

In this example, we see that the two main pathways leading to low molecular weight hydrocarbons are:
\begin{enumerate}
\item[-] the reaction of a neutral carbon atom with cationic H-bearing species. In the chemical network presented here, H$_3^+$, as an abundant cationic molecule containing H, is considered.
\item[-] the reaction of cationic carbon with H$_2$, which acts as a preferential partner for obvious abundance reasons.
\end{enumerate}
Of course, other potential partners may also be involved in such a process, but with probably weaker contributions. It is also important to note that no neutral-neutral reactions have been considered here because of their significantly lower reaction rates as compared to ion-molecule processes, as explained in Sect.\,\ref{sect-ion-mol}. It has also been emphasized in Sect.\,\ref{sect-der} that the dissociative electronic recombination is a very important process for polyatomic cationic species, leading to neutral compounds after the loss of a fragment. Here, several compounds may undergo such a process, and it is illustrated for instance in the case of cationic hydrocabons that are precursors of species such as methane, ethylene or acetylene. One should also note the important role of cosmic-rays, and secondary cations such as He$^+$, in the destruction of methane and other molecules in such molecular clouds. 

Alternatively, the successive addition of H starting from atomic C is a possible formation mechanism on dust grains. Indeed, in agreement with the principles developed in Sect.\,\ref{grain}, as mobile species, H atoms constitute valuable reaction partners for less mobile ones already adsorbed on dust grains, such as C$^+$, or hydrocarbons such as CH$_2$, or other molecules of the same series of hydrogenated cationic carbons. 

The spectroscopic signature of aliphatic C--H stretching viabrations in the near infrared has been reported in most directions, revealing the existence of mainly short aliphatic chains \citep[see e.g.][]{sandfordratio}. On the other hand, several unsaturated hydrocabons have been firmly identified. Among these molecules, one could mention detection of a few neutral unsaturated carbon chains such as C$_8$H \citep{c8h}, C$_7$H \citep{c7h} or C$_4$H$_2$ and C$_6$H$_2$ \citep{ism-benzene}. Additional examples can be found in the more scarce category of anions, with  C$_4$H$^-$ \citep{secondanion}, C$_6$H$^-$ \citep{firstanion} and C$_6$H$^-$ \citep{thirdanion}.\\

\begin{figure}[ht]
\begin{center}
\includegraphics[width=9cm,angle=0]{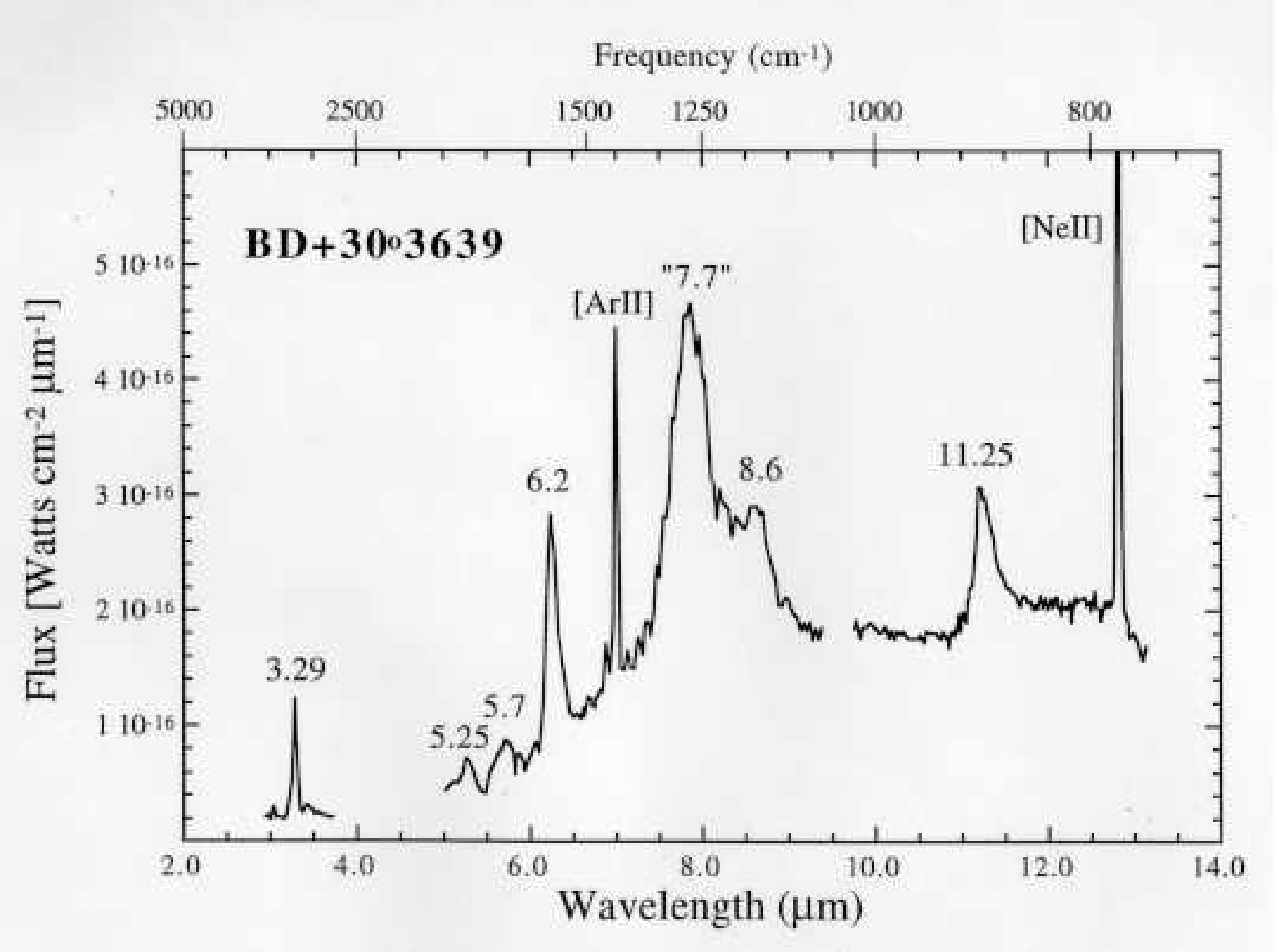}
\caption{Infrared spectrum in the direction of the planetary nebula BD\,+36$^\circ$3639. Most of the porminent spectral features are attributed to PAHs (\copyright NASA).\label{pahir}}
\end{center}
\end{figure}

Investigations of the Universe in the infrared revealed also a priori unexpected interstellar emission bands at 3.29, 6.2, 8.7 of 11.3\,$\mu$m forming the core of an important issue in interstellar astrophysics \citep[see e.g.][]{gfm-pah}. For instance, see the infrared spectrum shown in Fig.\,\ref{pahir}. Such spectral features have been found to be associated with many astrophysical objects such as planetary and reflection nebulae, H\,{\sc ii} regions, or even extragalactic sources. The ubiquity and the strength of these spectral lines showed that they were due to widespread and abundant interstellar carriers. It has first been proposed that the mechanism responsible for such emission bands was most likely infrared molecular fluorescence lines pumped by UV radiation \citep{allam-pah1}. Although the mechanism appears to be correct, the idea that large molecules such as aromatic compounds were the carriers of those lines came a bit later \citep{dw-pah,lp-pah,allam-pah2}. Since then, it is now admitted by the scientific community that polycyclic aromatic hydrocarbons (PAHs) constitute an abundant component of the interstellar medium, and that they play a significant role in its energy budget.

\begin{figure}[ht]
\begin{center}
\includegraphics[width=10cm,angle=0]{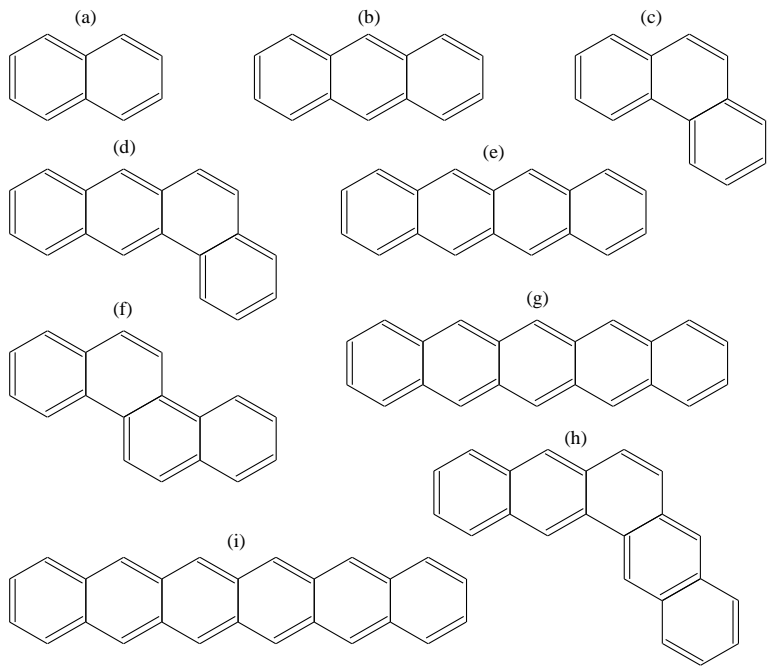}
\caption{Molecular structure of a few catacondensed PAHs: (a) naphtalene, (b) anthracene, (c) phenanthrene, (d) tetraphene, (e) naphtacene (or tetracene), (f) chrysene, (g) pentacene, (h) pentaphene and (i) hexacene. In each case, only one resonance form is illustrated.\label{pah-cata}}
\end{center}
\end{figure}

The most common aromatic molecule is benzene. This molecule has been discovered in the interstellar medium through infrared observation by \citet{ism-benzene}. Polycyclic aromatic hydrocarbons are mainly planar molecules allowing the delocalization of electron across more than one aromatic ring. Following this idea, the next aromatic compound starting from benzene contains 4 additional carbons providing each one additional $p$ electron: i.e. naphtalene. If we go on this way, adding 4 other carbons to increase linearly the aromatic compound, we obtain anthracene. The detection of cationic anthracene (C$_{10}$H$_{14}^+$) has recently been reported by \citet{anthracenecation}. These molecules start the series of {\it catacondensed} polycyclic aromatic hydrocarbons (see Fig.\,\ref{pah-cata}). In this series, no carbon belongs to more than two rings. In catacondensed PAHs, we can envisage two subclasses: the {\it acenes} which are linear, and the {\it phenes} which consists in bent rows of rings.

\begin{figure}[ht]
\begin{center}
\includegraphics[width=10cm,angle=0]{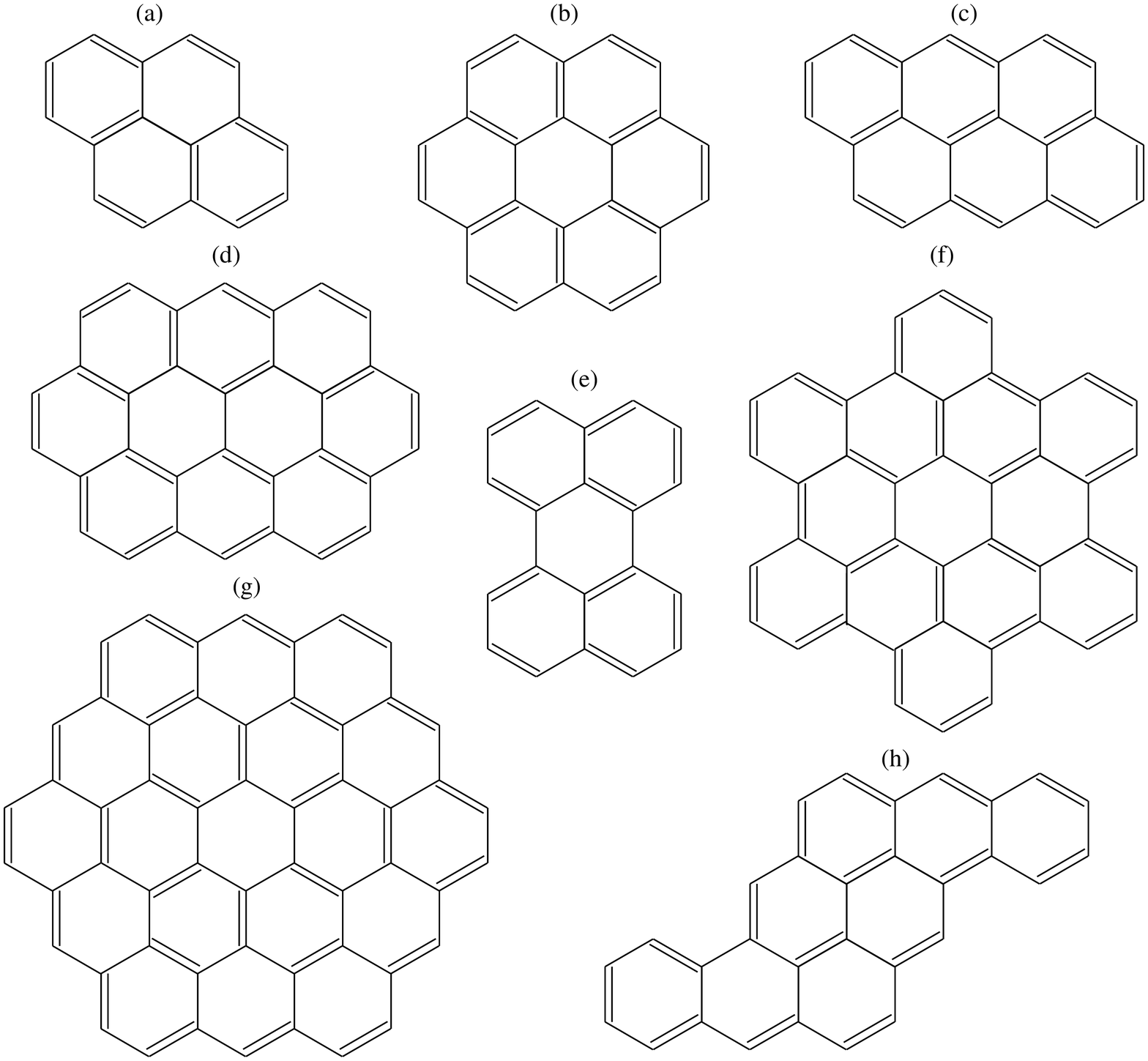}
\caption{Molecular structure of a few pericondensed PAHs: (a) pyrene, (b) coronene, (c) antanthrene, (d) ovalene, (e) perylene, (f) hexabenzocoronene, (g) circumcoronene and (h) pyranthrene. In each case, only one resonance form is illustrated.\label{pah-peri}}
\end{center}
\end{figure}

Beside catacondensed PAHs, we can consider also the class of {\it pericondensed} PAHs containing carbons that are members of three separate rings. The structure of some members of this class is shown in Fig.\,\ref{pah-peri}. Pericondensed PAHs are sometimes called superaromatics. Among pericondensed aromatics, centrally condensed ones are particularly stable because their structure generally allows an improved electron delocalization. The general formula of centrally condensed PAHs is C$_{6r^2}$H$_{6r}$, where r is an integer. For $r$ equal to 1, we retrieve benzene. For $r$ = 2, we obtain coronene, and so on. For such aromatics, there are $3r^2 - 3r + 1$ hexagonal cycles arranged in $r - 1$ rings around the central cycle. Considering a typical C-C bond length of about 1.4\,\AA\,., the surface area of one aromatic ring is about 5\,\AA\,. It should also be noted that PAHs containing a few tens of carbons can have sizes larger than 100\,\AA\,. Such sizes are typical of the smallest dust grains in the interstellar medium. The characteristic size of these molecules is therefore intermediate between small molecules and dust particles.\\

\begin{figure}[h]
\begin{center}
\includegraphics[width=9cm,angle=0]{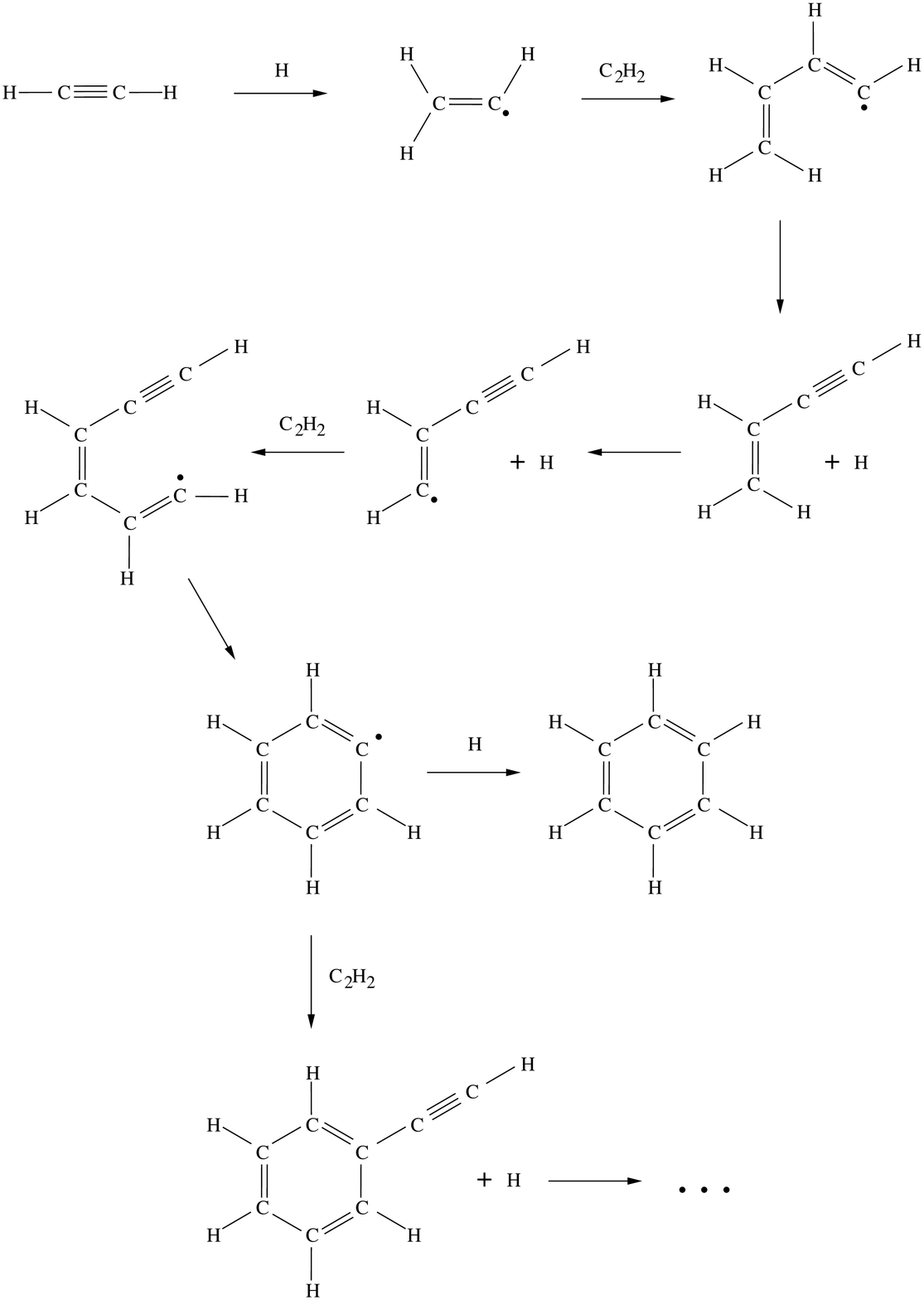}
\caption{Probable formation mechanism of benzene initiated by a radical species, i.e. hydrogenated acetylene, adapted from \citet{allam-pah3}. The process repeats until the formation of a phenyl radical. This latter radical can yield benzene through the addition of hydrogen, or can undergo additional reactions with C$_2$H$_2$ molecules leading to more extended PAHs.\label{pahform2}}
\end{center}
\end{figure}

The main process likely to lead to the formation of PAHs is the condensation of short carbon chains. Condensation is the name given to chemical processes leading to the formation of new C-C bondings in organic compounds. In the case of the formation of aromatic compounds, this process consists in the ion-molecule or in the radical-molecule addition of short unsaturated aliphatic molecules such as alcenes, allenes and alcynes. For instance, we can consider the reactions illustrated in Fig.\,\ref{pahform2} leading to the formation of the phenyl radical (C$_6$H$_5$) through the successive addition of C$_2$H$_2$ molecules. From such a phenyl radical, further additions of acetylene molecules are likely to increase the size of the aromatic molecule through successive formation of additional rings \citep[see][]{allam-pah3}. 

\begin{figure}[ht]
\begin{center}
\includegraphics[width=9cm,angle=0]{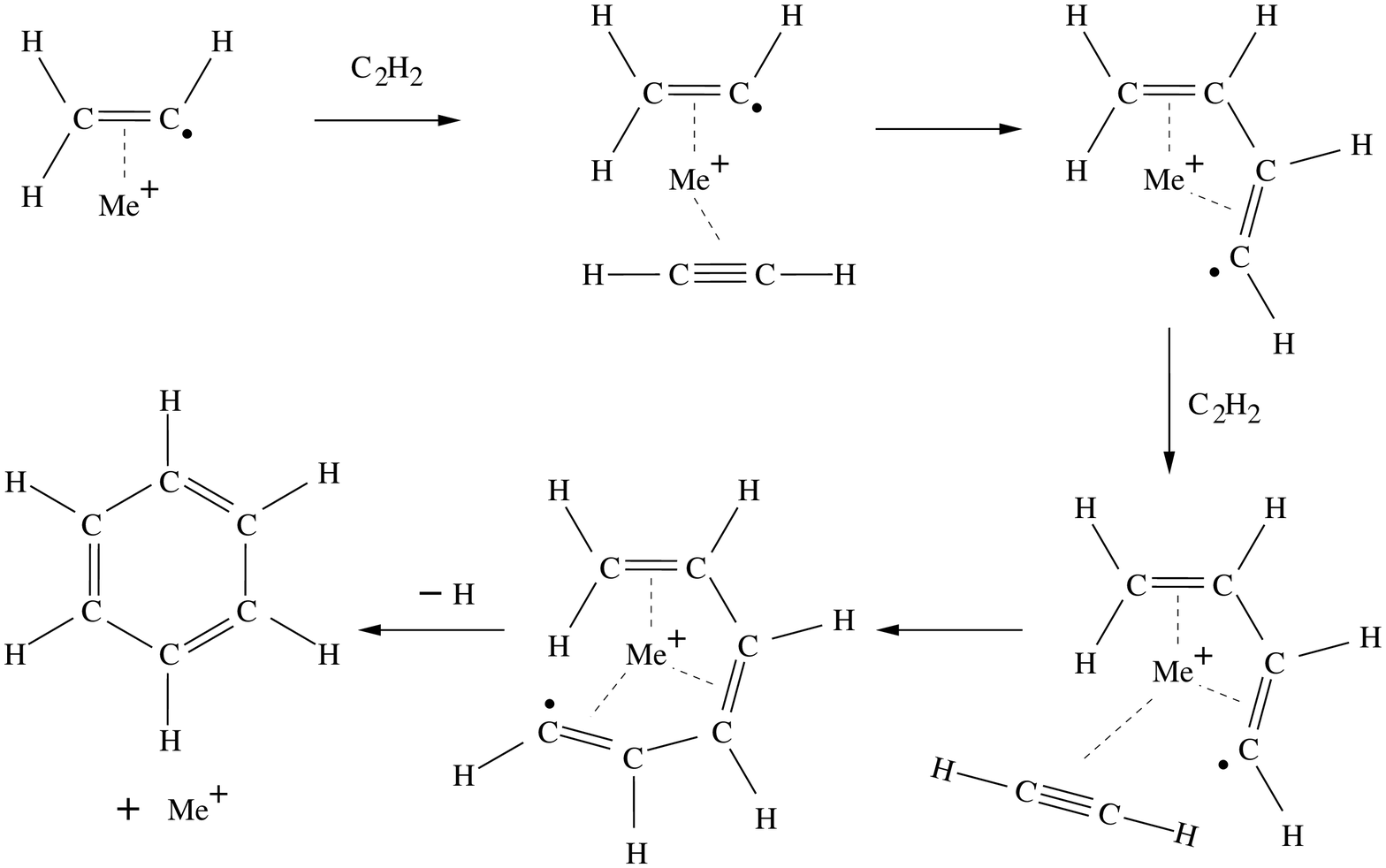}
\caption{Probable formation mechanism of benzene where a metallic cation acts as a catalyst. The positive charge favors the approach of acetylene molecules, and the presence of the cation constrains the conformation adopted by the radicalar intermediates. A similar mechanism as in Fig.\,\ref{pahform2} in the presence of the metallic cation can also lead to the formation of phenyl radical.\label{pahorgmet}}
\end{center}
\end{figure}

It should also be noted that we cannot reject a scenario where the addition of small unsaturated aliphatic chains is catalyzed by metallic ions, via the formation of organometallic compounds as illustrated in Fig.\,\ref{pahorgmet}. The interaction of metallic cations with $\pi$ electrons of alcenes can lead to the formation of adducts to catalize the approach of molecules likely to react to form aromatics \citep[see e.g.][]{pah-bohme}.

It has also been proposed that the extension of the size of PAHs may proceed through the concerted addition of monocyclic aromatics. A concerted addition occurs when bonds are formed and others are broken simultaneously. Such an approach is therefore more favorable from the activation point of view. For instance, \citet{pah-shock} considered the formation of PAHs by shock waves propagating in the interstellar medium. In this scenario, the energy required for the cycloaddition is taken from the shock. In the context of concerted cycloaddition, the product of the cycloaddition has to undergo an intramolecular rearrangement followed by the elimination of H$_2$ in order to yield planar PAHs.\\

A significantly important aspect of the chemistry of PAHs in the interstellar medium is {\it photochemistry}. As mentioned earlier in this section, the presence of PAHs in the Galaxy and beyond has been revealed by their infrared emission bands arising from the relaxation of their vibrational degrees of freedom, after excitation by UV photons. The high capability of PAH to relax progressively through cascades in their vibrational states is reponsible for their high stability upon photodissociation. However, some channels in these cascades may lead to the breaking of chemical bondings with subsequent loss of fragments \citep[see e.g.][]{tielens}.

The probability of photodissociation can be expressed by the following relation:
$$ \mathrm{p_d(E)} = \frac{\mathrm{k(E)}}{\mathrm{k(E) + k_{IR}(E)}} $$
where k(E) and k$_{\mathrm{IR}}$(E) are respectively the rate coefficients for the dissociation and for the infrared radiative relaxation channels.

In case of dissociation, two pathways can be envisaged:
\begin{enumerate}
\item[-] Side $\sigma$ bond rupture: the absorption of a photon can lead to the breaking of a C-X bond, where X is hydrogen or a sidegroup such as -OH, -CH$_3$, -NH$_2$. Alternatively, the bond rupture can occur in the sidegroup, with subsequent release of H or for instance of the methyl group of an ethyl sidegroup \citep{geballe-pah}.
\item[-] Concerted decondensation: as an alternative reaction channel leading to the breaking of $\sigma$ bonds, the possible concerted fragmentation of an aromatic molecule into acetylene molecules should also be considered. According to \citet{allam-pah3}, this process is the only pathway involving the breaking of C-C bonds in aromatics that may be envisaged, even though it has a higher energy threshold than the rupture of sidegroup bonds.
\end{enumerate}

Another process that is important for PAHs is an {\it electron attachment} process. These large molecules are indeed likely to give rise to radiative electronic recombinations such as
$$ \mathrm{PAH + e^- \rightarrow (PAH^-)^* \rightarrow PAH^- + h\nu} $$
The electronic capture ($\mathrm{k_c}$) leads to an activated complex which can itself relax radiatively ($\mathrm{k_r}$) to yield the negatively charged PAH, or it can undergo an electronic detachment ($\mathrm{k_d}$) leading back to the neutral PAH. Such a two-step process involving an activated complex is very similar to that of radiative association reactions described in Sect.\,\ref{rar}.

In the case of the electron capture of PAHs, the reaction rate coefficient is generally expressed the following way:
$$ \mathrm{k_{ra} = \big(\frac{k_c}{k_r + k_d}\big)\,k_r = \big(\frac{k_r}{k_r + k_d}\big)\,k_c = S_e\,k_c}$$
where $\mathrm{S_e}$ is defined as the sticking coefficient of the electron on the PAH. The same expression for the recombination of PAH cations holds, but in the latter case the radiative stabilization is very rapid and the sticking coefficient is essentially unity.\\

Processes such as {\it charge exchange} are also important for PAHs. The abundances of PAH anions in molecular clouds is regulated by the balance between electron capture and the recombination with metal or molecular cations:
$$ \mathrm{PAH^- + M^+ \rightarrow  PAH + M} $$
Such a process is likely the most important one leading to the neutralization of PAH anions in the ISM.\\

 Beside the charge transfer reaction described above, {\it ion-neutral reactions} involving PAHs can lead to the formation of adducts ions. For instance, the addition of Si$^+$ is mentioned by \citet{millar-pah}:
$$ \mathrm{PAH + Si^+ \rightarrow  SiPAH^+} $$
Another example of ion-molecule reaction that may be important in interstellar clouds is the interaction of PAHs with ionized helium \citep{pah-bohme}:
$$ \mathrm{PAH + He^+ \rightarrow  PAH^{+2} + He + e^-} $$
This reaction is a significant source of dications in dense clouds where He$^+$ is abundant because of the efficient ionization of neutral helium by cosmic-rays.\\

It should also be noted that {\it PAH-catalysed processes} should play a role in the chemistry of various astrophysical environments. As chemical species are likely to form adducts with PAHs (see above), it has been proposed that adsorption-like processes may occur involving PAHs and atomic hydrogen \citep[see e.g.][]{pah-bohme}. As a result, a process similar to that described in Sect.\,\ref{grain} for the formation of small molecules may take place on large PAHs. Considering the large abundance of PAHs in some regions of the interstellar medium, such a process may constitute a significant channel for the formation of molecular hydrogen, for instance. A brief discussion of this scenario is given by \citet{pah-bohme}. Following this idea, other PAH-catalyzed processes are likely to take place in the interstellar medium such as neutralization reactions, or other chemical reactions unlikely to take place efficiently in gas phase without the support of a third partner.\\

Finally, the last class of carbonaceous compounds worth mentioning is that of fullerenes. Since their discovery by \citet{fullerkroto}, this class of molecules has been considered to be likely present in interstellar clouds. The discovery of C$_{60}$ and C$_{70}$ in astrophysical environments was however confirmed rather recently \citep{fullercami,c60snellgren,gielen2011}, even though some hints of their presence were already reported a few years before \citep{c60first2004,c60first2007}. A formation mechanism invoking C-H bond dissociation and internal structural rearrangements through the action of UV light on large PAHs has been proposed by \citet{formc602011}.

\subsubsection{Organic compounds with functional groups}\label{sect-org}

A significant leap forward on the way of chemical complexity is achieved when functional groups are included in hydrocarbons. Molecules containing various functional groups have been detected in several astrophysical environments. A few examples are given below:
\begin{enumerate}
\item[-] Methanol. This is the most simple organic compound containing an hydroxyl group. It has been detected in the interstellar medium by \citet{ball-methanol}.
\item[-] Ethanol. Ethylic alcohol have been detected in the 70's \citep{zuckerman-ethanol}.
\item[-] Dimethylether. This is the simplest molecule containing an ether function. \citet{snyder-ether} reported on its presence in the Orion Nebula. The probable formation mechanism of this molecule has been discussed by \citet{peeters-dme}.
\item[-] Metanethiol. This is the simplest mercaptan, i.e. an organic molecule containing a thiol functional group. The detection of this molecule in the interstellar medium was first reported by \citet{linke-mercaptan}.
\item[-] Cyanomethane. This is the smallest organic molecule found in the ISM containing a nitrile group. Cyanomethane has been detected in molecular clouds by \citet{solomon-cyanome}.
\item[-] Cyanodecapentayne. The longest polyyine (HC$_{11}$N) has been firmly identified in the ISM by \citep{hc11n}.
\item[-] Formaldehyde. This is the simplest molecule containing an aldehyde functional group. It is found in many astrophysical environments, and its first detection was reported by \citet{snyder-formaldehyde}.
\item[-] Acetaldehyde. This is the equivalent of the formaldehyde, but with a hydrogen substituted by a methyl group. Several studies led to the detection of this molecule in the interstellar medium \citep{gottlieb-acetaldehyde,fourikis-acetaldehyde,gilmore-acetaldehyde}.
\item[-] Formic acid. This is the simplest carboxylic acid. It has been detected in the interstellar medium by \citet{zuckerman-formic} and \citet{winnewisser-formic}.
\item[-] Acetic acid. The discovery of acetic acid by \citet{mehringer-acetic} added a new chemical species from the class of carboxylic acids in the census of interstellar molecules.
\item[-] Acetone. This is the simplest organic molecule including a ketone functional group. This molecule has been reported in the interstellar medium by \citet{combes-acetone} and \citet{snyder-acetone}.
\item[-] Methylformate. This is the simplest organic compound including an ester functional group. Its presence has been reported in the direction of the Galactic Center by \citet{brown-meform} and \citet{churchwell-meform}.
\item[-] Formamide. This is the simplest organic molecule including an amide functional group. Its detection dates back in the early 70's \citep{rubin-formamide}.
\item[-] Methylamine. This is the simplest primary amine. Its presence has been reported in various places of the interstellar medium \citep{kaifu-meamine,fourikis-meamine}.
\item[-] Ketenimine. This molecule is simultaneously a representant of the classes of imines and ketenes. Its was discovered in a star forming region by \citet{lovas-ketenimine}.
\end{enumerate}

This short (and incomplete) census of interstellar organic molecules shows that, among the functional groups usually found in organic chemistry, many have already been identified in space. This is rather promising in the sense that most of the organic representatives of chemistry are present in the interstellar medium, and one may expect most of their chemistry to take place in various astrophysical environments. Their detection constitutes in addition a strong evidence that efficient chemical processes likely to produce complex molecules are at work in space. It is also interesting to note that some organic molecules with functional groups have also been firmly identified in comets, as discussed for instance by \citet{comets-ir}, \citet{comets-mm}, \citet{comets-bockelee} and \citet{charrodg2008}. \\

\begin{figure}[h]
\begin{center}
\includegraphics[width=12cm,angle=0]{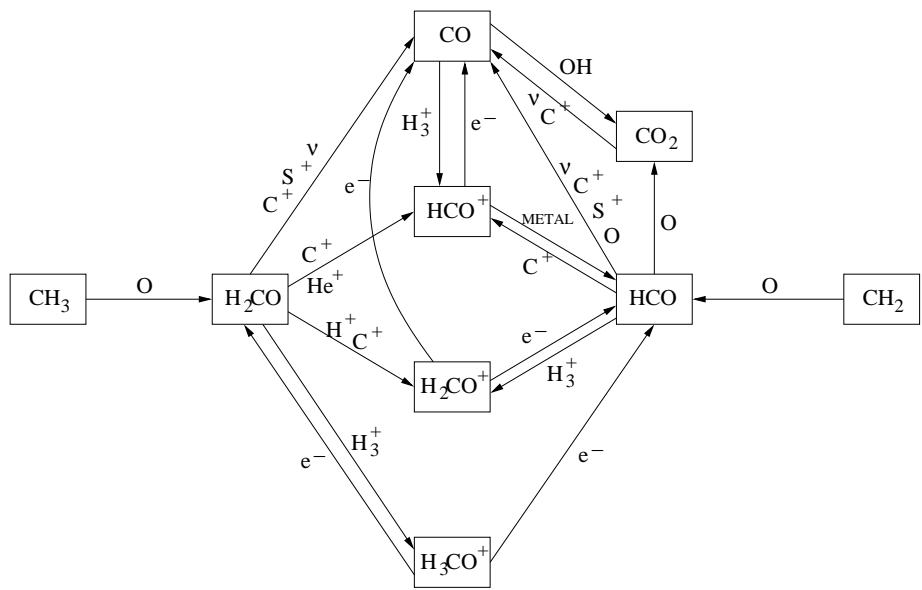}
\caption{Simplified chemical network including the main reactions involved in the formation of formaldehyde in interstellar clouds (adapted from \citealt{prasad-cno2}).\label{formaldehyde-cn}}
\end{center}
\end{figure}

A first illustration of the basic interstellar chemistry likely to lead to such compounds is shown in Fig.\,\ref{formaldehyde-cn}. This example shows the main processes leading to the formation and destruction of a rather abundant molecular species that is very important for chemistry: formaldehyde. 

The two main mechanisms for its formation are the following:
\begin{enumerate}
\item[-] the neutral-neutral reaction between methyl radical and an oxygen atom. As emphasized in Sect.\,\ref{sect-neu-neu}, chemical processes between neutral partners become really significant when radical species are involved, because of the lack of significant activation barrier likely to affect the kinetics of the reaction. This is typically what happens here. In addition, the formation of the new chemical bonding with the oxygen occurs as a H group dissociates, taking away the excess energy likely to destabilize the new molecule.
\item[-] the dissociative electronic recombination of a cationic species. Once again, provided the cationic precursor is sufficiently abundant, it leads fastly to the formation of formaldehyde.
\end{enumerate}
On the other hand, ionic processes may efficiently lead to its destruction. It should also be noted that, provided the astronomical environment is translucent enough to allow ultraviolet photons to interact with it, prohotodissociation is an additional efficient destruction mechanism.\\

An additional leap forward in the sense of increasing molecular complexity is achieved if more than one functional group is included in the same molecule. This is typically what is found in a few molecules that have been identified in space in the last past years. So far, five interstellar molecules belong to this catogory: glycolaldehyde \citep{firstsugar}, ethylene glycol \citep{antifreeze}, amino acetonitrile \citep{aminoacetonitrile}, cyanoformaldehyde \citep{cyanoform} and cyanomethanimine \citep{cyanomethanimine}. These molecules have been discovered in the Srg\,B2 molecular cloud complex located close to the center of the Galaxy. The presence of organic molecules of that complexity is the signature that a complex chemistry is at work in such molecular clouds.

\section{The top-down approach}\label{TD}

\subsection{Biomolecules}

\subsubsection{Carbohydrates}

It is interesting to emphasize that the first molecule of the {\it carbohydrate} series (i.e. glycolaldehyde) has been firmly identified in the ISM. This is especially important if one keeps in mind the important role played by sugars in our complex biochemistry. The exact mechanism likely to lead to its synthesis in the ISM is still not firmly determined. A polimerization-like formation process involving formaldehyde molecules is a possibility, considering the wealth of formaldehyde found in interstellar clouds. It has been proposed that a process similar to the formose reaction\footnote{The formose reaction is a polymerization reaction of formaldehyde catalyzed by water \citep{formosereaction}.} on grain surfaces in molecular clouds could be efficient \citep[see e.g.][]{firstsugar}. In this context, it is interesting to emphasize the clear discovery of many polyols (i.e. molecules containing several hydroxyl groups, presenting therefore properties very similar to carbohydrates) in meteorites \citep{murchison3}.
 
\subsubsection{Nitrogenated bases}

Cyclic molecules containing heterotatoms, i.e. atoms other than carbon, are also worth considering. In the context of the study of chemistry in space and of its relation with Earth's biochemistry, several {\it heterocyclic molecules} such as heterocyclic bases are of crucial interest. The latter molecules belong to the category of building blocks of the genetic material of living beings on Earth. To date, only a few heterocyclic molecules have been found in the interstellar medium: c-C$_2$H$_4$O \citep{dickens-ethdiox}, and presumably c-C$_2$H$_3$O and c-C$_2$H$_5$N. In addition, the analysis of meteoritic organic matter revealed the presence of various more complex heterocyclic compounds, containing one or more nitrogen atoms, including pyrimidines, purines, and derivatives of quinoline and isoquinoline \citep{stoks1,stoks2}. In the interstellar medium, recent investigations failed to detect purines and pyrimidines (see \citet{kuan-pyrimidine} for the case of pyrimidine in interstellar clouds). However, three points deserve to be mentioned here:
\begin{enumerate}
\item[-] the large abundance of carbon rich polycyclic molecules (PAHs) provides strong evidence for the fact that very efficient processes are at work in the interstellar medium to build cyclic and polycyclic organic compounds.
\item[-] the large abundance of compounds such as nitriles or cyanates allows us to expect some of these compounds to be involved in the processes responsible for the formation of unsaturated cyclic organic compounds.
\item[-] in the case of comets, for instance, large complexes of organic compounds made of carbon and nitrogen are likely present \citep{rettig-hcn,matthews-hcn,matthews-hcn2}. These complexes are expected to contain many cyclic units including simultaneously carbon and nitrogen, in a way that is reminiscent to the structure of nitrogen-bearing heterocyclic compounds. This fact constitutes an important hint for the existence of purine- and pyrimidine-like compounds in space.
\end{enumerate}
However, if a molecule such as adenine (C$_5$H$_5$N$_5$) is formed in the ISM, it is unlikely to proceed through the successive neutral-neutral addition of HCN molecules, as demonstrated by \citet{hcnoligomers}. It should also be kept in mind that the stability of pyrimidine, and other nitrogenated heterocycles against UV photolysis is rather limited and far below that of monocyclic and polycyclic aromatic hydrocarbons; and the photostability decreases with the increasing number of nitrogen atoms in the cyclic compounds \citep{nheterostab}. Any circumstellar N-heterocyclic molecule formed on dust grains would likely be destroyed by UV photons or cosmic rays after rejection in the gas phase of the ISM. Acoording to \citet{nheteroform}, such molecules would more probably be formed on small interplanetary bodies such as those found in the form of meteorites on planetary surfaces. Nucleobases such as uracile have indeed been found in a meteorite, with confirmation of the extraterrestrial origin thanks to isotopic enrichment measurements \citep{martins-nubasesmurchison}.

\subsubsection{Amino acids}

The identification of amino acetonitrile in the ISM deserves also a few more comments. The nitrile functional group is indeed likely to be converted into a carboxylic acid through a hydrolysis reaction, leading directly to the first natural {\it amino acid}: glycine. Amino acetonitrile can therefore be considered as a potential direct precursor of glycine.

The discovery of amino acids in Space is really a challenge for molecular spectroscopists, and scenarios to produce such molecules in astrophysical environments are investigated by astrochemists. For instance, the UV irradiation of ice analogues\footnote{Ice analogues are artifical ices prepared in laboratories, whose molecular composition is supposed to be close to that of interstellar or proto-planetary ices.} in laboratory leads easily to the formation of many different amino acids, as demonstrated notably by \citet{AAiceanalogues} and \citet{AAiceanalogues2}. Amino acids may also be produced in interstellar environments via a Strecker-like process, such as discussed for instance by \citet{strecker2010}. The discovery of amino acids in interstellar space would constitute a significant improvement of our view of the molecular content of the Universe, as it would contribute to fill the gap between currently confirmed interstellar molecules and Earth-like biochemistry. A few years ago, the announcement of the discovery of glycine in the interstellar medium by \citet{kuan-glycine} has been questioned. Several studies aimed at confirming this putative discovery but confirmation is still lacking \citep[see e.g.][]{snyder-glycine}. To date, the detection of glycine in the interstellar medium is still not admitted by the astrochemical community. However, more recently, glycine has been identified much closer, in the material ejected from a comet \citep{glystardust}. This detection is credited to the Stardust\footnote{Stardust was a NASA mission designed to collect interplanetary dust and to launch the sample in the Earth's atmosphere. For additional information, see {\tt http://stardustnext.jpl.nasa.gov/}} spacecraft that passes through gas and dust surrounding the icy nucleus of the comet Wild-2, in 2004. The confirmation of the presence of molecules such as amino acids out of the Earth's atmosphere lends significant support to the idea that astronomical environments could have played a pivotal role in prebiotic chemistry.

\subsection{The homochirality problem}
Provided chemical compounds have reached a significant level of complexity, new properties are likely to appear, notably in relation with the stereochemistry of molecules. Chiral compounds (especially amino acids) have indeed been identified in the molecular content of meteorites, in particular the Murchison meteorite which exploded into fragments in Australia, in 1969 \citep[see][]{murchison, murchison2}.

\begin{figure}[h]
\begin{center}
\includegraphics[width=8cm,angle=0]{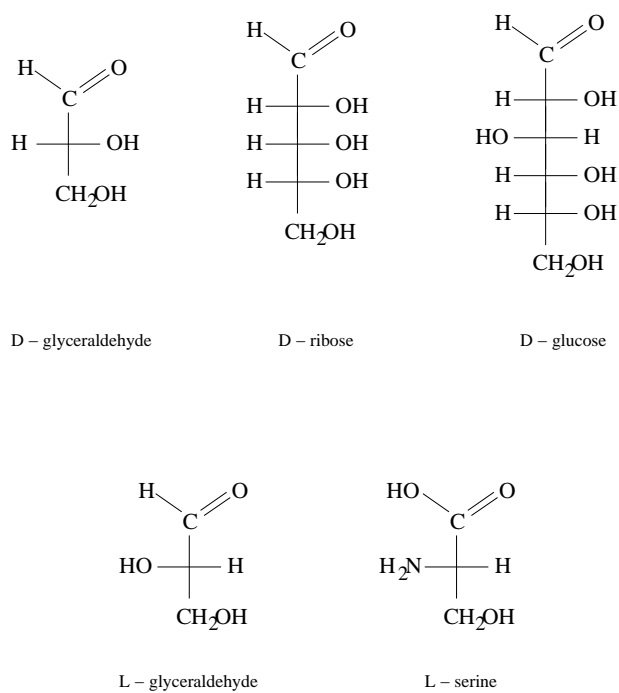}
\caption{Upper part: Fischer projections of D monosaccharides. The assignement of the D notation refers to the fact that the configuration of the last chiral carbon is R, i.e. with the hydroxyl group on the right. Lower part: Fischer projections of the amino acid serine, whose DL-notation is attributed by analogy with glyceraldehyde.\label{DLsugars}}
\end{center}
\end{figure}

In the case of sugars, stereoisomers can be distinguished using conventions leading to the distinction between D and L sugars. The simplest chiral monosaccharide is glyceraldehyde, whose natural enantiomer is denoted (+) as it rotates the plane polarized light in a clockwise direction. (+)-glyceraldehyde is known to have the $R$ configuration at its stereogenic center, and its Fischer projection is shown in Fig.\,\ref{DLsugars}. This enantiomer with the R configuration is conventionnaly called D-glyceraldehyde. For monosaccharides, the generally adopted conventions consider that the carbonyl is represented on top of the projection. In the projection of glyceraldehyde, one can see that the hydroxyl group attached to the chiral center is on the right side. By analogy with the case of glyceraldehyde, all sugars with the same configuration on the chiral carbon opposite to the carbonyl group will be referred to as D sugars. A few examples are provided in Fig.\,\ref{DLsugars}. It is important to note that D and L notations do not indicate the direction in which a given sugar rotates plane-polarized light. A D sugar may be dextrorotatory or levorotatory. The notation only refers to the stereochemistry of the last chiral carbon. The D and L notations are also used to distinguish enantiomers of amino acids. In the latter case, the stereochemical similarity of natural serine with L-glyceraldehyde (where the hydroxyl group is replaced by the amino group) leads to similar conventions: in Fischer projections, amino acids with the amino group on the left (S configuration) are referred to as L amino acids. Here again, the optical activity is not completely determined by the configuration of the chiral center in the amino acid. Among the 20 natural amino acids, 19 are chiral. And among these 19 L amino acids, 9 are dextrorotatory \citep[see e.g.][for a discussion]{mcmurry}!

Homochirality is a term that refers to a group of compounds with the same sense of chirality. In biology, homochirality is found inside living organisms. Active forms of amino acids are almost exclusively of the L-form\footnote{It should be noted that D amino acids are also used in living organisms, in particular in peptidoglycans, i.e. composite macromolecules made of crosslinked polysaccharides and peptides found in cell membranes of bacteria.} and most biologically relevant sugars are of the D-form. The stereochemistry of organic compounds is of crucial importance in biochemistry as it is responsible for the three-dimensional configuration of complex molecules such as proteins and (deoxy)ribonucleic acids. If the stereochemistry of the sugars in nucleotides and of amino acids in proteins was not strictly controlled, biologically important molecules would have very different properties. The fact that we find only L-type amino acids and D-type sugars is therefore a major element in the complex chemistry of living organisms.

The question here is the following: where did this homochirality come from? This question is really a challenge for chemistry, and maybe a question worth asking to astrochemistry.\\

\begin{figure}[ht]
\begin{center}
\includegraphics[width=9cm,angle=0]{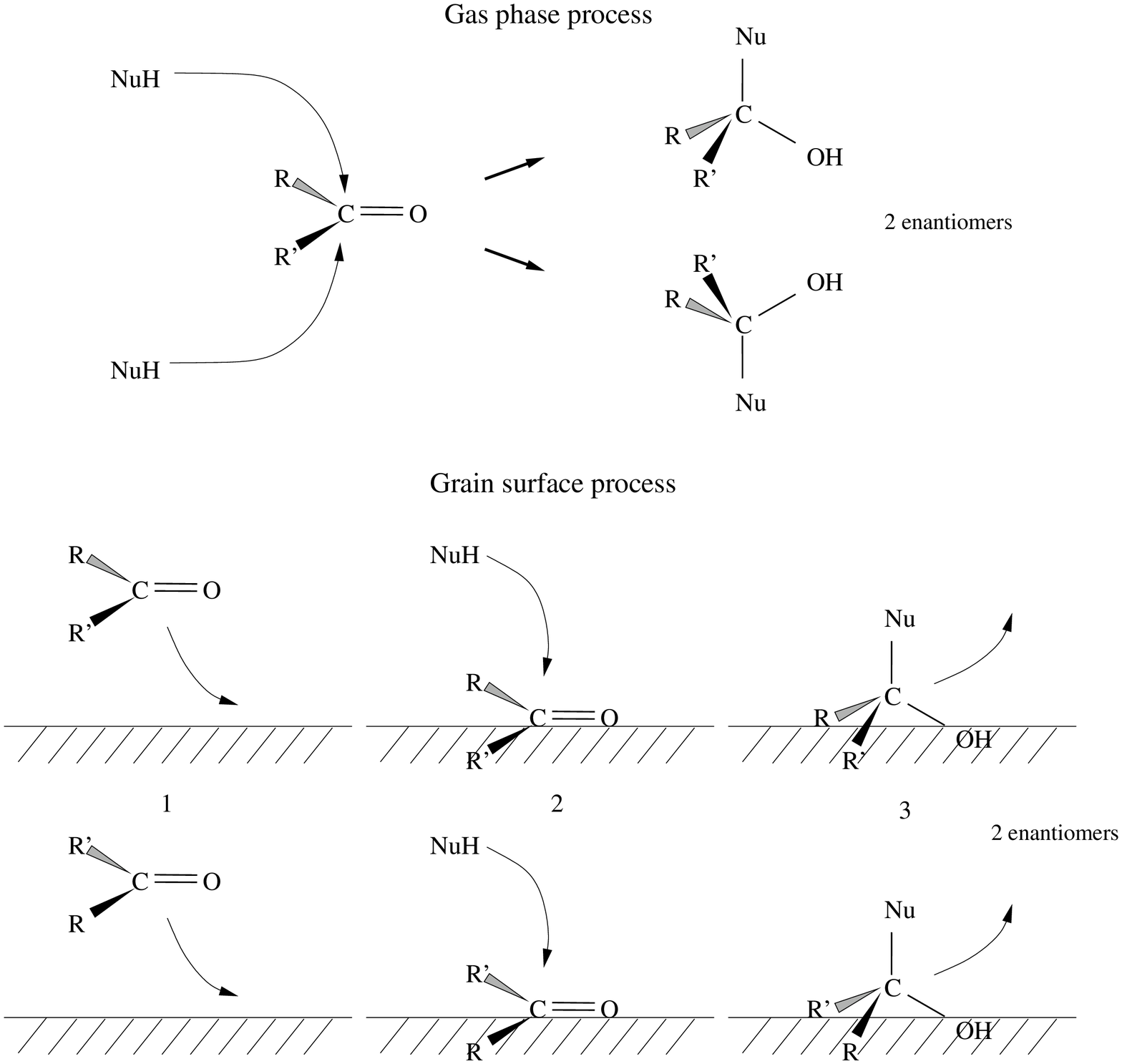}
\caption{Upper part: nucleophilic addition on a carbonyl group in gas phase. Lower part: nucleophilic addition on a carbonyl group on a grain surface.\label{racemate}}
\end{center}
\end{figure}

The problem comes from the fact that without any constraint, the synthesis of chiral molecules leads to racemic mixtures (1:1 ratio between both enantiomers). This can be illustrated in Fig.\,\ref{racemate} in the case of the nucleophilic addition on a carbonyl group. Two major cases can be considered:
\begin{enumerate}
\item[-] Gas phase process. If one considers a plane trigonal carbonyl group in gas phase, the approach of the nucleophilic partner is likely to occur with equal probability on any side of the plane. If a stereogenic center is created, this nucleophilic addition will lead to the two different enantiomers depending on the approach direction. 
\item[-] Grain surface process. The carbonyl bearing molecule will first adsorb on the grain surface. The molecule has a priori equal chance to adsorb on one side or the other. The approach of the nucleophilic partner will occur on the only side that faces the gas phase. In this case, the enantiomeric selection occurs when the molecule adsorbs on the grain surface, and not during the nucleophilic attack.
\end{enumerate}

In both cases, we see that such an addition leads to the two enantiomeric forms of the reaction product, without any selectivity. Both processes lead to a racemic mixture. Basic synthesis processes does therefore not suggest that any enantioselectivity can be expected in the production of chiral molecules.\\

\subsubsection{Symmetry breaking}

As basic synthesis processes do not seem to lead to enantiomeric excesses of chiral compounds, two scenarios may be considered to break the symmetry:
\begin{enumerate}
\item[-] The synthesis of chiral compounds may be constrained by enantioselective interactions in the context of rather complex processes. This is what occurs in living beings on Earth where for instance enzymes act as enantioselective catalysts. But in inorganic environments, such a selectivity is unlikely. As biological systems make use of specific enantiomers, where did the first microorganism find the enantiomeric excess that triggered the homochirality of our biochemistry?
\item[-] That is not during the synthesis of chiral molecules that the enantiomeric symmetry is broken, but during their destruction. This is where the impact of the astrophysical environment is likely to play a role.
\end{enumerate}

\begin{figure}[ht]
\begin{center}
\includegraphics[width=10cm,angle=0]{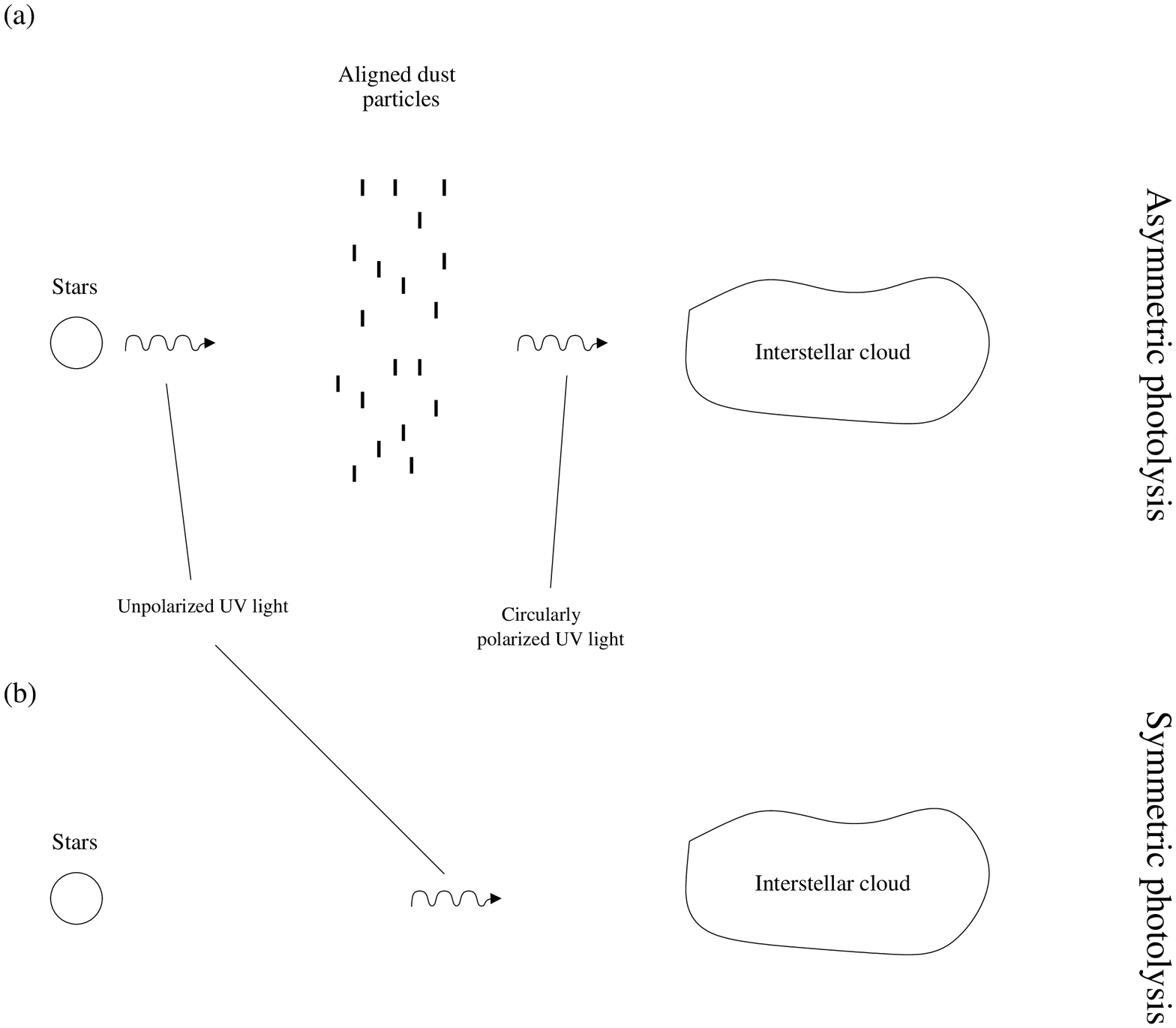}
\caption{Illustration of the impact of an asymmetry on the astronomical scale on the asymmetry on the molecular scale, through circular polarization of light due to magnetically aligned dust particles. (a) asymmetric photolysis due to circularly polarized light. (b) symmetric photolysis in the presence of unpolarized light. \label{asymphot}}
\end{center}
\end{figure}

The latter scenario is worth considering in the context of the interstellar medium permeated by UV light likely to break chemical bonding, and therefore destroy molecules. As already discussed, photodissociation of chemical species is a common process in the interstellar medium. In addition, it is interesting to note that circularly polarized\footnote{Light from neighboring stars can be circularly polarized after crossing a region of the interstellar medium where non-spherical dust grains are aligned under the influence of a large scale magnetic field.} light may constitute the asymmetry seed needed to break the symmetry between enantiomeric forms of a given compounds (i.e. D and L amino acids or sugars). Ultraviolet circularly polarized light does indeed generate a small enantiomeric excess in an initially racemic mixture through asymmetric photochemical destruction of L and D enantiomers at different rates, i.e. asymmetric photolysis. In other words, the kinetic constants for the photolysis of L and D enantiomers through irradiation by circularly polarized light are slightly different. Both enantiomers are therefore destroyed, but one of them with a rate that is slightly lower than the other. In this model, asymmetry on the molecular scale results from asymmetry on the astronomical scale. According to such a scenario, the homochirality should affect the solar neighborhood (and not only the Earth) in the same sense. However, an opposite homochirality cannot be completely ruled out in other regions of our Galaxy, or elsewhere in the Universe. The general scenario is summarized in Fig.\,\ref{asymphot}. Useful references for this scenario are \citet{jorissen1} and \citet{jorissen2}, along with \citet{nuevo2006}, \citet{takano2007} and \citet{chiralcpl2011} for reports on laboratory studies aiming at producing small enantiomeric excesses through the action of circularly polarized light.\\

Once the racemic mixture is converted into a scalemic mixture (i.e. with a small enantiomeric excess), we are still very far from homochirality. It has been proposed that the enantiomeric asymmetry may be amplified by some kind of autocatalytic process, favoring the synthesis of a particular enantiomer in the presence of a given enantiomeric excess. Such an asymmetric autocatalysis has been described for instance by \citet{asymautocata}. It is indeed known -- and in several cases understood -- that the incorporation of a chiral agent in a process can lead to an enantioselective synthesis. To some extent, we could say that {\it asymmetry is contagious}. On the other hand, one can also envisage that the incorporation of small enantiomeric excesses in living organisms may lead to a progressive selectivity in the incorporation of one particular enantiomer. The latter scenario is also worth considering in order to investigate the complex question of homochirality, but it is still a matter of debate.\\

\subsubsection{Racemization processes}
Beside the existence of processes potentially responsible for the breaking of the racemic symmetry, it should be mentioned that processes may also contribute to restitute this symmetry: i.e. {\it racemization processes}. The importance of racemization processes is discussed by \citet{cataldo-racem}. Typically, racemization denotes any reaction which converts enantiomers until an equimolar mixture of the two enantiomeric forms is achieved. In other words, it is a process that converts a scalemic mixture into a racemic mixture. In this context, we may consider mainly \citep{cataldo-racem}:
\begin{enumerate}
\item[-] action of heat, even though heat has only little chance to be high enough to redistribute chemical bondings in most astrophysical environments.
\item[-] action of light, i.e. photoracemization, through UV or X-ray photons
\item[-] action of high energy particles, i.e. radioracemization. This is typically what is likely to occur through the interaction of cosmic-rays with molecules. In the particular case of obscured interstellar clouds, this process is more likely than photoracemization.
\end{enumerate}
Schematically, racemization can be illustrated by the following relations
$$ 2\,\mathrm{R} \longrightarrow \mathrm{R} + \mathrm{S} $$
or
$$ 2\,\mathrm{S} \longrightarrow \mathrm{R} + \mathrm{S} $$
where R and S\footnote{R and S configurations are usually used to differentiate the two enantiomeric forms of a given chiral molecule, according to the Cahn-Ingold-Prelog sequence rules \citep{CIP}.} represent respectively the two enantiomers of a given chiral molecule presenting one stereogenic center.

On Earth, in aqueous solution, chiral molecules such as most of the natural amino acids are likely to undergo a unimolecular transformation called tautomerization. If one considers the natural chiral amino acids, the chiral carbon is bonded to one hydrogen. The tautomerization takes place when this $\alpha$-hydrogen\footnote{The $\alpha$ position indicates the carbon adjacent to the carbonyl group.} is removed before the breaking of the double bond and the formation of a new double bond between the $\alpha$ and $\beta$ carbons to yield an enol. This enol is not chiral anymore, and can convert to the carboxylic acid with equal probability to give a $R$ or $S$ configuration at the stereogenic center on the $\alpha$ position. This tautomerization process has thus the net result to convert a scalemic mixture into a racemic one. The occurrence of this process should not be ruled out in astrophysical environments as well, e.g. on dust grains or in icy mantles of dust grain, or even in comets. However, it should be noted that the carbonyl form is a priori intrinsically much more stable, and actually in solution the tautomerization requires a basic catalysis. The enol form may however be more stable if the double C-C bonding participates in a conjugated system. In such a case, tautomerization is not expected to be a very efficient racemization process.

It should be noted that, as far as amino acids are concerned, tautomerization may only be effective if one is dealing with $\alpha$-H-amino acids, such as natural amino acids. In the case of $\alpha$-methyl amino acids where there is no hydrogen bonded to the chiral carbon, the tautomerization cannot occur. This is the main reason why $\alpha$-Me-amino acids are likely to retain their configuration more efficiently than $\alpha$-H-amino acids. This might be the a likely reason to explain the rather high enantiomeric excesses that have been found in $\alpha$-methyl amino acids in meteorites by \citet{eemeteorites2009}. However, one should be very carefull when addressing the issue of chirality, and enantiomeric excesses in meteorites: the main concern here is the question of contamination by Earth material, and chemical alteration after the deposit of the meteoritic material on our planet's surface \citep[see e.g.][for a discussion of this issue in the particular case of the Tagish Lake Meteorite]{herdscience}.\\

The {\it homochirality problem} is a complex issue, that is far from being solved by modern science. However, it seems that some relevant elements may be found in interstellar regions, and not necessarily only on Earth even if prebiotic ages are considered. This longstanding issue still needs to be investigated in detail, and most probably following a pluridisciplinary approach.

\section{Concluding remarks}

The evolution of molecular material from the simplest species to the more complex ones is still a matter of debate and of intensive research. However, wealth of discoveries and theoretical developments allow to sketch a first global view of the physico-chemical processes at work in various astrophysical environments, and this was briefly summarized in the present paper. The so-called astro-molecules span a wide range of structural and chemical properties, from the simplest homonucleear species to the more complex polyatomic organic molecules presenting more than one functional group. Many rungs of the molecular complexity ladder have been climbed. 

Even though one may consider at first sight that a huge part of the effort has been accomplished to understand the filiation between molecules of various complexity, the trek has just begun. The investigation of the issue of molecular complexity will, fortunately, benefit of significant advances in the forthcoming years. First of all, observational facilities likely to reveal the signature of many interstellar and interplanetary molecules are being developed, with unprecedented performances. As a striking example, let us consider the Atacama Large Millimeter Array (ALMA\footnote{\tt http://almascience.eso.org/}), an international project in partnership between the European Southern Observatory (ESO), North America and Japan (with a contribution from Chile). ALMA will in some way open the eyes of Humanity on the molecular Universe, with much more clarity and sensitivity than previous observational facilities. On the other hand, substantial developments in the modelling of gas-phase and grain surface chemistry in astrophysical environments, with ceaseless and relentless enhancement of the sophistication level, lead almost every day to improvements in our understanding of astrochemistry. Finally, one should not neglect the interest of laboratory experiments (either ground-based or space-borne) aiming at reproducing physico-chemical processes of astrochemical interest to improve our understanding of molecular transformation in space environments.

The exploration of molecular complexity in astrophysical environments is a hard and demanding mission, that was triggered by the first interstellar indentification of an interstellar molecule in 1930's. And since the epoch of this first discovery, progress made by the scientific community to understand astrochemical processes never ceased to accelerate, and the adventure is far from having come to its end. 

 \section*{Acknowledgements}
This is a pleasure to thank the students of the Master in Space Sciences, and a few students of the Master in Chemistry and of the Master in Geology, who followed courses devoted to astrochemistry at the University of Li\`ege. Their participation in the lectures and fruitful discussions significantly contributed to drive a strong motivation to explore more deeply the fascinating world of astrochemistry. The ADS database was used for the bibliography, and the UMIST database was consulted to provide the reaction rate constant examples.


\begin{thebibliography}{10}

\bibitem[{{Adams} {et~al.}(1984){Adams}, {Smith}, \& {Millar}}]{ASM-kinexc}
{Adams}, N.~G., {Smith}, D., \& {Millar}, T.~J. 1984, \mnras, 211, 857

\bibitem[{{Ag\'undez} {et~al.}(2010){Ag\'undez}, {Cernicharo}, {Gu\'elin}, {Kahane}, {Roueff}, {K{\l}os}, {Aoiz}, {Lique}, {Marcelino}, {Goicoechea}, {Gonz\'alez Garc\'{\i}a}, {Gottlieb}, {McCarthy}, \& {Thaddeus}}]{cnanion}
{Ag\'undez}, M., {Cernicharo}, J., {Gu\'elin}, M., {et~al.} 2010, \aap, 517, L2

\bibitem[{{Aharonian} {et~al.}(2005){Aharonian}, {Akhperjanian}, {Bazer-Bachi},
  {et~al.}}]{aharonian2005}
{Aharonian}, F., {Akhperjanian}, A.G., {Bazer-Bachi}, A.R. {et~al.} 2005, \aap,
  437, L7

\bibitem[{{Allamandola} \& {Norman}(1978)}]{allam-pah1}
{Allamandola}, L.~J. \& {Norman}, C.~A. 1978, \aap, 66, 129

\bibitem[{{Allamandola} {et~al.}(1985){Allamandola}, {Tielens}, \&
  {Barker}}]{allam-pah2}
{Allamandola}, L.~J., {Tielens}, A.~G.~G.~M., \& {Barker}, J.~R. 1985, \apjl,
  290, L25

\bibitem[{{Allamandola} {et~al.}(1989){Allamandola}, {Tielens}, \&
  {Barker}}]{allam-pah3}
{Allamandola}, L.~J., {Tielens}, G.~G.~M., \& {Barker}, J.~R. 1989, \apjs, 71,
  733
  
\bibitem[{{Ball} {et~al.}(1970){Ball}, {Gottlieb}, {Lilley}, \&
  {Radford}}]{ball-methanol}
{Ball}, J.~A., {Gottlieb}, C.~A., {Lilley}, A.~E., \& {Radford}, H.~E. 1970,
  \apjl, 162, L203
  
  \bibitem[{{Bell} {et~al.}(1997){Bell}, {Feldman}, {Travers}, {McCarthy},
  {Gottlieb}, \& {Thaddeus}}]{hc11n}
{Bell}, M.~B., {Feldman}, P.~A., {Travers}, M.~J., {et~al.} 1997, \apjl, 483,
  L61
  
\bibitem[{{Belloche} {et~al.}(2008){Belloche}, {Menten}, {Comito},
  {M{\"u}ller}, {Schilke}, {Ott}, {Thorwirth}, \& {Hieret}}]{aminoacetonitrile}
{Belloche}, A., {Menten}, K.~M., {Comito}, C., {et~al.} 2008, \aap, 482, 179

\bibitem[{{Bern{\'e}} \& {Tielens}(2011)}]{formc602011}
{Bern{\'e}}, O. \& {Tielens}, A.~G.~G.~M. 2011, PNAS, 109, 401

\bibitem[{{Bernstein} {et~al.}(2002){Bernstein}, {Dworkin}, {Sandford}, {Cooper}, \& {Allamandola}}]{AAiceanalogues}
{Bernstein}, M.~P., {Dworkin}, J.~P., {Sandford}, S.~A., {et~al.} 2002, Nature, 416, 401

\bibitem[{{Bockel{\'e}e-Morvan} {et~al.}(2000){Bockel{\'e}e-Morvan}, {Lis},
  {Wink}, {Despois}, {Crovisier}, {Bachiller}, {Benford}, {Biver}, {Colom},
  {Davies}, {G{\'e}rard}, {Germain}, {Houde}, {Mehringer}, {Moreno}, {Paubert},
  {Phillips}, \& {Rauer}}]{comets-bockelee}
{Bockel{\'e}e-Morvan}, D., {Lis}, D.~C., {Wink}, J.~E., {et~al.} 2000, \aap,
  353, 1101
  
\bibitem[{{Bohme}(1992)}]{pah-bohme}
{Bohme}, D.~K. 1992, Chem. Rev., 92, 1487

\bibitem[{{Brown} {et~al.}(1975){Brown}, {Crofts}, {Godfrey}, {Gardner},
  {Robinson}, \& {Whiteoak}}]{brown-meform}
{Brown}, R.~D., {Crofts}, J.~G., {Godfrey}, P.~D., {et~al.} 1975, \apjl, 197,
  L29
  
\bibitem[{{Cahn} {et~al.}(1966){Cahn}, {Ingold}, \&
  {Prelog}}]{CIP} 
{Cahn}, R.~S., {Ingold}, C., \& {Prelog}, V. 1966, Angewandte Chemie, 5, 385

\bibitem[{{Cami} {et~al.}(2010){Cami}, {Bernard-Salas}, {Peeters}, \&
  {Malek}}]{fullercami}
{Cami}, J., {Bernard-Salas}, J., {Peeters}, E., \& {Malek}, S.~E. 2010,
  Science, 329, 1180
  
\bibitem[{{Capriotti} \& {Kozminski}(2001)}]{capriotti2001}
{Capriotti}, E.~R. \& {Kozminski}, J.~F. 2001, \pasp, 113, 677  

\bibitem[{{Cataldo} {et~al.}(2005){Cataldo}, {Brucato}, \&
  {Keheyan}}]{cataldo-racem}
{Cataldo}, F., {Brucato}, J.~R., \& {Keheyan}, Y. 2005, Journal of Physics:
  Conf. Series, 6, 139

\bibitem[{{Cazaux} \& {Tielens}(2004)}]{cazaux}
{Cazaux}, S. \& {Tielens}, A.~G.~G.~M. 2004, \apj, 604, 222

\bibitem[{{Cerf} \& {Jorissen}(2000)}]{jorissen1}
{Cerf}, C. \& {Jorissen}, A. 2000, Space Science Reviews, 92, 603

\bibitem[{{Cernicharo} \& {Guelin}(1996)}]{c8h}
{Cernicharo}, J. \& {Guelin}, M. 1996, \aap, 309, L27

\bibitem[{{Cernicharo} {et~al.}(2001){Cernicharo}, {Heras}, {Tielens}, {Pardo},
  {Herpin}, {Gu{\'e}lin}, \& {Waters}}]{ism-benzene}
{Cernicharo}, J., {Heras}, A.~M., {Tielens}, A.~G.~G.~M., {et~al.} 2001, \apjl,
  546, L123
  
\bibitem[{{Cernicharo} {et~al.}(2007){Cernicharo}, {Gu{\'e}lin}, {Ag{\"u}ndez},
  {Kawaguchi}, {McCarthy}, \& {Thaddeus}}]{secondanion}
{Cernicharo}, J., {Gu{\'e}lin}, M., {Ag{\"u}ndez}, M., {et~al.} 2007, \aap,
  467, L37
  
\bibitem[{{Charnley} \& {Rodgers}(2008)}]{charrodg2008}
{Charnley}, S.~B. \& {Rodgers}, S.~D. 2008, Space Science Reviews, 138, 59

\bibitem[{{Cheung} {et~al.}(1968){Cheung}, {Rank}, {Townes}, {Thornton}, \&
  {Welch}}]{cheung-nh3}
{Cheung}, A.~C., {Rank}, D.~M., {Townes}, C.~H., {Thornton}, D.~D., \& {Welch},
  W.~J. 1968, Physical Review Letters, 21, 1701
  
\bibitem[{{Churchwell} \& {Winnewisser}(1975)}]{churchwell-meform}
{Churchwell}, E. \& {Winnewisser}, G. 1975, \aap, 45, 229

\bibitem[{{Combes} {et~al.}(1987){Combes}, {Gerin}, {Wootten}, {Wlodarczak},
  {Clausset}, \& {Encrenaz}}]{combes-acetone}
{Combes}, F., {Gerin}, M., {Wootten}, A., {et~al.} 1987, \aap, 180, L13

\bibitem[{{Cooper} {et~al.}(2001){Cooper}, {Kimmich}, {Belisle}, {Sarinana},
  {Brabham}, \& {Garrel}}]{murchison3}
{Cooper}, G., {Kimmich}, N., {Belisle}, W., {et~al.} 2001, \nat, 414, 879

\bibitem[{{Cronin} \& {Pizzarello}(1997)}]{murchison}
{Cronin}, J.~R. \& {Pizzarello}, S. 1997, Science, 275, 951

\bibitem[{{Crovisier}(1997)}]{comets-ir}
{Crovisier}, J. 1997, Earth Moon and Planets, 79, 125

\bibitem[{{Crovisier} \& {Bockel{\'e}e-Morvan}(1997)}]{comets-mm}
{Crovisier}, J. \& {Bockel{\'e}e-Morvan}, D. 1997, in ESA Special Publication,
  Vol. 401, The Far Infrared and Submillimetre Universe., ed. A.~{Wilson},
  45

\bibitem[{{De Becker}(2007)}]{debeckerreview}
{De Becker}, M. 2007,  A\&A Rev., 14, 171

\bibitem[{{de Marcellus} {et~al.}(2011){de Marcellus}, {Meinert}, {Nuevo},
  {Filippi}, {Danger}, {Deboffle}, {Nahon}, {Le Sergeant d'Hendecourt}, \&
  {Meierhenrich}}]{chiralcpl2011}
{de Marcellus}, P., {Meinert}, C., {Nuevo}, M., {et~al.} 2011, \apjl, 727, L27

\bibitem[{{Dickens} {et~al.}(1997){Dickens}, {Irvine}, {Ohishi}, {Ikeda},
  {Ishikawa}, {Nummeli[n}, \& {Hjalmarson}}]{dickens-ethdiox}
{Dickens}, J.~E., {Irvine}, W.~M., {Ohishi}, M., {et~al.} 1997, in Bulletin of
  the American Astronomical Society, Vol.~29, 1245
  
\bibitem[{{Dickman}(1978)}]{dickman-h2-co}
{Dickman}, R.~L. 1978, \apjs, 37, 407

\bibitem[{{Douglas} \& {Herzberg}(1941)}]{ch+-douglas}
{Douglas}, A.~E. \& {Herzberg}, G. 1941, \apj, 94, 381

\bibitem[{{Duley} \& {Williams}(1981)}]{dw-pah}
{Duley}, W.~W. \& {Williams}, D.~A. 1981, \mnras, 196, 269

\bibitem[{{Drury}(2012)}]{drury2012}
{Drury}, L.~O. 2012, Astroparticle Physics, 39, 52

\bibitem[{{Elsila} {et~al.}(2009){Elsila}, {Glavin}, \&
  {Dworkin}}]{glystardust}
{Elsila}, J.~E., {Glavin}, D.~P., \& {Dworkin}, J.~P. 2009, Meteoritics and
  Planetary Science, 44, 1323

\bibitem[{{Engel} \& {Macko}(1997)}]{murchison2}
{Engel}, M.~H. \& {Macko}, S.~A. 1997, \nat, 389, 265

\bibitem[{{Fourikis} {et~al.}(1974{\natexlab{a}}){Fourikis}, {Sinclair},
  {Robinson}, {Godfrey}, \& {Brown}}]{fourikis-acetaldehyde}
{Fourikis}, N., {Sinclair}, M.~W., {Robinson}, B.~J., {Godfrey}, P.~D., \&
  {Brown}, R.~D. 1974{\natexlab{a}}, Australian Journal of Physics, 27, 425

\bibitem[{{Fourikis} {et~al.}(1974{\natexlab{b}}){Fourikis}, {Takagi}, \&
  {Morimoto}}]{fourikis-meamine}
{Fourikis}, N., {Takagi}, K., \& {Morimoto}, M. 1974{\natexlab{b}}, \apjl, 191,
  L139

\bibitem[{{Freyer} {et~al.}(2003){Freyer}, {Hensler}, \&
  {Yorke}}]{freyer2003}
{Freyer}, T., {Hensler}, G., \& {Yorke}, H.W. 2003, \apj, 594, 888

\bibitem[{{Geballe} {et~al.}(1989){Geballe}, {Tielens}, {Allamandola},
  {Moorhouse}, \& {Brand}}]{geballe-pah}
{Geballe}, T.~R., {Tielens}, A.~G.~G.~M., {Allamandola}, L.~J., {Moorhouse},
  A., \& {Brand}, P.~W.~J.~L. 1989, \apj, 341, 278
  
\bibitem[{{Gielen} {et~al.}(2011){Gielen}, {Cami}, {Bouwman}, {Peeters}, \&
  {Min}}]{gielen2011}
{Gielen}, C., {Cami}, J., {Bouwman}, J., {Peeters}, E., \& {Min}, M. 2011,
  \aap, 536, A54

\bibitem[{{Gillett} {et~al.}(1973){Gillett}, {Forrest}, \& {Merrill}}]{gfm-pah}
{Gillett}, F.~C., {Forrest}, W.~J., \& {Merrill}, K.~M. 1973, \apj, 183, 87

\bibitem[{{Gilmore} {et~al.}(1976){Gilmore}, {Morris}, {Palmer}, {Johnson},
  {Lovas}, {Turner}, \& {Zuckerman}}]{gilmore-acetaldehyde}
{Gilmore}, W., {Morris}, M., {Palmer}, P., {et~al.} 1976, \apj, 204, 43

\bibitem[{{Glavin} \& {Dworkin}(2009)}]{eemeteorites2009}
  {Glavin}, D.~P., \& {Dworkin}, J.~P. 2009, Meteoritics and
  Planetary Science Suppl., 5009

\bibitem[{{Gottlieb}(1973)}]{gottlieb-acetaldehyde}
{Gottlieb}, C.~A. 1973, in Molecules in the Galactic Environment, ed. M.~A.
  {Gordon} \& L.~E. {Snyder}, 181
  
\bibitem[{{Guelin} {et~al.}(1997){Guelin}, {Cernicharo}, {Travers}, {McCarthy},
  {Gottlieb}, {Thaddeus}, {Ohishi}, {Saito}, \& {Yamamoto}}]{c7h}
{Guelin}, M., {Cernicharo}, J., {Travers}, M.~J., {et~al.} 1997, \aap, 317, L1  
  
\bibitem[{{Gu{\'e}lin} {et~al.}(2000){Gu{\'e}lin}, {Muller}, {Cernicharo},
  {Apponi}, {McCarthy}, {Gottlieb}, \& {Thaddeus}}]{guelin-sicn}
{Gu{\'e}lin}, M., {Muller}, S., {Cernicharo}, J., {et~al.} 2000, \aap, 363, L9

\bibitem[{{Helder} {et~al.}(2012){Helder}, {Vink}, {Bykov}, {Ohira}, {Raymond},
  \& {Terrier}}]{helderpasnr2012}
{Helder}, E.~A., {Vink}, J., {Bykov}, A.~M., {et~al.} 2012, Space Sci. Rev., 173, 369

\bibitem[{{Herd} {et~al.}(2011){Herd}, {Blinova}, {Simkus},
  {Huang}, {Tarozo}, {Alexander}, {Gyngard}, {et~al.}}]{herdscience}
{Herd}, C.~D.~K., {Blinova}, A., {Simkus}, D.~N., {et~al.} 2011, Science, 332, 1304

\bibitem[{{Hillas}(1984)}]{hillas1984}
{Hillas}, A.~M. 1984, \araa, 22, 425
  
\bibitem[{{Hollis} {et~al.}(2000){Hollis}, {Lovas}, \& {Jewell}}]{firstsugar}
{Hollis}, J.~M., {Lovas}, F.~J., \& {Jewell}, P.~R. 2000, \apjl, 540, L107

\bibitem[{{Hollis} {et~al.}(2002){Hollis}, {Lovas}, {Jewell}, \&
  {Coudert}}]{antifreeze}
{Hollis}, J.~M., {Lovas}, F.~J., {Jewell}, P.~R., \& {Coudert}, L.~H. 2002,
  \apjl, 571, L59

\bibitem[{{Iglesias-Groth} {et~al.}(2011){Iglesias-Groth}, {Manchado}, {Rebolo}, {Gonzalez-Hernandez},
  {Garcia-Hernandez}, \& {Lambert}}]{anthracenecation}
{Iglesias-Groth}, S., {Manchado}, A., {Rebolo}, R., {et~al.} 2011, \mnras, 407, 2157 

\bibitem[{{Jorissen} \& {Cerf}(2002)}]{jorissen2}
{Jorissen}, A. \& {Cerf}, C. 2002, Origins of Life and Evolution of the
  Biosphere, 32, 129

\bibitem[{{Kaifu} {et~al.}(1974){Kaifu}, {Morimoto}, {Nagane}, {Akabane},
  {Iguchi}, \& {Takagi}}]{kaifu-meamine}
{Kaifu}, N., {Morimoto}, M., {Nagane}, K., {et~al.} 1974, \apjl, 191, L135

\bibitem[{{Koyama} {et~al.}(1995){Koyama}, {Petre}, {Gotthelf}, {Hwang},
  {Matsuura}, {Ozaki} \& {Holt}}]{koyama1995}
{Koyama}, K., {Petre}, R., {Gotthelf}, E.V., {et~al.} 1995, Nature, 378, 255

\bibitem[{{Kroto} {et~al.}(1985){Kroto}, {Heath}, {Obrien}, {Curl}, \&
  {Smalley}}]{fullerkroto}
{Kroto}, H.~W., {Heath}, J.~R., {Obrien}, S.~C., {Curl}, R.~F., \& {Smalley},
  R.~E. 1985, \nat, 318, 162

\bibitem[{{Kuan} {et~al.}(2003{\natexlab{a}}){Kuan}, {Charnley}, {Huang},
  {Tseng}, \& {Kisiel}}]{kuan-glycine}
{Kuan}, Y.-J., {Charnley}, S.~B., {Huang}, H.-C., {Tseng}, W.-L., \& {Kisiel},
  Z. 2003{\natexlab{a}}, \apj, 593, 848

\bibitem[{{Kuan} {et~al.}(2003{\natexlab{b}}){Kuan}, {Yan}, {Charnley},
  {Kisiel}, {Ehrenfreund}, \& {Huang}}]{kuan-pyrimidine}
{Kuan}, Y.-J., {Yan}, C.-H., {Charnley}, S.~B., {et~al.} 2003{\natexlab{b}},
  \mnras, 345, 650

\bibitem[{{Lacy} {et~al.}(1991){Lacy}, {Carr}, {Evans}, {Baas}, {Achtermann},
  \& {Arens}}]{lacy-ch4}
{Lacy}, J.~H., {Carr}, J.~S., {Evans}, II, N.~J., {et~al.} 1991, \apj, 376, 556

\bibitem[{{Larralde} {et~al.}(1995){Larralde}, {Robertson},
  \& {Miller}}]{formosereaction}
{Larralde}, R., {Robertson}, M.~P., \& {Miller}, S. 1995, Proc. Natl. Acad. Sci, 92, 8158

\bibitem[{{Leger} \& {Puget}(1984)}]{lp-pah}
{Leger}, A. \& {Puget}, J.~L. 1984, \aap, 137, L5

\bibitem[{{Linke} {et~al.}(1979){Linke}, {Frerking}, \&
  {Thaddeus}}]{linke-mercaptan}
{Linke}, R.~A., {Frerking}, M.~A., \& {Thaddeus}, P. 1979, \apjl, 234, L139

\bibitem[{{Lovas} {et~al.}(2006){Lovas}, {Hollis}, {Remijan}, \&
  {Jewell}}]{lovas-ketenimine}
{Lovas}, F.~J., {Hollis}, J.~M., {Remijan}, A.~J., \& {Jewell}, P.~R. 2006,
  \apjl, 645, L137
  
\bibitem[{{Martins} {et~al.}(2008){Martins}, {Botta}, {Fogel}, {Sephton},
  {Glavin}, {Watson}, {Dworkin}, {Schwartz}, \&
  {Ehrenfreund}}]{martins-nubasesmurchison}
{Martins}, Z., {Botta}, O., {Fogel}, M.~L., {et~al.} 2008, Earth and Planetary
  Science Letters, 270, 130
  
\bibitem[{{Matthews}(1995{\natexlab{a}})}]{matthews-hcn}
{Matthews}, C. 1995{\natexlab{a}}, in Astronomical Society of the Pacific
  Conference Series, Vol.~74, Progress in the Search for Extraterrestrial
  Life., ed. G.~S. {Shostak}, 95

\bibitem[{{Matthews}(1995{\natexlab{b}})}]{matthews-hcn2}
{Matthews}, C.~N. 1995{\natexlab{b}}, \planss, 43, 1365
  
\bibitem[{{McCarthy} {et~al.}(2006){McCarthy}, {Gottlieb}, {Gupta}, \&
  {Thaddeus}}]{firstanion}
{McCarthy}, M.~C., {Gottlieb}, C.~A., {Gupta}, H., \& {Thaddeus}, P. 2006,
  \apjl, 652, L141  

\bibitem[{{McKellar}(1940)}]{ism-cn}
{McKellar}, A. 1940, \pasp, 52, 187

\bibitem[{{McMurry}(1992)}]{mcmurry}
{McMurry}, J. 1992, {Organic chemistry, 3rd ed.}, Brooks/Cole Publishing Company, 1992

\bibitem[{{Mehringer} {et~al.}(1997){Mehringer}, {Snyder}, {Miao}, \&
  {Lovas}}]{mehringer-acetic}
{Mehringer}, D.~M., {Snyder}, L.~E., {Miao}, Y., \& {Lovas}, F.~J. 1997, \apjl,
  480, L71

\bibitem[{{Mil'man}(1989)}]{milman}
{Mil'man}, B.~L. 1989, Journal of structural chemistry, 29, 957

\bibitem[{{Millar}(1992)}]{millar-pah}
{Millar}, T.~J. 1992, \mnras, 259, 35P

\bibitem[{{Mimura}(1995)}]{pah-shock}
{Mimura}, K. 1995, \gca, 59, 579

\bibitem[{{Munoz Caro} {et~al.}(2002){Munoz Caro}, {Meierhenrich}, {Schutte}, {Barbier}, {Arcones Segovia}, {Rosenbauer}, {Thiemann}, {Brack} \& {Greenberg}}]{AAiceanalogues2}
{Munoz Caro}, G.~M., {Meierhenrich}, U.~J., {Schutte}, W.~A., {et~al.} 2002, Nature, 416, 403

\bibitem[{{Nuevo} {et~al.}(2006){Nuevo}, {Meierhenrich}, {Mu{\~n}oz Caro},
  {Dartois}, {D'Hendecourt}, {Deboffle}, {Auger}, {Blanot}, {Bredeh{\"o}ft}, \&
  {Nahon}}]{nuevo2006}
{Nuevo}, M., {Meierhenrich}, U.~J., {Mu{\~n}oz Caro}, G.~M., {et~al.} 2006,
  \aap, 457, 741
  
\bibitem[{{Pagani} {et~al.}(2010){Pagani}, {Steinacker}, {Bacmann}, {Stutz}, \&
  {Henning}}]{pagani2010}
{Pagani}, L., {Steinacker}, J., {Bacmann}, A., {Stutz}, A., \& {Henning}, T.
  2010, Science, 329, 1622

\bibitem[{{Peeters} {et~al.}(2003){Peeters}, {Botta}, {Charnley}, {Ruiterkamp}, \& {Ehrenfreund}}]{nheteroform}
{Peeters}, Z., {Botta}, O., {Charnley}, S.~B., {et~al.} 2003, \apj, 593, L129
  
\bibitem[{{Peeters} {et~al.}(2005){Peeters}, {Botta}, {Charnley}, {Kisiel}, {Kuan}, \& {Ehrenfreund}}]{nheterostab}
{Peeters}, Z., {Botta}, O., {Charnley}, S.~B., {et~al.} 2005, \aap, 433, 583

\bibitem[{{Peeters} {et~al.}(2006){Peeters}, {Rodgers}, {Charnley},
  {Schriver-Mazzuoli}, {Schriver}, {Keane}, \& {Ehrenfreund}}]{peeters-dme}
{Peeters}, Z., {Rodgers}, S.~D., {Charnley}, S.~B., {et~al.} 2006, \aap, 445,
  197

\bibitem[{{Philips} \& {Huggins}(1981)}]{philhugg}
{Philips}, T.~G. \& {Huggins}, P.~J. 1981, \apj, 251, 533

\bibitem[{{Prasad} \& {Huntress}(1980{\natexlab{a}})}]{prasad-cno}
{Prasad}, S.~S. \& {Huntress}, Jr., W.~T. 1980{\natexlab{a}}, \apjs, 43, 1

\bibitem[{{Prasad} \& {Huntress}(1980{\natexlab{b}})}]{prasad-cno2}
{Prasad}, S.~S. \& {Huntress}, Jr., W.~T. 1980{\natexlab{b}}, \apj, 239, 151

\bibitem[{{Prasad} \& {Tarafdar}(1983)}]{PT-CR}
{Prasad}, S.~S. \& {Tarafdar}, S.~P. 1983, \apj, 267, 603

\bibitem[{{Reimer} {et~al.}(2006){Reimer}, {Pohl}, \&
  {Reimer}}]{RPR2006}
{Reimer}, A., {Pohl}, M., \& {Reimer}, O. 2006, \apj,
  644, 1118

\bibitem[{{Remijan} {et~al.}(2007){Remijan}, {Hollis}, {Lovas}, {Cordiner},
  {Millar}, {Markwick-Kemper}, \& {Jewell}}]{thirdanion}
{Remijan}, A.~J., {Hollis}, J.~M., {Lovas}, F.~J., {et~al.} 2007, \apjl, 664,
  L47

\bibitem[{{Remijan} {et~al.}(2008){Remijan}, {Hollis}, {Lovas}, {Stork},
  {Jewell}, \& {Meier}}]{cyanoform}
{Remijan}, A.~J., {Hollis}, J.~M., {Lovas}, F.~J., {et~al.} 2008, \apjl, 675,
  L85
  
\bibitem[{{Rettig} {et~al.}(1992){Rettig}, {Tegler}, {Pasto}, \&
  {Mumma}}]{rettig-hcn}
{Rettig}, T.~W., {Tegler}, S.~C., {Pasto}, D.~J., \& {Mumma}, M.~J. 1992, \apj,
  398, 293
  
\bibitem[{{Rimola} {et~al.}(2010){Rimola}, {Sodupe}, \&
  {Ugliengo}}]{strecker2010}
{Rimola}, A., {Sodupe}, M., \& {Ugliengo}, P. 2010, Physical Chemistry Chemical
  Physics (Incorporating Faraday Transactions), 12, 5285
  
\bibitem[{{Rubin} {et~al.}(1971){Rubin}, {Swenson}, {Benson}, {Tigelaar}, \&
  {Flygare}}]{rubin-formamide}
{Rubin}, R.~H., {Swenson}, Jr., G.~W., {Benson}, R.~C., {Tigelaar}, H.~L., \&
  {Flygare}, W.~H. 1971, \apjl, 169, L39
  
\bibitem[{{Sandford} {et~al.}(1991){Sandford}, {Allamandola}, {Tielens},
  {Sellgren}, {Tapia}, \& {Pendleton}}]{sandfordratio}
{Sandford}, S.~A., {Allamandola}, L.~J., {Tielens}, A.~G.~G.~M., {et~al.} 1991,
  \apj, 371, 607  
  
  \bibitem[{{Sellgren} {et~al.}(2007){Sellgren}, {Uchida}, \&
  {Werner}}]{c60first2007}
{Sellgren}, K., {Uchida}, K.~I., \& {Werner}, M.~W. 2007, \apj, 659, 1338

\bibitem[{{Sellgren} {et~al.}(2010){Sellgren}, {Werner}, {Ingalls}, {Smith},
  {Carleton}, \& {Joblin}}]{c60snellgren}
{Sellgren}, K., {Werner}, M.~W., {Ingalls}, J.~G., {et~al.} 2010, \apjl, 722,
  L54
  
\bibitem[{{Shibata} {et~al.}(1998){Shibata}, {Yamamoto}, {Matsumoto},
  {Yonekubo}, \& {Osanai}}]{asymautocata}
{Shibata}, T., {Yamamoto}, J., {Matsumoto}, N., {Yonekubo}, S., \& {Osanai}, S.
  a.~K. 1998, J. Am. Chem. soc., 120, 12157

\bibitem[{{Smith} {et~al.}(2001){Smith}, {Talbi}, \& {Herbst}}]{hcnoligomers}
{Smith}, I.~W.~M., {Talbi}, D., \& {Herbst}, E. 2001, \aap, 369, 611

\bibitem[{{Snyder} \& {Buhl}(1971)}]{snyder-hcn}
{Snyder}, L.~E. \& {Buhl}, D. 1971, \apjl, 163, L47
  
\bibitem[{{Snyder} {et~al.}(1974){Snyder}, {Buhl}, {Schwartz}, {Clark},
  {Johnson}, {Lovas}, \& {Giguere}}]{snyder-ether}
{Snyder}, L.~E., {Buhl}, D., {Schwartz}, P.~R., {et~al.} 1974, \apjl, 191, L79+

\bibitem[{{Snyder} {et~al.}(1969){Snyder}, {Buhl}, {Zuckerman}, \&
  {Palmer}}]{snyder-formaldehyde}
{Snyder}, L.~E., {Buhl}, D., {Zuckerman}, B., \& {Palmer}, P. 1969, Physical
  Review Letters, 22, 679
  
\bibitem[{{Snyder} {et~al.}(2002){Snyder}, {Lovas}, {Mehringer}, {Miao},
  {Kuan}, {Hollis}, \& {Jewell}}]{snyder-acetone}
{Snyder}, L.~E., {Lovas}, F.~J., {Mehringer}, D.~M., {et~al.} 2002, \apj, 578,
  245

\bibitem[{{Snyder} {et~al.}(2005){Snyder}, {Lovas}, {Hollis}, {Friedel},
  {Jewell}, {Remijan}, {Ilyushin}, {Alekseev}, \& {Dyubko}}]{snyder-glycine}
{Snyder}, L.~E., {Lovas}, F.~J., {Hollis}, J.~M., {et~al.} 2005, \apj, 619, 914


\bibitem[{{Solomon} {et~al.}(1971){Solomon}, {Jefferts}, {Penzias}, \&
  {Wilson}}]{solomon-cyanome}
{Solomon}, P.~M., {Jefferts}, K.~B., {Penzias}, A.~A., \& {Wilson}, R.~W. 1971,
  \apjl, 168, L107
  
\bibitem[{{Stoks} \& {Schwartz}(1981)}]{stoks1}
{Stoks}, P.~G. \& {Schwartz}, A.~W. 1981, \gca, 45, 563

\bibitem[{{Stoks} \& {Schwartz}(1982)}]{stoks2}
{Stoks}, P.~G. \& {Schwartz}, A.~W. 1982, \gca, 46, 309

\bibitem[{{Str\"omgren}(1937)}]{stromgren}
{Str\"omgren}, B. 1937, \apj, 89, 526

\bibitem[{{Swings} \& {Rosenfeld}(1937)}]{ch-swings}
{Swings}, P. \& {Rosenfeld}, L. 1937, \apj, 86, 483

\bibitem[{{Takano} {et~al.}(2007){Takano}, {Takahashi}, {Kaneko}, {Marumo}, \&
  {Kobayashi}}]{takano2007}
{Takano}, Y., {Takahashi}, J., {Kaneko}, T., {Marumo}, K., \& {Kobayashi}, K.
  2007, Earth and Planetary Science Letters, 254, 106

\bibitem[{{Tielens}(2005)}]{tielens}
{Tielens}, A.~G.~G.~M. 2005, {The Physics and Chemistry of the Interstellar
  Medium}, Cambridge University Press, 2005
  
\bibitem[{{Turner} {et~al.}(1994){Turner}, {Steimle}, \&
  {Meerts}}]{turner-nacn}
{Turner}, B.~E., {Steimle}, T.~C., \& {Meerts}, L. 1994, \apjl, 426, L97

\bibitem[{{Werner} {et~al.}(2004){Werner}, {Uchida}, {Sellgren}, {Marengo},
  {Gordon}, {Morris}, {Houck}, \& {Stansberry}}]{c60first2004}
{Werner}, M.~W., {Uchida}, K.~I., {Sellgren}, K., {et~al.} 2004, \apjs, 154,
  309

\bibitem[{{Wilson} {et~al.}(1970){Wilson}, {Jefferts}, \& {Penzias}}]{WJP-co}
{Wilson}, R.~W., {Jefferts}, K.~B., \& {Penzias}, A.~A. 1970, \apjl, 161, L43
  
\bibitem[{{Winnewisser} \& {Churchwell}(1975)}]{winnewisser-formic}
{Winnewisser}, G. \& {Churchwell}, E. 1975, \apjl, 200, L33
  
\bibitem[{{Woodall} {et~al.}(2007){Woodall}, {Ag{\'u}ndez}, {Markwick-Kemper},
  \& {Millar}}]{umist}
{Woodall}, J., {Ag{\'u}ndez}, M., {Markwick-Kemper}, A.~J., \& {Millar}, T.~J.
  2007, \aap, 466, 1197
  
\bibitem[{{Zaleski} {et~al.}(2013){Zaleski}, {Seifert}, {Steber}, {Muckle},
  {Loomis}, {Corby}, {Martinez}, {Crabtree}, {Jewell}, {Hollis}, {Lovas},
  {Vasquez}, {Nyiramahirwe}, {Sciortino}, {Johnson}, {McCarthy}, {Remijan}, \&
  {Pate}}]{cyanomethanimine}
{Zaleski}, D.~P., {Seifert}, N.~A., {Steber}, A.~L., {et~al.} 2013, \apjl, 765,
  L10
  
\bibitem[{{Ziurys} {et~al.}(1995){Ziurys}, {Apponi}, {Guelin}, \&
  {Cernicharo}}]{ziurys-mgcn}
{Ziurys}, L.~M., {Apponi}, A.~J., {Guelin}, M., \& {Cernicharo}, J. 1995,
  \apjl, 445, L47
  
\bibitem[{{Zuckerman} {et~al.}(1971){Zuckerman}, {Ball}, \&
  {Gottlieb}}]{zuckerman-formic}
{Zuckerman}, B., {Ball}, J.~A., \& {Gottlieb}, C.~A. 1971, \apjl, 163, L41

\bibitem[{{Zuckerman} {et~al.}(1975){Zuckerman}, {Turner}, {Johnson}, {Lovas},
  {Fourikis}, {Palmer}, {Morris}, {Lilley}, {Ball}, \&
  {Clark}}]{zuckerman-ethanol}
{Zuckerman}, B., {Turner}, B.~E., {Johnson}, D.~R., {et~al.} 1975, \apjl, 196,
  L99

    
\end{thebibliography}


\end{document}